\documentclass[aps,prd,nofootinbib,twocolumn,superscriptaddress,preprintnumbers,balancelastpage,longbibliography]{revtex4-2}
\usepackage{aas_macros}

\usepackage{placeins}
\usepackage{amsmath,amssymb,mathtools,bm}
\usepackage{graphicx, color, hepunits}
\usepackage[dvipsnames]{xcolor}
\usepackage{cancel}
\usepackage[normalem]{ulem}
\usepackage{float}
 \usepackage{filecontents}
\usepackage{multirow}
 \usepackage{hyperref} 
\hypersetup{
    colorlinks=true,       
    linkcolor=blue,        
    citecolor=blue,        
    filecolor=magenta,     
    urlcolor=blue          
}
\usepackage[utf8]{inputenc}
\usepackage[english]{babel}
\usepackage{array}

\usepackage{tikz}
\usepackage{tikz-feynman}
\tikzfeynmanset{compat=1.1.0}

\newcommand{\gagg}{g_{a \gamma \gamma}}

\newcommand{\ganngagg}{g_{ann}\times g_{a \gamma \gamma}}

\newcommand{\gann}{g_{ann}}
\newcommand{\gapp}{g_{app}}

\newcommand{\es}[2] {\begin{equation} \label{#1} \begin{split} #2 \end{split} \end{equation}}

\begin{document}

\title{Cosmological Neutron Stars Produce Diffuse Axion X-Ray Signatures}

\author{Orion Ning}
\affiliation{Leinweber Institute for Theoretical Physics, University of California, Berkeley, CA 94720, U.S.A.}
\affiliation{Theoretical Physics Group, Lawrence Berkeley National Laboratory, Berkeley, CA 94720, U.S.A.}

\author{Kailash Raman}
\affiliation{Leinweber Institute for Theoretical Physics, University of California, Berkeley, CA 94720, U.S.A.}
\affiliation{Theoretical Physics Group, Lawrence Berkeley National Laboratory, Berkeley, CA 94720, U.S.A.}

\author{Benjamin R. Safdi}
\affiliation{Leinweber Institute for Theoretical Physics, University of California, Berkeley, CA 94720, U.S.A.}
\affiliation{Theoretical Physics Group, Lawrence Berkeley National Laboratory, Berkeley, CA 94720, U.S.A.}

\date{\today}

\begin{abstract}
Axion-like particles can be abundantly produced through scattering processes in the cores of neutron stars (NSs). If they are ultralight ($m_a \lesssim 10^{-4}$ eV), then they can efficiently convert to detectable photons in the external NS magnetospheres, and if they are heavy ($m_a \gtrsim 1$ eV), then they can decay into photons before reaching Earth. In this work, we search for the resulting X-ray signatures from both of these channels summing over the \textit{cosmological} NS population.  We compare the predicted axion-induced X-ray signal to the cosmic X-ray background today as measured by a number of instruments such as NuSTAR, HEAO, Swift, and INTEGRAL. We model the axion-induced signal using NS cooling simulations and magnetic field evolution models. We find no evidence for axions and derive strong constraints for both ultralight and heavy axion scenarios, covering new parameter space for the axion-photon and axion-nucleon couplings. Our results rule out the axion-explanation of the Magnificent Seven X-ray excess from nearby isolated NSs.
\end{abstract}
\maketitle

\begin{figure}[!htb]
\centering
\includegraphics[width=0.49\textwidth]{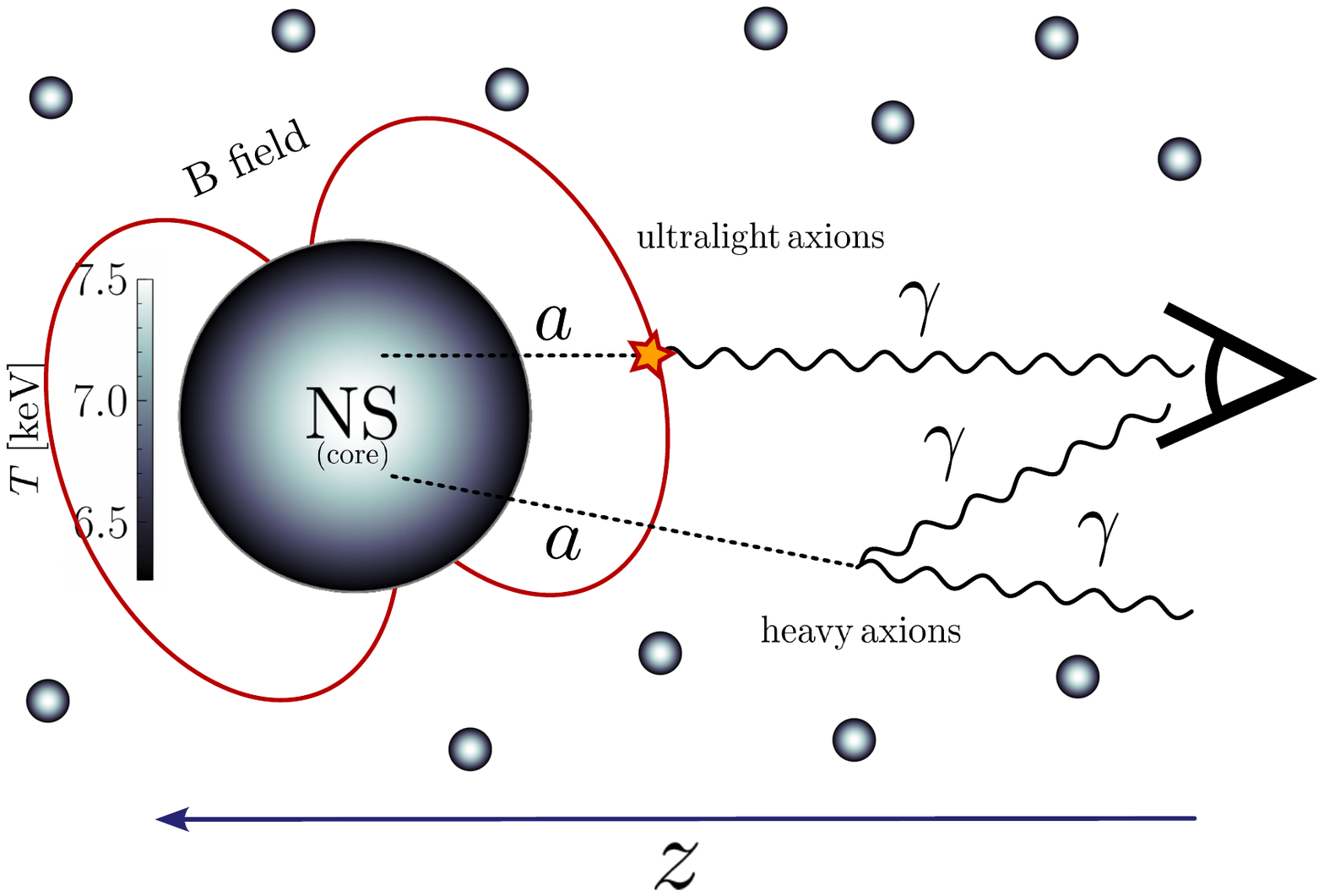}
\vspace{-0.4cm}
\caption{Axions are produced from the cores of NSs in our cosmological NS population. If the axions are ultralight, they can convert to photons in the magnetospheres of the host NS (upper process), and if the axions are heavy, they can spontaneously decay to photons at some later cosmic epoch (lower process). We search for both of these channels in cosmic X-ray background data, summing over the collective contributions from the cosmological NS population. The NS temperature distribution is illustrated for an example $t \sim 10^5$ yr NS.}
\label{fig:fig1}
\end{figure}

\noindent
{\bf Introduction.---}Neutron stars (NSs) are well-known laboratories for fundamental physics. Some of the strongest constraints on the quantum chromodynamics (QCD) axion and axion-like particles come from studies of NS cooling and X-ray and gamma-ray signals that could be induced from these systems due to axions~\cite{Iwamoto:1984ir,Brinkmann:1988vi,Keller:2012yr,Leinson:2014ioa,Sedrakian:2015krq,Sedrakian:2018kdm,Hamaguchi:2018oqw,Buschmann:2019pfp, Buschmann:2021juv, Leinson:2021ety,Jaeckel:2017tud,Hoof:2022xbe,Muller:2023vjm,Lella:2024dmx,Benabou:2024jlj,Raffelt:1996wa,Brockway:1996yr,Grifols:1996id,Payez:2014xsa,Hoof:2022xbe,Caputo:2024oqc,Manzari:2024jns,Fiorillo:2025gnd}.  For example, it is well known that axions could be produced in abundance in the cores of cooling NSs through nucleon bremsstrahlung and Cooper pair breaking and recombination processes. Given the feeble interaction strengths of axions, they then escape the cooling star, providing a source of energy loss in the same way as neutrinos. However, unlike neutrinos, ultralight axions may efficiently convert to observable photons in the strong magnetic fields surrounding the NSs or, in the case of heavier axions, undergo delayed decays to photon pairs. To-date, studies have focused on the search for axion-induced X-ray signals from nearby isolated NSs; for example, Ref.~\cite{Buschmann:2019pfp} showed that an excess of hard X-rays observed by the Chandra and XMM-Newton telescopes towards multiple of the nearby Magnificent Seven (M7) NSs~\cite{Dessert:2019dos} could be explained by ultralight axion production in the NS cores and the subsequent conversion in the NS magnetospheres. In this work, we search for the same fundamental set of physical processes as in~\cite{Buschmann:2019pfp} except applied to the cosmological NS population, which produces a contribution to the isotropic cosmic X-ray background (see Fig.~\ref{fig:fig1}). We find no evidence for such a signal and robustly rule out the axion interpretation of the M7 X-ray excess.  

Axion-like particles are motivated extensions of the Standard Model of particle physics that emerge generically in {\it e.g.} string theory constructions~\cite{Witten:1984dg, Choi:1985je, Barr:1985hk, Svrcek:2006yi, Arvanitaki:2009fg, Demirtas:2018akl, Halverson:2019cmy, Mehta:2021pwf, Gendler:2023kjt,Benabou:2025kgx,Fallon:2025lvn,Agrawal:2025rbr}.  Axion-like particles are related to the quantum chromodynamics (QCD) axion, which may solve the strong {\it CP} problem of the neutron electric dipole moment and can explain the dark matter of the Universe~\cite{Peccei:1977hh, Peccei:1977ur, Weinberg:1977ma, Wilczek:1977pj, Preskill:1982cy, Abbott:1982af, Dine:1982ah}.  String theory constructions may produce the QCD axion in addition to a multitude of axion-like-particles through the dimensional reduction of higher-dimensional gauge fields on cycles of the compact manifold~\cite{Witten:1984dg,Choi:1985je,Barr:1985hk,Svrcek:2006yi,Arvanitaki:2009fg}. Some of these axion-like-particles may be ultralight, with masses much less than that of the QCD axion, while others could be much heavier than the QCD axion. The ultralight axion-like-particles could have derivative couplings to fermions and couple to electromagnetism (but see~\cite{Agrawal:2022lsp}), while axions heavier than the QCD axion may also couple directly to QCD (see, {\it e.g.},~\cite{Benabou:2024jlj} for a discussion).  We refer to all such particles in this work simply as axions, and we probe new parameter space for ultralight and heavy axions.

The idea of a cosmologically-produced diffuse axion-induced photon background has been explored and constrained in the context of supernovae (SNe)~\cite{Calore:2020tjw, Caputo:2021rux, Benabou:2024jlj, Candon:2025fnb} and even main sequence stars~\cite{Nguyen:2023czp}, though our fiducial results, which use NuSTAR hard X-ray data~\cite{Krivonos:2020qvl}, provide leading sensitivity to axions for certain coupling combinations and axion masses. Our work builds on an extensive literature studying axion production in stars and the subsequent conversion to X-rays in astrophysical magnetic fields, though most previous works study nearby stars in the Milky Way~\cite{Dessert:2020lil,Xiao:2021,Xiao:2022rxk,Dessert:2019sgw, Dessert:2021bkv, Ning:2024ozs} or nearby galaxies and galaxy clusters~\cite{Ning:2024eky,Candon:2024eah,Ning:2025tit,Ning:2025kyu}.  

\noindent
{\bf Axion Production in Neutron Stars.---}\label{sec:axions_NS}Axions can be thermally produced in the cores of NSs, whose temperatures typically range from $T\sim \mathcal{O}(1) - \mathcal{O}(100)$ keV, depending primarily on the age as we discuss below. 
In this work we focus on axion production through interactions with nucleons. The axion-nucleon couplings are parametrized by
\es{eq:axion_L_nuc}{
{\mathcal L} \supset \sum_f {C_{aNN} \over 2 f_a } \partial_{\mu} a \bar N \gamma^{\mu} \gamma_5 N   \,,
}
where $a$ is the axion field, $f_a$ is the axion decay constant, and $N$ is the nucleon field (proton $p$ or neutron $n$).  The coefficient $C_{aNN}$ is dimensionless and characterizes the ultraviolet (UV) completion of the theory. (Note that it is convenient to define a dimensionless coupling $g_{aNN} = C_{aNN} m_N /f_a$, with $m_N$ the nucleon mass.) Ultralight axions with masses well below that of the equivalent QCD axion with the same $f_a$ must not couple to QCD through the operator $a G \tilde G / f_a$, with $G$ the QCD field strength, unless allowing for fine tuning or additional dynamics ({\it e.g.},~\cite{Hook:2018jle}). Thus, contributions to $C_{aNN}$ for ultralight axions arise from UV derivative couplings of the axion to quarks, which are also induced under the renormalization group~\cite{Dessert:2019sgw, Srednicki:1985xd, Bauer:2017ris}.  

Below the electroweak symmetry breaking scale, axions couple to gauge bosons through the dimension-5 interactions 
\es{}{
{\mathcal L} \supset -{1 \over 4} g_{a\gamma\gamma}a F \tilde F + {a \over 8 \pi f_a} G^a \tilde G^a \,,
}
where the second term describes the interaction of the axion with the QCD field strength $G^a$; this term is absent for ultralight axions, as discussed above. The first term gives the coupling of the axion to quantum electrodynamics (QED), with $F$ the QED field strength and with Lorentz indices suppressed. When discussing massive axions, we are particularly interested in the mass range between the eV and MeV scales. We can then integrate out the axion coupling to gluons to derive the axion effective field theory (EFT) at the energy scales relevant for thermal physics in NSs, inducing the irreducible contribution to $g_{a\gamma\gamma}$ denoted by $g_{a\gamma\gamma}^{\rm QCD} \approx -1.92 \alpha_{\rm EM} / (2 \pi f_a)$, which adds together with UV contributions to $g_{a\gamma\gamma}$. For $m_a \ll m_\pi$, with $m_\pi$ the pion mass, integrating out the axion-QCD coupling generates axion nucleon couplings with $C_{app} \approx -0.50$ and $C_{ann} \approx -0.02$~\cite{diCortona:2015ldu}; these contributions to $C_{aNN}$ add to those induced from UV axion-quark couplings, if present.

Inside a NS core, axions are produced through nucleon bremsstrahlung, \textit{i.e.} $N + N \to N + N + a$, in the degenerate nucleon environment. (If neutron superfluidity or proton superconductivity are present, then Cooper pair breaking and recombination processes (PBF) can also produce axions, as we discuss later in this work.) The axion emission spectrum from bremsstrahlung is a modified thermal spectrum, $dF/dE \sim z^3 (z^2+4\pi^2)/(e^z -1)$, where $z=E/T$, $E$ is the local axion energy, and $T$ the local NS core temperature. The exact formulae for the emissivities, in the cases of proton bremsstrahlung, neutron bremsstrahlung, and proton-neutron bremsstrahlung, are taken from~\cite{Buschmann:2021juv}. We discuss further details of the bremsstrahlung calculation in the Supplementary Material (SM). In our fiducial analysis for ultralight axions we set $\gann = \gapp$ and then quote sensitivity in terms of the coupling combination $g_{ann} \times g_{a\gamma\gamma}$, with $g_{a\gamma\gamma}$ required to convert the axions to photons in the NS magnetosphere, as we discuss more below. In the heavy axion case, we assume a model consistent with a grand unification theory (GUT) with no UV tree-level couplings to quarks, which implies that the axions have $g_{a\gamma\gamma} = (E/N) \alpha_{\rm EM} / (2 \pi f_a) + g_{a\gamma\gamma}^{\rm QCD}$ and $E/N = 8/3$ as appropriate for the standard embedding of the Standard Model in the GUT gauge group (see, {\it e.g.},~\cite{Agrawal:2022lsp}), in addition to the axion-nucleon couplings quoted before.  

To compute the axion emission rates from the NS cores, we use NS profiles produced from the code package \texttt{NSCool}~\cite{2016ascl.soft09009P}, which simulates and evolves NSs in spherically symmetric general relativity
given parameter choices such as the NS mass and equation of state (EOS). This results in radial profiles of the local temperature at each timestep in the NS lifetime. The temperatures, in combination with the densities and Fermi momenta, are then used to compute the axion emission rates described earlier. For our fiducial NS model, we assume a BSk24 EOS~\cite{Pearson:2018tkr} (consistent with mass-radius data as discussed in~\cite{Buschmann:2021juv}) and no superfluidity (as supported by~\cite{Buschmann:2021juv}, but see the SM where we incorporate and discuss the systematic uncertainties associated with superfluid models). We also do not consider light elements in the envelope (in the SM we explain why this does not strongly affect our results).  The NS masses are drawn from an initial mass distribution, discussed more below.

\noindent
{\bf The Neutron Star Cosmological Population.---}\label{sec:NS_cosmo}The formalism described above allows us to compute the axion luminosity over the lifetime of a NS, given its initial mass.  To describe the axion luminosity from the cosmic NS population, however, we need models for the NS formation rate as a function of cosmic time, as well as the NS mass distribution, and in the case of ultralight axions we also need a model for the magnetic field profiles of the cosmic NSs as a function of the NS age. We begin by detailing our formalism for the ultralight axion scenario, as it is slightly more involved than the case of heavy axions, since in the latter case the magnetic field profiles of the NSs do not affect the signal.

The axion-induced cosmological NS flux signal $d\phi/dE$ is evaluated as a redshift ($z$) integral as follows: 
\begin{multline}
\label{eq:NS_cosmo}
\frac{d\phi (E)}{dE} = \int_0^{\infty} dz (1+z) \frac{1}{H(z)(1+z)} \\ \times \frac{dN (E(1+z))}{dE} p_{a\to \gamma}(E(1+z)) [R_{\rm NS}(z)]  \, ,
\end{multline}
with \mbox{$H = H_0\sqrt{\Omega_m ( 1 + z)^3 + \Omega_{\Lambda}}$}.  We take $H_0 \approx 70$ km/s/Mpc, $\Omega_m \approx 0.3$, and $\Omega_{\Lambda} \approx 0.7$. $R_{\rm NS}(z)$ is the NS formation rate, $p_{a\to \gamma}$ is the axion-to-photon conversion probability, and $dN/dE$ is the axion emission spectrum. Note that $dN/dE$ and $p_{a\to \gamma}$, at a given $E$, are averaged over the NS populations, with the averages taken over NS ages and other properties such as initial magnetic field and mass.

The quantity $R_{\rm NS}(z)$ can be derived from the rate of SN formation, $R_{\rm SN}(z)$. Ref.~\cite{Priya:2017bmm} adopts the form of $R_{\rm SN}(z)$ from the star formation history (SFH) inferred from multiwavelength observations~\cite{2006ApJ...651..142H}, resulting in a piecewise fit whose pieces scale as $(1+z)^{\alpha}$ where $\alpha$ is positive before $z=1$ and increasingly negative after, with a broad peak roughly spanning the region $z=1$ to $z=3$. This fit also implies an overall normalization at $z=0$ of $R_{\rm SN}(z=0) = (1.25 \pm 0.5) \times 10^{-4}$ yr$^{-1}$ Mpc$^{-3}$~\cite{Lien:2010yb, 2006ApJ...651..142H}, corresponding to the estimated rate of SN formation today. From $R_{\rm SN}$, we extract the approximate fraction of SN which result in NSs, roughly $\sim$0.8, which then gives the rate of NS formation $R_{\rm NS}(z)$, and is explicitly shown in the SM (we additionally explore alternate models of $R_{\rm NS}$ in the SM, although they only have a minor effect on our final results).

Our fiducial calculation of the population-average axion emission flux $dN/dE$ is taken over NS ages spanning $\sim 10$ years to $\sim 10^7$ years. The lower bound comes from a conservative estimate of the minimum age of a NS that would be unaffected by the initial stellar debris ejecta (see, \textit{e.g.},~\cite{Igoshev:2021ewx, Yakovlev:2004iq}), which is important in order to not impede the axion-to-photon conversion in the magnetosphere. The upper bound comes from the typical timescale at which the axion-induced flux from a NS becomes negligible---a NS with an age of $10^7$ years has an $\mathcal{O}({\rm eV})$ core temperature. This also implies that a single NS in our fiducial scenario emits axions on a timescale of $t_{\rm emit} \sim 10^7$ years, after which we cease the production of axions. Variations on these parameters, as well as illustrations, are explored in the SM.

Finally, our NS population is drawn from a mass distribution, for which we adopt the preferred unimodal model in~\cite{You:2024bmk} as our fiducial model. This model, inferred by a compilation of NS observations spanning radio pulsars, gravitational waves, and X-ray binaries, describes a broad distribution with a turn-on at $1.1$ $M_{\odot}$, a peak around $1.3$ $M_{\odot}$, and a power law following the peak extending to NS masses even above $2$ $M_{\odot}$. Notably, this distribution seems preferred over canonical double Gaussian models of NSs taken in other literature, \textit{e.g.}~\cite{Alsing:2017bbc, Antoniadis:2016hxz}, and is reasonably consistent with the common assumption of $1.4$ $M_{\odot}$ typically taken for NS masses. We explore and discuss the Gaussian model in the SM.

\noindent
{\bf Ultralight Axion Searches using Magnetosphere Conversions.---}\label{sec:ultralight}For ultralight axions, we consider their conversion in NS magnetospheres after their production in NS cores. The axion-to-photon conversion probability, $p_{a\to \gamma}$, crucially depends on the magnetic field profile of the NS, with an overall strength parameterized by $\gagg$.

We adopt the formalism for calculating the axion-to-photon conversion probability in compact object magnetospheres from~\cite{Dessert:2019sgw, Buschmann:2019pfp}, which also takes into account the Euler-Heisenberg term which can suppress conversion due to strong field QED~\cite{Raffelt:1987im}. Note that the axion-to-photon conversion probability is generally efficient roughly when the phase mismatch $\delta k$ of the axion (with mass $m_a$ and energy $E$) and photon is small over the magnetic field coherence length $L$. That is, the conversion becomes suppressed roughly when $\delta k L \sim \frac{m_a^2}{2E} L > 1$.  For low $m_a$, such that $\delta k L \ll 1$, the conversion probability becomes independent of $m_a$; in our analysis, $E \sim 10 \, \, {\rm keV}$ and $L \sim 10 \, \, {\rm km}$, being the typical extent of the NS magnetosphere.  Thus, we expect our sensitivity to axions to be independent of $m_a$ for $m_a \lesssim {\rm few} \times 10^{-4}$ eV. 

Following~\cite{Buschmann:2019pfp}, we assume the NS magnetic fields are dipoles with polar strength $B_0$.  At birth the initial $B_0$ are drawn from a broad distribution, and then during the subsequent evolution of the NSs the field strengths are expected to decrease; however, the magnetic field evolution is highly uncertain (see, {\it e.g.},~\cite{Faucher-Giguere:2005dxp,Popov:2009jn}). To account for the wide variety of magnetic fields available for NS magnetospheres across the cosmic population, we create a characteristic ensemble of NSs, drawing initial $B_0$ values from the NS population model named  `Model 2' in~\cite{Safdi:2018oeu}, which consists of initial magnetic fields log-normally distributed with central value $\log{B_0 / {\rm G}} \approx 13.25$ and standard deviation $\sigma_{\log{B_0 / {\rm G}}} \approx 0.6$.  Note that this magnetic field model was tuned to reproduce the magnetic field distribution of observed pulsars in the ATNF pulsar catalog~\cite{Manchester:2004bp}.  The magnetic field decay in this model arises from effects such as ambipolar diffusion, Hall drift, and Ohmic heating; in particular, the magnetic field is evolved alongside the core temperature which we extract from \texttt{NSCool} (see the SM for further details and a discussion of results using an alternate magnetic field model), which then allows us to solve the axion-photon mixing equations to compute $p_{a\to\gamma}$. 

\begin{figure}[!htb]
\centering
\includegraphics[width=0.49\textwidth]{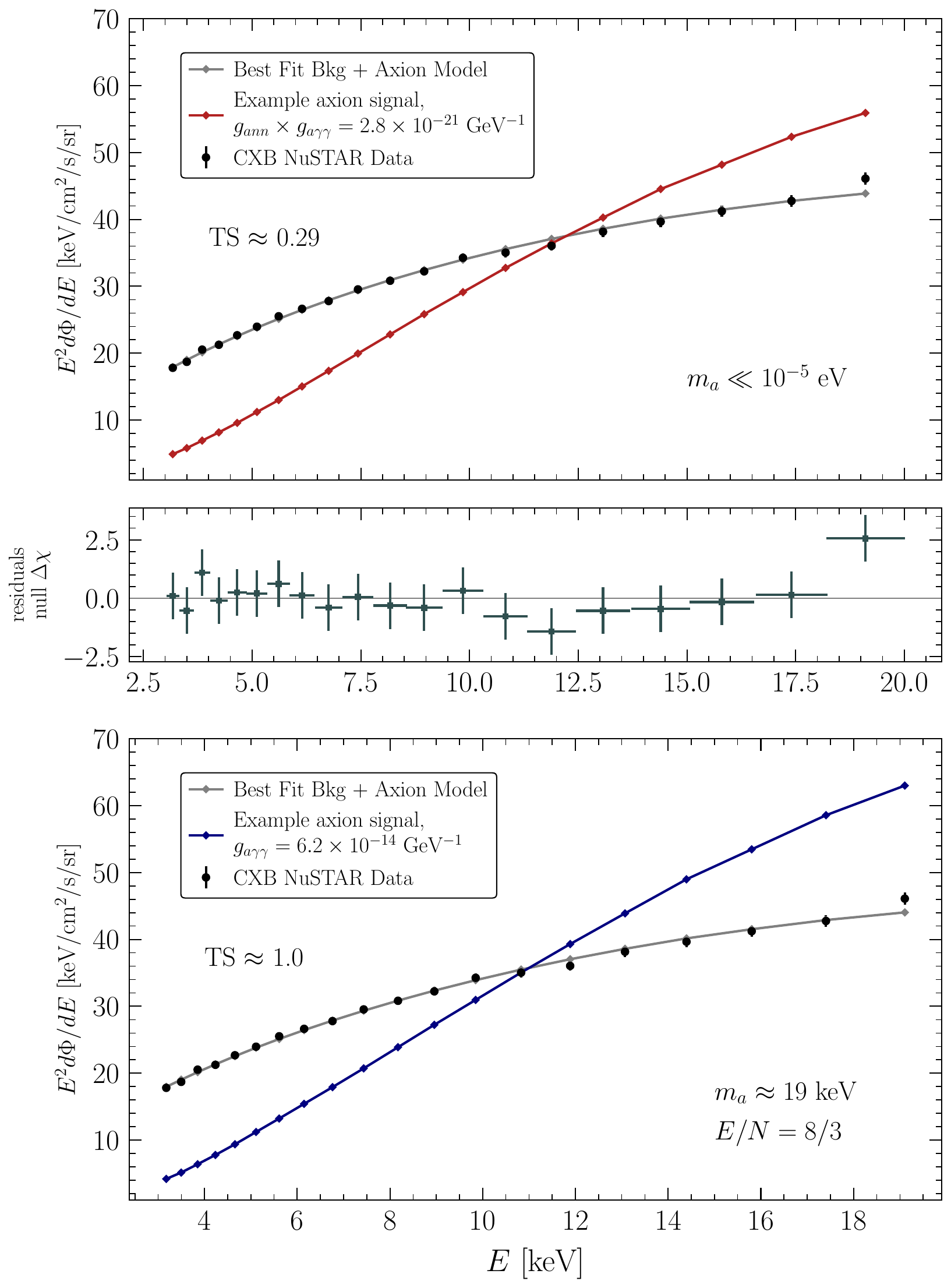}
\vspace{-0.4cm}
\caption{An illustration of how our axion-induced signal compares to the measured NuSTAR CXB data, which extends up to $\sim$20 keV. (Top) We show our best-fit axion and CXB background model (gray), as well as an example axion signal with the indicated coupling, all in the low $m_a$ limit. The residuals of the data compared to the fitted background model under the null hypothesis are shown directly below. (Bottom) The same but for a heavy axion with the indicated mass, and indicated $E/N$. We see that in both cases the axion spectra is morphologically different than the CXB, and that the data are largely consistent with the null hypothesis.}
\label{fig:data_fid}
\end{figure}

In Fig.~\ref{fig:data_fid} (upper panel) we illustrate the differential axion-induced X-ray signal, shown in units of keV/cm$^2$/s/sr. We illustrate the spectrum assuming an ultralight axion ($m_a \ll 10^{-5}$ eV) and the indicated coupling constant combination.  
Note that this spectrum peaks around $50$ keV and extends to even higher energies (see the SM), though our fiducial analysis using NuSTAR observations is restricted to energies below $20$ keV, as we describe later. In Fig.~\ref{fig:data_fid} we compare the predicted signal to the cosmic X-ray background (CXB) as inferred from NuSTAR~\cite{Krivonos:2020qvl}; the CXB is thought to primarily arise from active galactic nuclei (AGN). We remark that in this hard X-ray regime, there is no significant attenuation from intergalactic or interstellar medium absorption, \textit{i.e.} we are in the optically thin limit.

Ref.~\cite{Krivonos:2020qvl} used $\sim$7 Ms of extragalactic deep field observations to measure the CXB. These measurements are made up to $\sim$20 keV; higher energies were not included because they are inhibited by instrumental systematics such as charged particle interactions, activation lines, and stray light sources (see, \textit{e.g.},~\cite{Wik:2014}), as well as Earth's own hard X-ray emission from CXB reflection and cosmic ray interactions~\cite{2007MNRAS.377.1726S, 2008MNRAS.385..719C}.  We emphasize that the 68\% uncertainties in Fig.~\ref{fig:data_fid} are provided by Ref.~\cite{Krivonos:2020qvl} and are dominated by systematic uncertainties, which we assume are uncorrelated from energy bin to energy bin, though in reality there are likely correlations. We also conduct analyses using data from the older instruments HEAO-1~\cite{1999ApJ...520..124G}, Swift/BAT~\cite{Ajello:2008xb}, and INTEGRAL~\cite{Churazov:2006bk}, which extend to higher energies $\sim$200 keV. These analyses yield similar though slightly stronger results relative to those using NuSTAR data, though with less well-controlled systematic uncertainties.  The details of these analyses are described in the SM. (We also examine Chandra CXB data which only go up to $\sim$7 keV~\cite{Cappelluti:2017miu}, resulting in subleading constraints which we do not consider further.)

We search for evidence in favor of the axion model in the NuSTAR data by using a Gaussian likelihood joint over energy bins, assuming uncorrelated and normally distributed uncertainties between energy bins, with the signal model added together with a null-hypothesis astrophysical model. For our fiducial NuSTAR analysis, the null hypothesis model is taken to be a 2-parameter power-law with a fixed exponential cut-off: $d \phi_{\rm null} / dE = A (E / \, {\rm keV})^{-n} e^{-E / 41.13 \, {\rm keV}}$, taken from the canonical parameterization of the CXB~\cite{1999ApJ...520..124G}. Our analysis of the null hypothesis model alone yields best-fit values $A \approx 8.82$ keV/keV cm$^{-2}$ s$^{-1}$ sr$^{-1}$ and $n \approx 0.34$, consistent with~\cite{Krivonos:2020qvl}. In the central sub-panel of Fig.~\ref{fig:data_fid} we show the residuals of the null-hypothesis-only fit; no significant structure is apparent in the residuals. We use standard frequentist techniques (see, {\it e.g.},~\cite{Safdi:2022xkm}) to search for evidence in favor of the axion model and to set one-sided 95\% upper limits on the coupling combination $|g_{ann} \times g_{a\gamma\gamma}|$ at fixed $m_a$ profiling over the nuisance parameters; note that we analytically continue the flux formulae to allow for negative $|g_{ann} \times g_{a\gamma\gamma}|$ so that we may apply Wilks' theorem. 

\begin{figure}[!htb]
\centering
\includegraphics[width=0.49\textwidth]{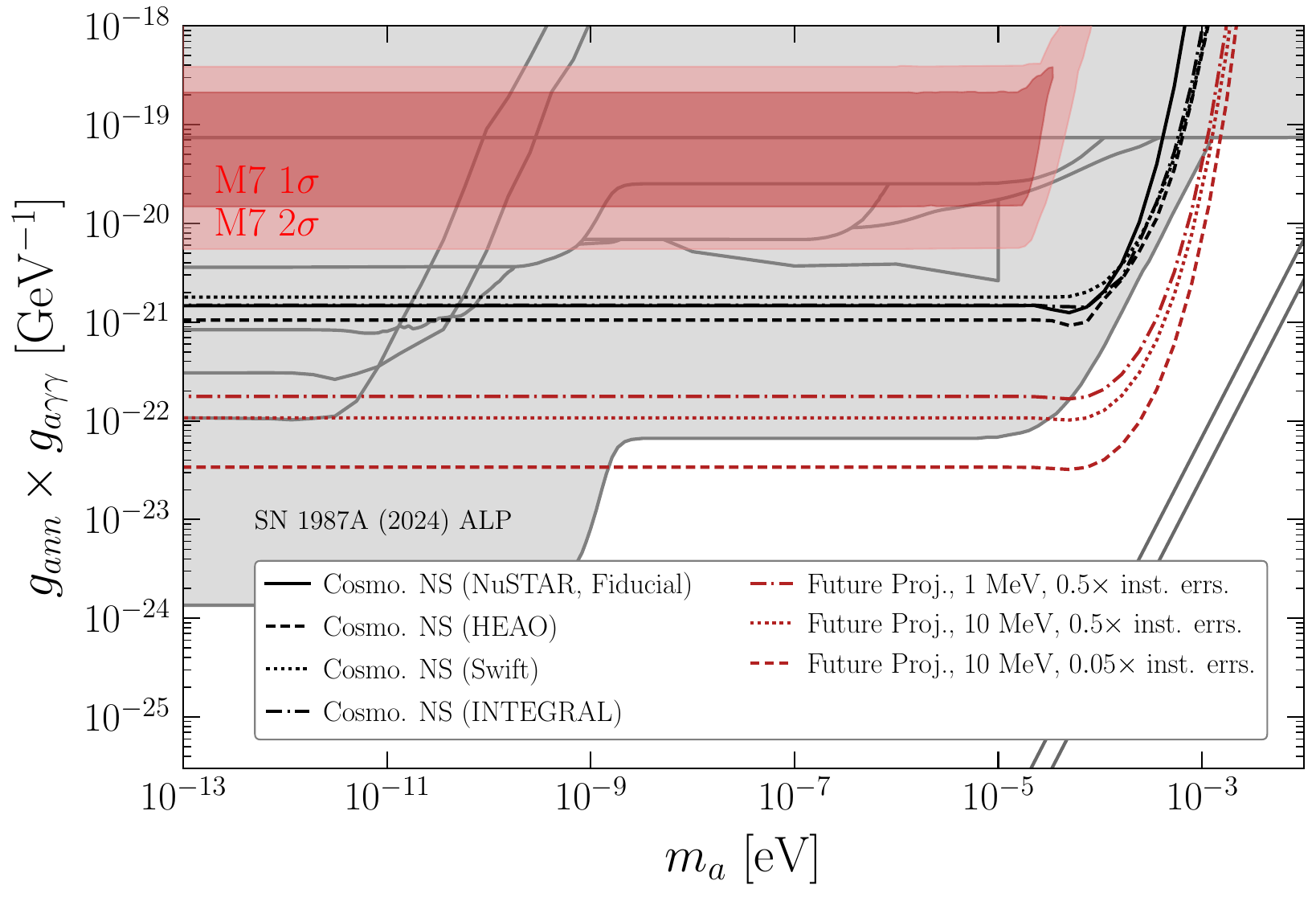}
\vspace{-0.4cm}
\caption{The 95\% upper limits on $|\ganngagg|$ obtained in this work from the non-observation of ultralight axions from cosmological NS populations. Our fiducial result (black, solid), uses observations of NuSTAR CXB data, while we also show constraints using CXB measurements from the HEAO instrument, Swift, and INTEGRAL. We illustrate projections for constraints on $\ganngagg$ from future telescopes which have smaller systematic uncertainties and reach to higher energies than those probed with NuSTAR (see text for details). Existing constraints are in gray, noting stringent constraints from SN 1987A~\cite{Manzari:2024jns} in particular, and the parameter space favored to explain the M7 X-ray excess is in red~\cite{Buschmann:2019pfp,Dessert:2019dos}, which is completely excluded by this work.
}
\label{fig:ultralight_limits}
\end{figure}

We find no evidence for ultralight axions, and so we set 95\% power-constrained~\cite{Cowan:2011an} upper limits on $\ganngagg$ as a function of $m_a$, using Wilks' theorem~\cite{Cowan:2010js, Cowan:2011an}. In the massless axion limit, we constrain the axion-nucleon axion-photon combined coupling at the level of $|\ganngagg| \lesssim 1.47 \times 10^{-21}$ GeV$^{-1}$, assuming $g_{ann} = g_{app}$. Also in that limit, we find a best-fit coupling of $|\ganngagg| \approx 8.81 \times 10^{-22}$ GeV$^{-1}$, and the significance in favor of the axion model is $\sim$$0.54\sigma$. Our constraints are illustrated in Fig.~\ref{fig:ultralight_limits} alongside current constraints (notably~\cite{Manzari:2024jns,Ning:2025kyu,Buschmann:2021juv,Noordhuis:2022ljw,Ruz:2024gkl,Dolan:2022kul}). While the parameter space we constrain is closely competitive with constraints from SN 1987A searches for axions converting to gamma-rays in external magnetic fields~\cite{Manzari:2024jns}, our analysis here relies on an entirely different set of systematic uncertainties. We detail and explore additional variations on our fiducial analysis in the SM, including a projection (see Fig.~\ref{fig:ultralight_limits}) for a more optimistic telescope going to higher energies with fewer instrumental systematics; that instrument is assumed to have an effective area of $0.1$ m$^2$ up to 1 MeV or 10 MeV and reduced systematic uncertainties, as indicated in Fig.~\ref{fig:ultralight_limits}, with background model described in the SM. 

\noindent
{\bf Heavy Axion Searches using Decays.---}\label{sec:heavy}In the case of heavy axions, we consider their decay into two X-ray photons after their production in our cosmological NS population. Since the axion masses we consider are below the threshold for pion decays, the axions must decay to two photons with decay rate $\Gamma_{a\gamma \gamma} = \gagg^2 m_a^3 / (64\pi)$. We follow the formalism in~\cite{Caputo:2021rux, Benabou:2024jlj} to compute the induced X-ray flux---the axion-induced photon signal is determined as
\begin{multline}
\label{eq:NS_cosmo_heavy}
\frac{d\phi(E)}{dE} = \int_0^{\infty} dz (1+z) \frac{1}{H(z)(1+z)} R_{\rm NS}(z) \\ \times \int_{E_z,\rm min}^{\infty} dE_z f_D(E_z) \frac{2}{p_a(E_z)}\frac{dN(E_z)}{dE} \,,
\end{multline}
where the kinematic quantities are $E_{z,\rm min} = (E' + m_a^2/4E')$, $E' = E(1+z)$, and $p_a(E_z) = \sqrt{E_z^2 - m_a^2}$. The fraction of axions with emission energy $E_z$ that have already decayed between the emission epoch $z$ and today is denoted by $f_D(E_z)$.

\begin{figure}[!t]
\centering
\includegraphics[width=0.49\textwidth]{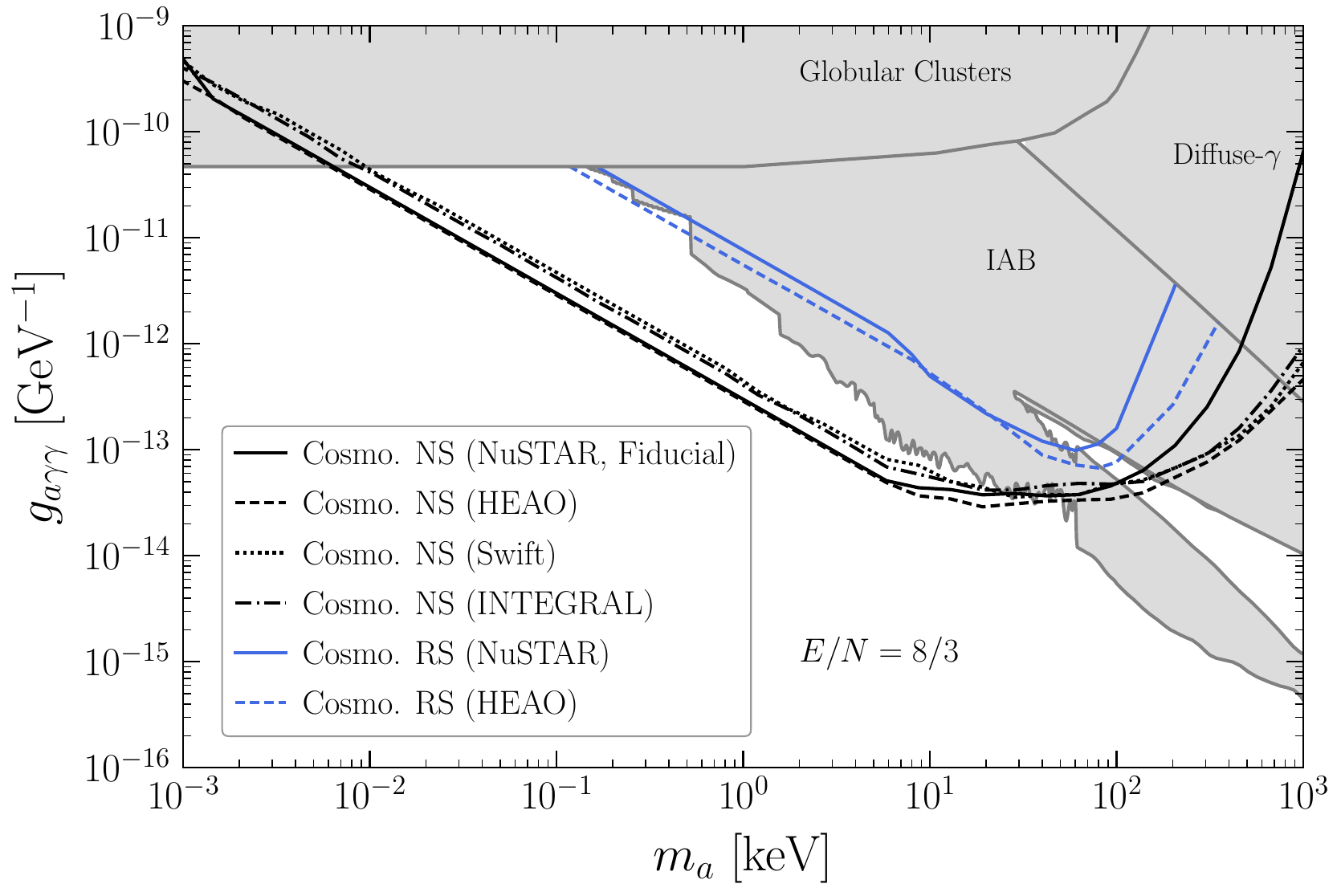}
\vspace{-0.4cm}
\caption{The 95\% upper limits on $\gagg$ obtained in this work from the non-observation of heavy axions from cosmological NS populations, assuming a GUT anomaly coefficient ratio as indicated (see main text), and for the observed NuSTAR (fiducial, black, solid), HEAO, Swift, and INTEGRAL-measured CXBs. Our fiducial result can also be compared to the subdominant constraints obtained from $\gagg$-only processes occurring in the cosmological regular star (RS) population (see text). We compare these constraints to existing constraints in gray, valid for our fiducial choice of $E/N = 8/3$. 
}
\label{fig:heavy_limits}
\end{figure}

The total photon flux $d\phi/dE$ is then used in the same way as in the ultralight case earlier, with the additional fiducial stipulation that, as discussed before, we assume $E/N = 8/3$ (following, {\it e.g.},~\cite{Benabou:2024jlj}). We compare the heavy axion signal with the NuSTAR observed CXB in Fig.~\ref{fig:data_fid} (we also show the results using HEAO, Swift, and INTEGRAL data, described further in the SM). We find no evidence for heavy axions, and set 95\% power-constrained upper limits on $|g_{a\gamma\gamma}|$ as a function of $m_a$ exactly analogous to the methods outlined in the ultralight axion case. Under the GUT benchmark $E/N = 8/3$, our limits, which are shown in Fig.~\ref{fig:heavy_limits}, exclude new parameter space across 4 eV $\lesssim m_a \lesssim 200$ keV (relevant existing constraints being~\cite{Dolan:2022kul,Langhoff:2022bij,Caputo:2021rux}), with our most stringent limit being $|\gagg| \lesssim 4.32 \times 10^{-14}$ GeV$^{-1}$ at $m_a \approx 19$ keV, which has a significance in favor of the axion model of $1.01\sigma$.  Variations on our fiducial analysis are explored in the SM. 

Also shown in Fig.~\ref{fig:heavy_limits} are our 95\% upper limits on heavy axions from another search, in which we look for processes involving the axion-photon coupling $\gagg$ only, in the estimated cosmological \textit{regular} stellar population, which do not include compact objects like NSs but do include high-temperature massive stars crucial for abundant axion emission through processes like Primakoff emission (see, \textit{e.g.},~\cite{Ning:2024eky, Dessert:2020lil}). The constraints are subdominant to the NS constraints, although they assume a different set of systematics about a complementary stellar population and do not rely on any particular axion UV completion. We provide more details about this search in the SM.

\noindent
{\bf Discussion.---}In this work we search for both ultralight and heavy axions produced from the cosmological NS population, which contributes a diffuse axion-induced X-ray background that can be compared to the CXB. In the case of ultralight axions, NS magnetospheres (which can be modeled over populations) facilitate axion-to-photon conversion. We find no evidence for axions and set stringent constraints on the combined axion-nucleon times axion-photon coupling, $\ganngagg$, competitive with current leading SN1987A bounds of the same couplings. In the case of heavy axions, which decay to photons outside the NS, we also find no evidence of axions, and in the context of our benchmark GUT model for heavy axions set the strongest constraints to-date for axion masses within a few orders of magnitude of a keV. In the SM we consider a number of systematics, such as the possibility of nucleon superfluidity, which has the potential to significantly boost the axion luminosity through PBF processes. A better understanding of nucleon superfluidity is crucial for improving this search in the future.  Additionally, our work serves as strong motivation for improving the measurements of the CXB, extending accurate measurement to higher energies and decreasing systematic uncertainties. 

\begin{acknowledgements}
{
{\it
We thank Josh Benabou, Malte Buschmann, Andrea Caputo, Chris Dessert, Josh Foster, Maurizio Giannotti, Andrew Long, Giuseppe Lucente, Yujin Park, and Kerstin Perez for helpful conversations and comments. The authors are supported in part by the DOE award
DESC0025293. B.R.S. acknowledges support from the
Alfred P. Sloan Foundation.  The work of O.N. is supported in part by the NSF Graduate Research Fellowship Program under Grant DGE2146752.
This research used resources of the National Energy Research Scientific Computing Center (NERSC), a U.S. Department of Energy Office of Science User Facility located at Lawrence Berkeley National Laboratory, operated under Contract No. DE-AC02-05CH11231 using NERSC award HEP-ERCAP0023978.   
}
}
\end{acknowledgements}

\bibliography{refs}

\clearpage

\onecolumngrid
\begin{center}
  \textbf{\large Supplementary Material for Cosmological Neutron Stars Produce Diffuse Axion X-Ray Signatures}\\[.2cm]
  \vspace{0.05in}
  {Orion Ning, Kailash Raman, and Benjamin R. Safdi}
\end{center}

\twocolumngrid

\setcounter{equation}{0}
\setcounter{figure}{0}
\setcounter{table}{0}
\setcounter{section}{0}
\setcounter{page}{1}
\makeatletter
\renewcommand{\theequation}{S\arabic{equation}}
\renewcommand{\thefigure}{S\arabic{figure}}
\renewcommand{\thetable}{S\arabic{table}}

\onecolumngrid

This Supplementary Material (SM) includes details about our NS axion production, our data analysis across our telescopes of interest, a survey of systematic uncertainties associated with our fiducial analyses, and a discussion of our search for axions in the cosmological regular star population.

\section{NS Axion Production}
\label{app:axion_production}

In this section, we detail our computation of axion emission rates from individual NSs. As discussed in the main text, we simulate the evolution of a NS in spherically symmetric general relativity using the package \texttt{NSCool}~\cite{2016ascl.soft09009P}. The NS EOS is used to compute relevant quantities such as local densities, neutron and proton Fermi momenta, effective nucleon masses, and the local metric. In cooling simulations including superfluidity, \texttt{NSCool} additionally computes the density-dependent nucleon condensation temperatures. Each \texttt{NSCool} run then outputs radial profiles of local temperature over the NS lifetime. 

Axions are produced primarily through two mechanisms in the NS core: nucleon-nucleon bremsstrahlung and PBF processes (note that we neglect the subdominant contributions from the NS crust, including electron-ion bremsstrahlung and PBF processes in the crust). Within NS matter, if there exists an attractive interaction between nucleons, then there exists a density-dependent critical temperature $T_c$ below which nucleons can condense into a superfluid of Cooper pairs. This condensation introduces a temperature-dependent energy gap $\Delta(T)$ into the nucleon dispersion relation. Superfluidity affects the axion luminosity through two mechanisms. First, condensation into Cooper pairs reduces the free nucleon phase space and suppresses nucleon bremsstrahlung emission. Second, since the thermal interactions can break Cooper pairs, nucleons can emit axions as they recondense into Cooper pairs. Neutrons can form both singlet ($^1S_0$) and triplet ($^3P_2$) state pairings, while protons form only singlet state pairings. 

The nucleon bremsstrahlung emissivity from degenerate NS matter is computed in~\cite{Iwamoto:1984ir, Brinkmann:1988vi, Iwamoto:1992jp}, and, as mentioned in the main Letter, follows a modified thermal spectrum, $dF/dE \sim z^3(z^2 + 4\pi^2)/(e^z - 1)$ with $z = E/T$. To take into account the variety of in-medium effects and suppression of the free nucleon phase space by Cooper pair formation, we use the formulae from Ref.~\cite{Buschmann:2021juv}. In-medium effects importantly also modify neutrino emission (primarily from modified URCA, nucleon bremsstrahlung, and PBF processes), which we incorporate self-consistently in our \texttt{NSCool} framework. These in-medium effects are parameterized by phenomenological suppression factors. (Alternatively, Ref.~\cite{Springmann:2024mjp} computes in-medium modifications to axion-nucleon couplings under the framework of heavy baryon chiral perturbation theory, while Ref.~\cite{Bottaro:2024ugp} recomputes the neutrino emissivities from Ref.~\cite{Friman:1979ecl} to take into account effects such as effective $\rho$-meson exchanges.) We compute superfluidity suppression factors from the formulae in  Ref.~\cite{1995A&A...297..717Y} and follow Ref.~\cite{Buschmann:2019pfp} and the \texttt{NSCool} code~\cite{2016ascl.soft09009P} in approximating the triplet pairing suppression factors by the singlet pairing suppression factors. We note that the results of Refs.~\cite{Iwamoto:1984ir,Brinkmann:1988vi,Iwamoto:1992jp} differ from Refs.~\cite{Ishizuka:1989ts,Stoica:2009zh} in their computation of pion mass effects. However, these differences are at the percent level at the relevant temperatures and densities for this work and are subdominant to other uncertainties. 

There exist at least three distinct processes of PBF emission in a NS: emission from neutron $^1S_0$ pairings, emission from proton $^1S_0$ pairings, and emission from neutron $^3P_2$ pairings. The neutron $P$-wave superfluid can occur in a total of 13 distinct phases~\cite{Khodel:2001yi}. The neutron superfluid will settle to a single distinct phase, but we detail the calculation for two possible phases of the $P$-wave superfluid with differing phase space anisotropies in the pairing gap, denoted $P^A$ and $P^B$. The pairing gaps for the two $P$-wave cases are given by
\begin{align}
    \Delta^A(T,\theta) &= \Delta_0^A(T)\sqrt{1+3\cos^2{\theta}} \\
    \Delta^B(T,\theta) &= \Delta_0^A(T)\sin{\theta} \,,
\end{align}
where $\theta$ is the angle of the nucleon momentum with respect to the quantization axis of the $P$-wave pairing. We note that the $P^A$ gap is always nonzero, while the $P^B$ gap goes to zero at the poles of the Fermi sphere.

Our convention for referring to superfluidity models will follow the format \texttt{A-B-C}, where \texttt{A}, \texttt{B}, \texttt{C} refer to the models used for the neutron $^1S_0$, the neutron $^3P_2$, and proton $^1S_0$ pairing gaps, respectively, with $\texttt{0}$ indicating that that channel is turned off (\textit{i.e.} our model without any superfluidity is $\texttt{0-0-0}$).

Since neutron $^1S_0$ pairing is attractive at NS crust densities and repulsive at core densities, the emission from this process is suppressed relative to nucleon bremsstrahlung emission and emission from other PBF processes. On the other hand, the proton $^1S_0$ and neutron $^3P_2$ pairings can be active in the core and are the dominant processes in PBF axion production when they occur. We use the formulae from Ref.~\cite{Buschmann:2021juv} to compute the PBF axion emissivity and the formulae from Refs.~\cite{Sedrakian:2015krq,Buschmann:2019pfp} for the emission spectra. We note that there exist some possible inconsistencies between the inclusion of $P$-wave superfluidity and isolated NS cooling data (see, \textit{e.g.},~\cite{Buschmann:2021juv}). However, with that caveat, we still examine the expected axion emission from $P$-wave PBF processes, which, as we show later, could have the potential to enhance the total axion flux from a given NS. Here, we extend the results of Ref.~\cite{Buschmann:2019pfp} and compute analytic expressions for the axion emission spectrum from the $P$-wave PBF processes.  The axion emissivity from PBF processes, unmodified by in-medium effects, is given by~\cite{Buschmann:2019pfp}
\begin{align}
    \epsilon_{a,{\rm PBF}}^P &= \frac{2g_{ann}^2}{3\pi m_N^2} \nu_n(0) T^5 I_{an}^P (z_T) \\
    I_{an}^P(z_T) &= \int \frac{d\Omega}{4\pi} z_{\theta}^5 \int_1^{\infty} dy \frac{y^3}{\sqrt{y^2-1}}\left[ f_F(z_{\theta}y)\right]^2 \,,
\end{align}
where $\nu_N(0)=m_Np_F(N)/\pi^2$ is the density of states at the Fermi surface, $z_{\theta}(T, \theta) \equiv \frac{\Delta(T,\theta)}{T} \equiv z_T(T)K(\theta)$, and  $f_F(x) = \frac{1}{e^x+1}$. Identifying the axion energy as $\omega = \frac{y}{2\Delta(T,\theta)}$ and taking the derivative of the emissivity with respect to the axion energy, the axion emission spectrum is given by

\begin{equation}
    J_{a,{\rm PBF}}^P(x) = \mathcal{N}_{a,{\rm PBF}}^P(z_T) \int \frac{\sin{\theta}d\theta}{2}K^2(\theta)\frac{x^3}{\sqrt{x^2-K^2(\theta)}}\left[f_F(z_T x) \right]^2 \,,
\end{equation}
where $x=\omega/(2\Delta_0(T))$ and the normalization factor $N_{a,{\rm PBF}}^P=\frac{\epsilon_{a,{\rm PBF}}^Pz_T^4}{I_{an}^P(z_T)T}$. Before taking the derivative, one must swap the integration order, yielding angular integral bounds that vary according to the value of $x$ and the type of pairing gap anisotropy. In case $P^A$, the integral is taken over the domain $\left[\theta_{\rm min}, \pi - \theta_{\rm min}\right]$ where
\begin{equation}
    \theta_{\rm min} = \begin{cases} \arccos{\sqrt{\frac{x^2-1}{3}}},&1<x<2\\ 0,&x>2 \end{cases} \,.
\end{equation}
In case $P^B$, the integral is taken over the domain  $\left[0,\theta_{\rm max}\right] \cup \left[\pi -\theta_{\rm max},\pi\right]$. Here,
\begin{equation}
    \theta_{\rm max} = \begin{cases} \arcsin{x},&x<1\\ \frac{\pi}{2},&x>1 \end{cases} \,.
\end{equation}
These integrals yield
\begin{align}
    J_{a,{\rm PBF}}^{P^A}(x) &= \mathcal{N}_{a,{\rm PBF}}^{P^A}(z_T) x^3 \left[f_F(z_T x) \right]^2 \times \begin{cases}
         \frac{\pi}{2\sqrt{3}}(1+x^2),& 1<x<2\\
        \frac{1+x^2}{\sqrt{3}}\arctan{\frac{\sqrt{3}}{\sqrt{x^2 - 4}}} - \sqrt{x^2-4},&x>2
    \end{cases} \\
    J_{a,{\rm PBF}}^{P^B}(x) &= \mathcal{N}_{a,{\rm PBF}}^{P^B}(z_T) x^3 \left[f_F(z_T x) \right]^2 \times \begin{cases}
         \frac{1+x^2}{2}\log{\frac{1+x}{1-x}} - x,&x<1\\
        \frac{1+x^2}{2}\log{\frac{x+1}{x-1}} - x,&x>1
    \end{cases} \,.
\end{align}
In this work, we only show the $P^A$ pairing case since the total emissivity and spectral shapes are similar in both cases.

\begin{figure}[!htb]
\centering
\includegraphics[width=0.49\textwidth]{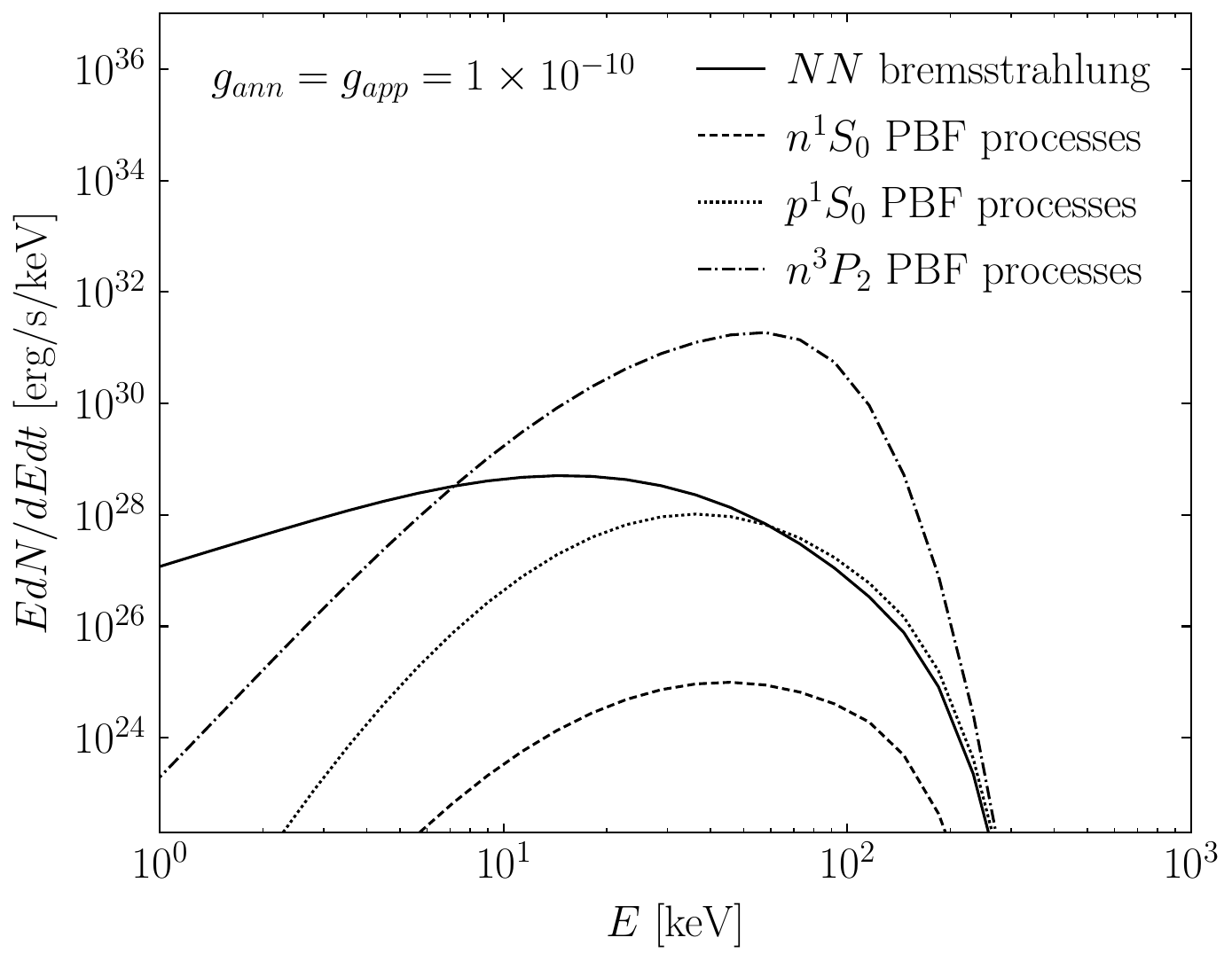}
\vspace{-0.4cm}
\caption{The instantaneous massless axion emissivity from a 1.4 $M_{\odot}$ NS with the BSk24 EOS at an age of $10^4$ years with the indicated axion-nucleon couplings. Results are shown for nucleon bremsstrahlung and PBF processes under superfluid model \texttt{SFB-SCGF-BS07}. In this model, the neutron $^1S_0$ pairing is computed using the model of Ref.~\cite{Schwenk:2002fq}, the proton $^1S_0$ pairing is computed using the model of Ref.~\cite{Baldo:2007jx}, and the neutron $^3P_2$ pairing is computed using the model of Ref.~\cite{Ding:2016oxp}.}
\label{fig:PBF_example}
\end{figure}

\begin{figure}[!htb]
\centering
\includegraphics[width=0.49\textwidth]{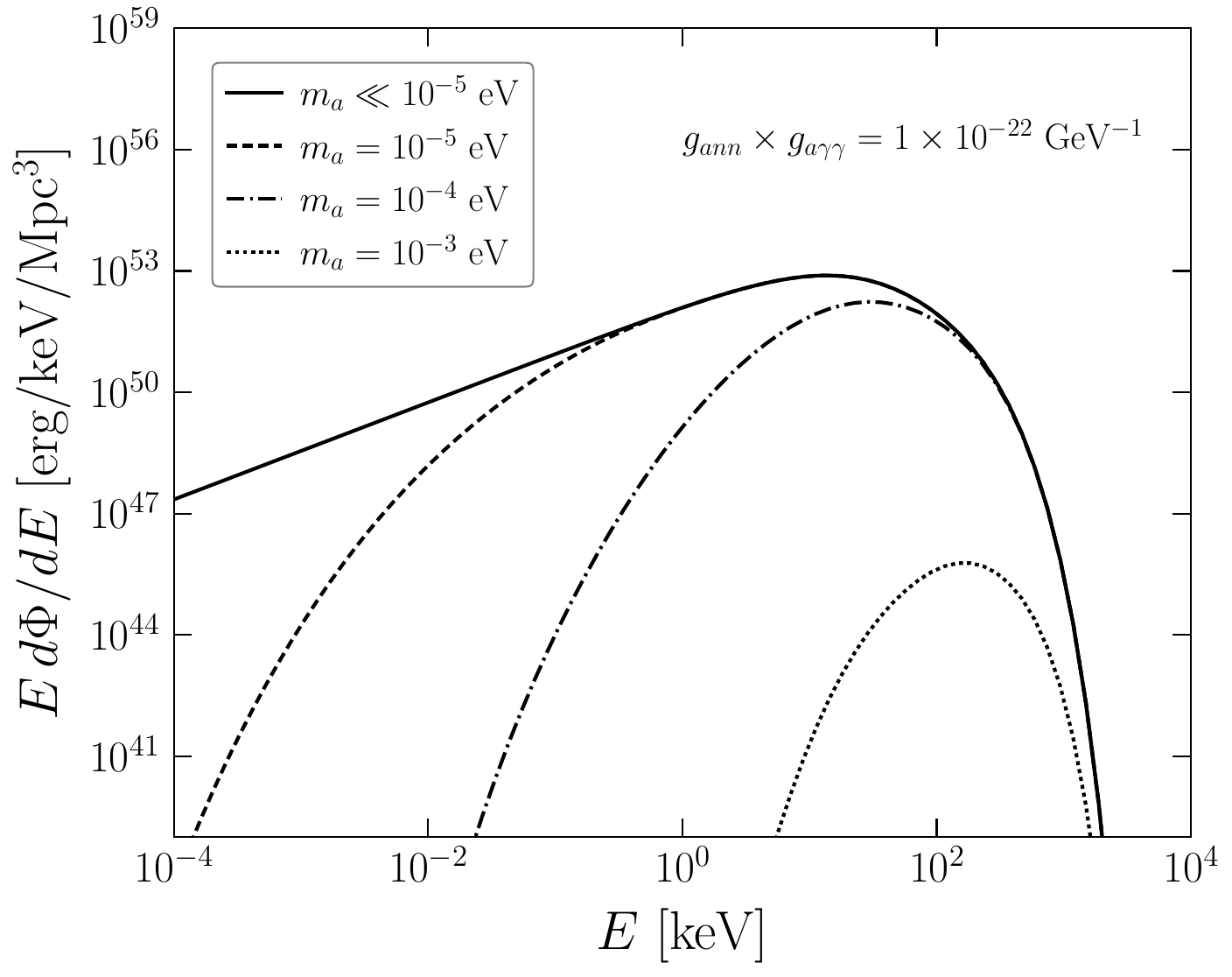}
\includegraphics[width=0.49\columnwidth]{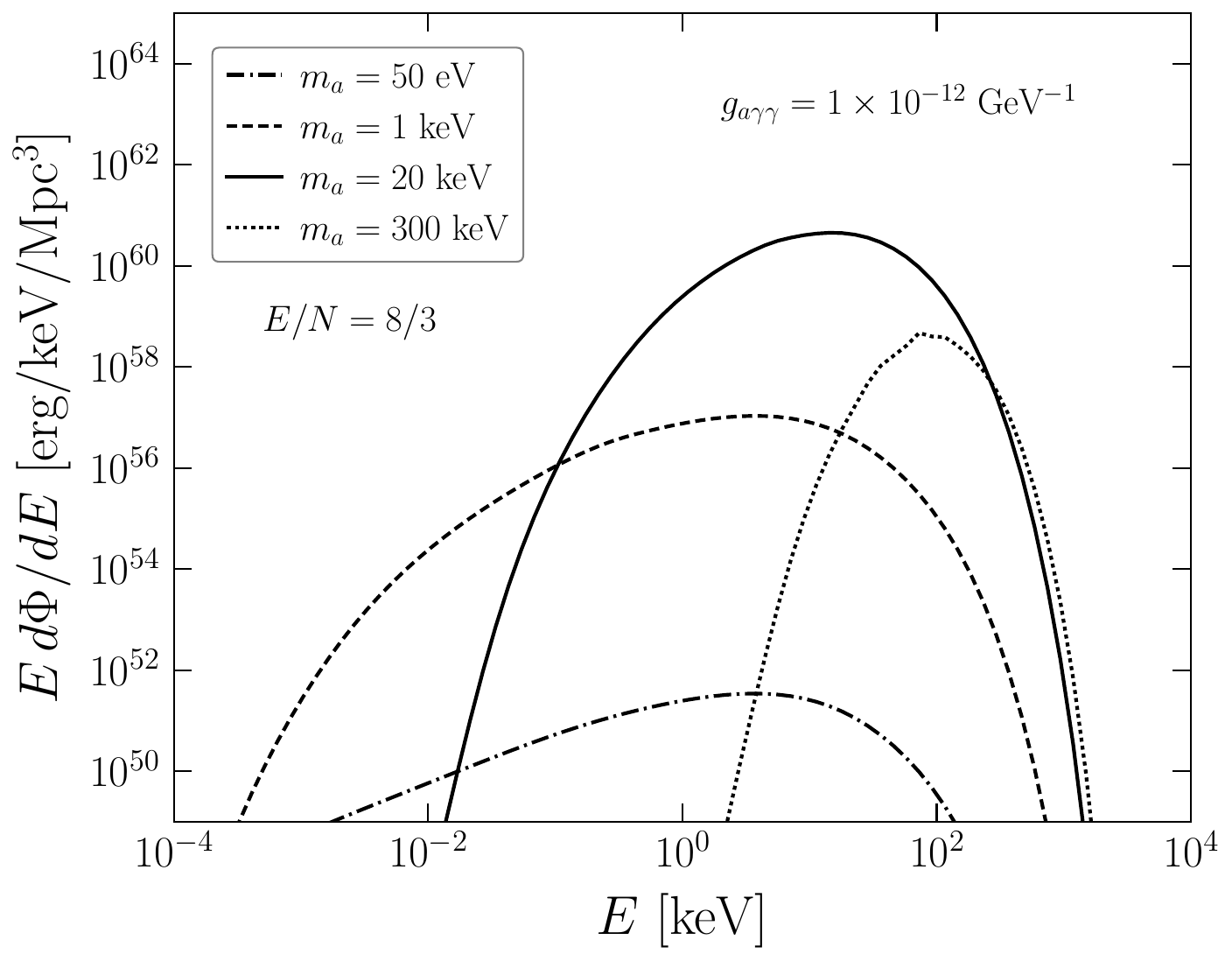}
\vspace{-0.4cm}
\caption{(Left) The fiducial cosmological NS photon energy density at the present epoch, for various ultralight axion masses, with the indicated $\ganngagg$ coupling. Note that our fiducial model assumes no superfluidity (\textit{i.e.} \texttt{0-0-0}). (Right) The same but for heavy axions with the indicated $\gagg$ coupling.}
\label{fig:cosmo_axion_spec}
\end{figure}

In Fig.~\ref{fig:PBF_example}, we illustrate the instantaneous axion emissivity from both nucleon bremsstrahlung and the PBF processes under superfluidity model \texttt{SFB-SCGF-BS07}, where all three pairing gaps are turned on, for the stated NS mass and EOS. Since the scale of PBF emission is set by the pairing gap, the PBF spectra tend to peak at higher energies than the bremsstrahlung spectra, although this is model-dependent. Ultimately, the axion emission discussed here at the level of individual NSs is combined with our NS ensemble representing the cosmological population following the prescription in the main Letter. In Fig.~\ref{fig:cosmo_axion_spec}, we illustrate explicit examples of the resulting cosmological NS photon energy density for both ultralight and heavy axion masses, for the present epoch, and for our fiducial (no superfluidity, \texttt{0-0-0}) model.

\section{NS Magnetosphere Distribution}
\label{app:NS_magneto}
In this section, we describe our prescription for assigning magnetosphere parameters to our NS population, which is used for the cosmological population analyzed in the main Letter. As our fiducial model for the magnetospheres of the NSs, we assume the Model 2 distribution detailed in~\cite{Safdi:2018oeu} (originally derived from~\cite{1990ApJ...356..359H}), which is tuned to accommodate the observations of pulsar magnetic fields in the ATNF pulsar catalog~\cite{Manchester:2004bp}. This model prescribes an initial magnetic field following a log-normal distribution with central value $\langle \log{B_0 / G} \rangle = 13.25$ and standard deviation $\sigma_{\log{(B_0/G)}} = 0.6$. More exactly, our probability distribution to find a NS with magnetic field (at the magnetic pole) of initial value $B_0$ is
\es{eq:GC_NS_B_dist}{
p(B_0) = \frac{1}{\sigma_{\log{B_0}} \sqrt{2 \pi}} \exp{-\frac{(\log{B_0} - \langle \log{B_0} \rangle)^2}{2 \sigma^2_{\log{B_0}}}} \,,
}
which is used to calculate the magnetosphere properties of our NS ensemble. Importantly, we also incorporate magnetic field decay over time; briefly, following~\cite{Safdi:2018oeu}, this involves effects such as ambipolar diffusion (most relevant for high $T_{\rm core}$ and high $B$ NSs), Hall drift, and Ohmic heating, which are incorporated into the magnetic field evolution through the differential equation
\es{eq:GC_NS_B_diffeq}{
\frac{dB}{dt} = -B \left( \frac{1}{\tau_{\rm H,O}} + \left( \frac{B}{B_0}\right)^2 \frac{1}{\tau_{\rm ambip}}\right) \,,
}
where $\tau_{\rm H,O}$ and $\tau_{\rm ambip}$ are the timescales for the combined Hall draft/Ohmic heating and ambipolar diffusion, respectively, with the initial condition $B = B_0$ (see details in~\cite{Safdi:2018oeu}). This equation is solved for each NS in our ensemble and taken at the desired age for that NS sample. Note that the ambipolar diffusion timescale $\tau_{\rm ambip}$ depends on the NS core temperature, and unlike in~\cite{Safdi:2018oeu}, which infers this quantity from approximate cooling laws~\cite{1983bhwd.book.....S}, here we take NS core temperatures directly from our simulated \texttt{NSCool} NSs, with all sources of cooling incorporated.

As a check on the sensitivity of our analysis to the choice of magnetic field model, we additionally implement the magnetic field model of Ref.~\cite{Graber:2023jgz}, which found an initial magnetic field distribution with central value $\langle \log B_0/G\rangle = 13.10$ and standard deviation $\sigma_{\log B_0/G} = 0.45$. For the magnetic field decay, we adopt their broken power-law model, detailed in Appendix A of Ref.~\cite{Graber:2023jgz}, with the choice of $a_{\rm late} = -3$. (See also Ref.~\cite{Witte:2025ilt} for discussions of alternate magnetic field models.)

\begin{figure}[!htb]
\centering
\includegraphics[width=0.49\textwidth]{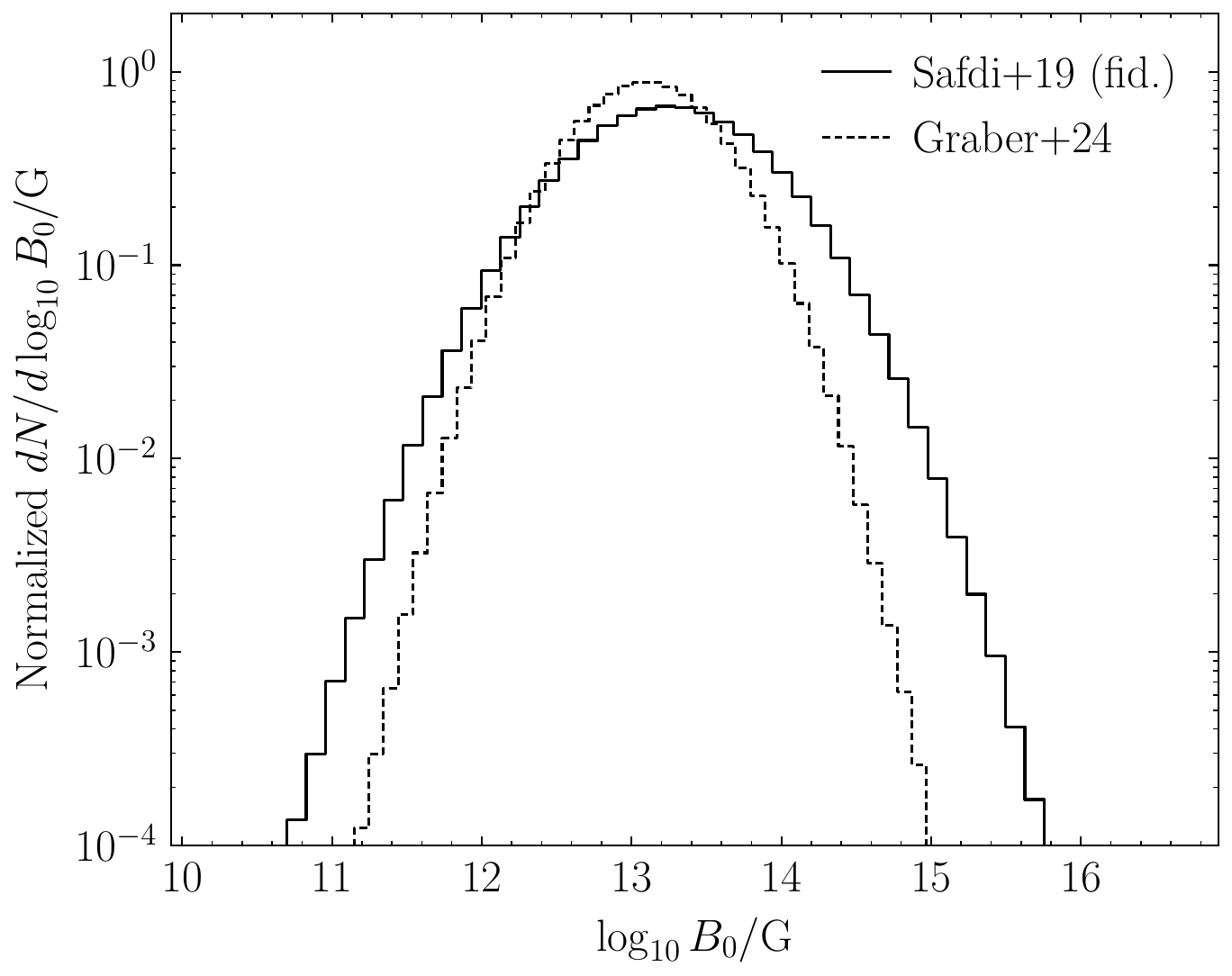}
\includegraphics[width=0.49\columnwidth]{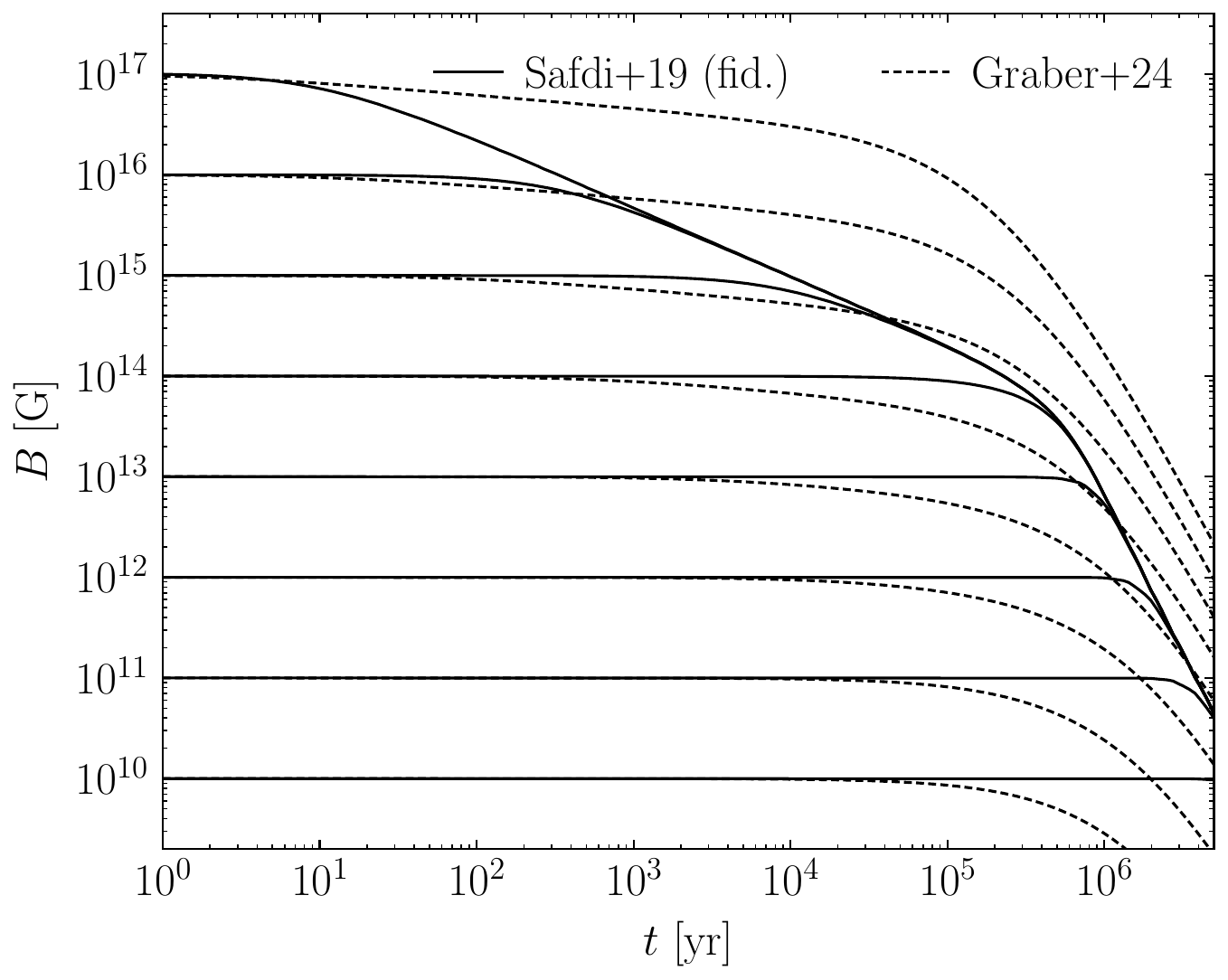}
\vspace{-0.4cm}
\caption{(Left) The normalized distribution of the initial magnetosphere magnetic fields $B_0$ across our NS ensemble for both the Model 2 distribution as detailed in Ref.~\cite{Safdi:2018oeu} (Safdi+19 (fid.)) and the model from Ref.~\cite{Graber:2023jgz} (Graber+24). (Right) The evolution of the magnetic field $B(t)$ as a function of time for various initial $B_0$. This evolution is evolved according to~\eqref{eq:GC_NS_B_diffeq} for the Safdi+19 model and according to a broken power law in the case of the Graber+24 model (see details in main text).}
\label{fig:GC_B}
\end{figure}

\begin{figure}[!htb]
\centering
\includegraphics[width=0.49\textwidth]{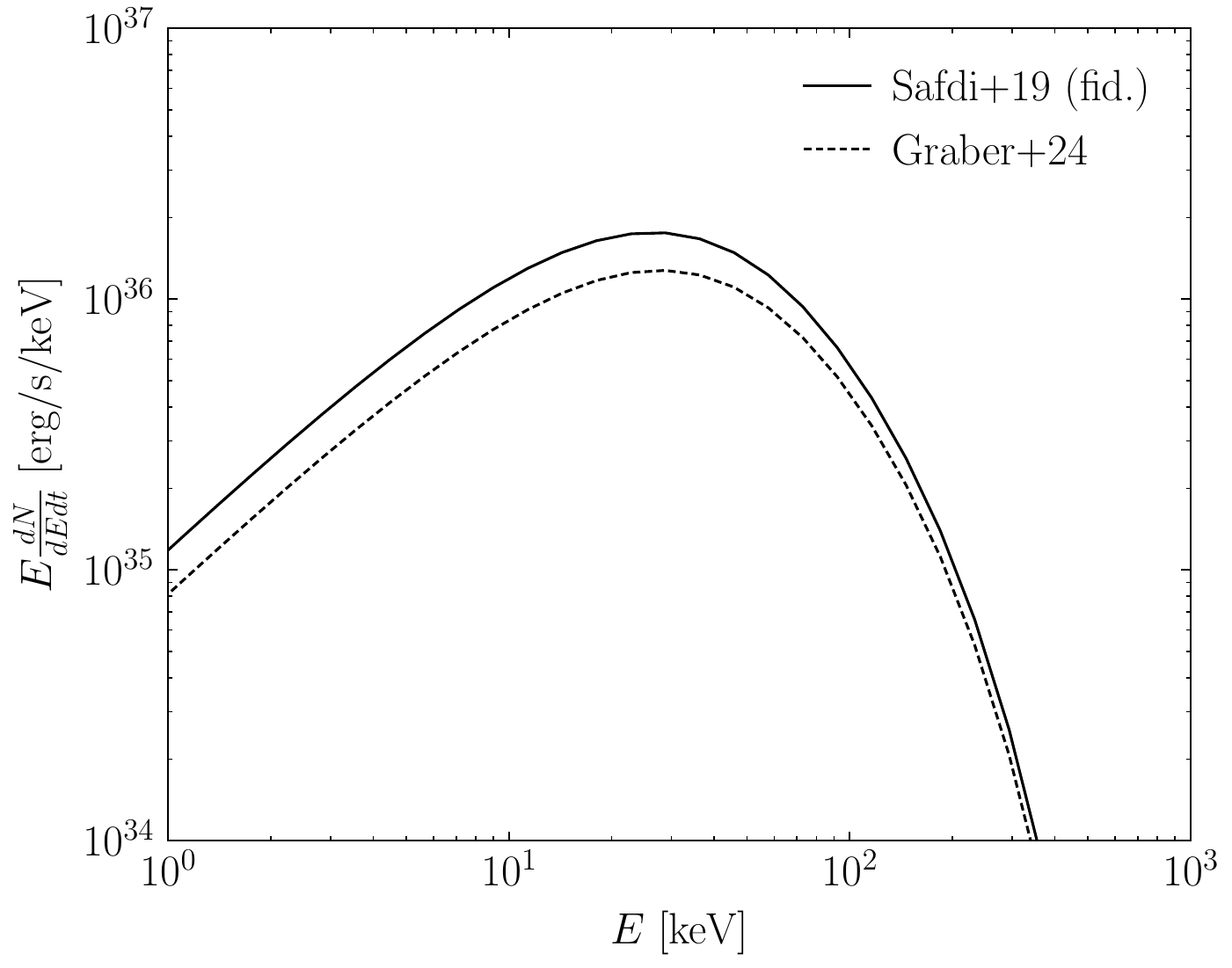}
\includegraphics[width=0.49\textwidth]{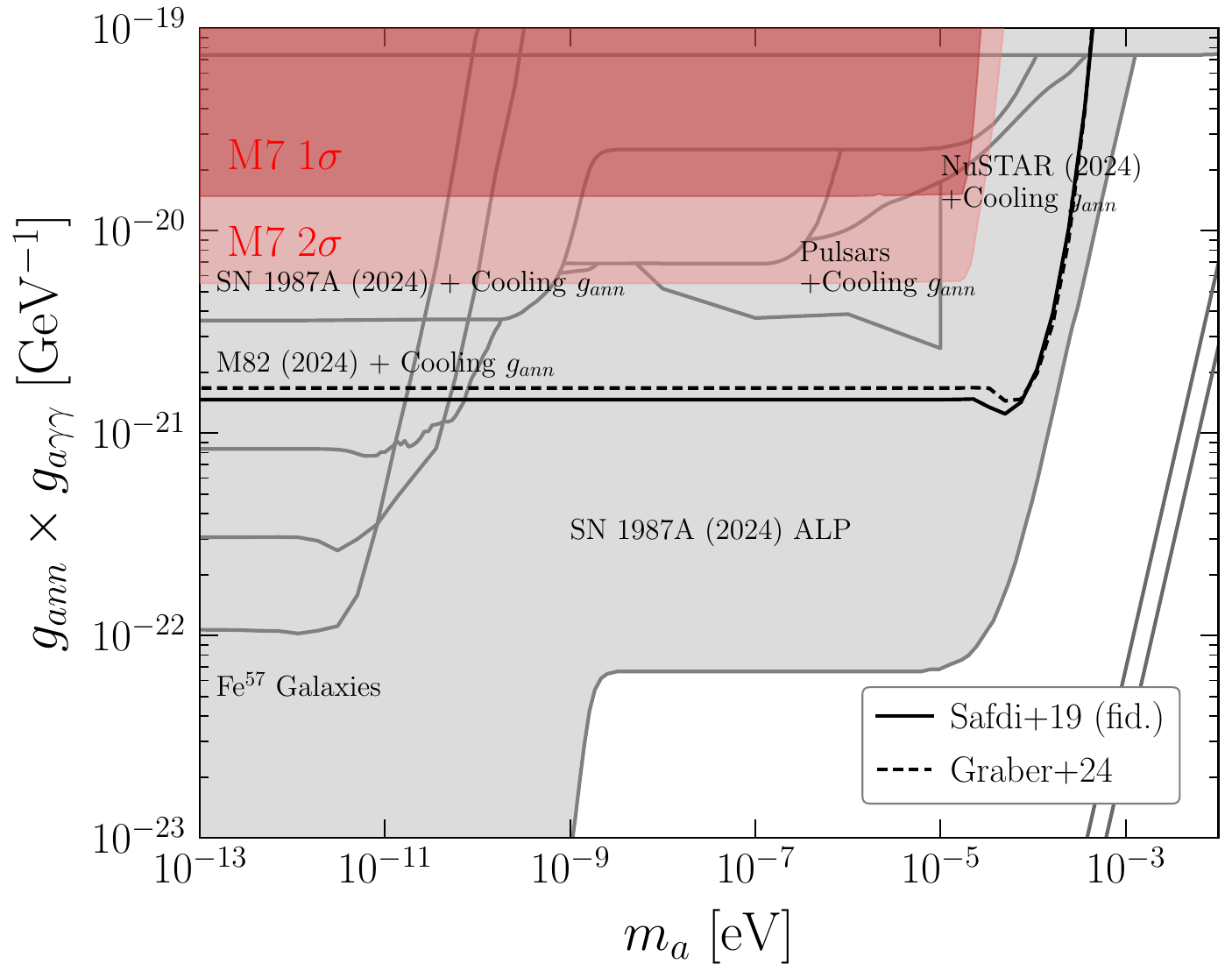}
\vspace{-0.4cm}
\caption{(Left) The lifetime integrated and NS mass and $B_0$-averaged photon flux from axion-to-photon conversions. We show results using the initial magnetic field distribution and decay model detailed in Model 2 of~\cite{Safdi:2018oeu} (Safdi+19) and the initial magnetic field and decay model of Ref.~\cite{Graber:2023jgz} (Graber+24). (Right) The resulting 95\% upper limits on $\ganngagg$ from these two models.}
\label{fig:B_model_limits}
\end{figure}

This entire prescription allows us to construct the complete NS ensemble relevant for ultralight axion searches. Note that we assume these parameter distributions for NS magnetospheres do not change appreciably over cosmic time. We illustrate the $B_0$ distribution for the two models, as well as the evolution of $B(t)$ over time, in Fig.~\ref{fig:GC_B}. In Fig.~\ref{fig:B_model_limits}, we additionally show the lifetime integrated and NS mass and $B_0$-averaged photon flux from axion-to-photon conversions for the two magnetic field models, along with the corresponding constraints on $g_{ann}\times g_{a\gamma\gamma}$ using fiducial NuSTAR CXB data, illustrating only minor differences in the final upper limits. For both magnetic field models, we assume the same fiducial axion production model.

\section{Analysis Procedure and Data Variations}
\label{app:analysis}
In this section, we briefly discuss our analysis procedure and the variations of data that we use to set constraints and future projections for axion coupling parameters. We construct our predicted axion-induced signal model, $S_i(m_a, \, g_{\rm coupling})$, where $m_a$ is the axion mass and $g_{\rm coupling}$ is equal to $\ganngagg$ for our ultralight axion analysis, and $\gagg$ for our heavy axion analysis. The index $i$ runs over energy bins, and $S_i$, which we take as a flux quantity, is in units of, \textit{e.g.}, keV/cm$^2$/s/sr, which can be directly compared to CXB data.

As discussed in the main Letter, we combine our signal model with a profiled astrophysical background model to construct our total signal as follows:
\es{eq:analysis_model}{
\mu_i(m_a, g_{\rm coupling}, \boldsymbol{\theta}) = S_i(m_a, g_{\rm coupling}) + B_i(\boldsymbol{\theta}) \,,
}
where here $\boldsymbol{\theta}$ refers to the set of nuisance parameters used in the astrophysical background modeling. For our fiducial NuSTAR analysis, $\boldsymbol{\theta} = \{ A, n \}$, \textit{i.e.} the two parameters of the power law portion of the canonical parameterization in~\cite{Krivonos:2020qvl} (note that though there is a high energy exponential cutoff scale $E_{\rm exp}$, as explained in~\cite{Krivonos:2020qvl} we fix this value to the canonical value of 41.13 keV given the overall weakness in constraining this parameter using the NuSTAR data below 20 keV). In our analysis of HEAO, Swift, and INTEGRAL up to energies $\sim$100 keV, we follow~\cite{1999ApJ...520..124G} and set $\boldsymbol{\theta} = \{ A, n_1, n_2, E_{\rm break},E_{\rm exp}\}$, which adds a broken power law component at higher energies (\textit{e.g.} $\gtrsim$ 60 keV) compared to the model used for NuSTAR.

\begin{figure}[!htb]
\centering
\includegraphics[width=0.49\textwidth]{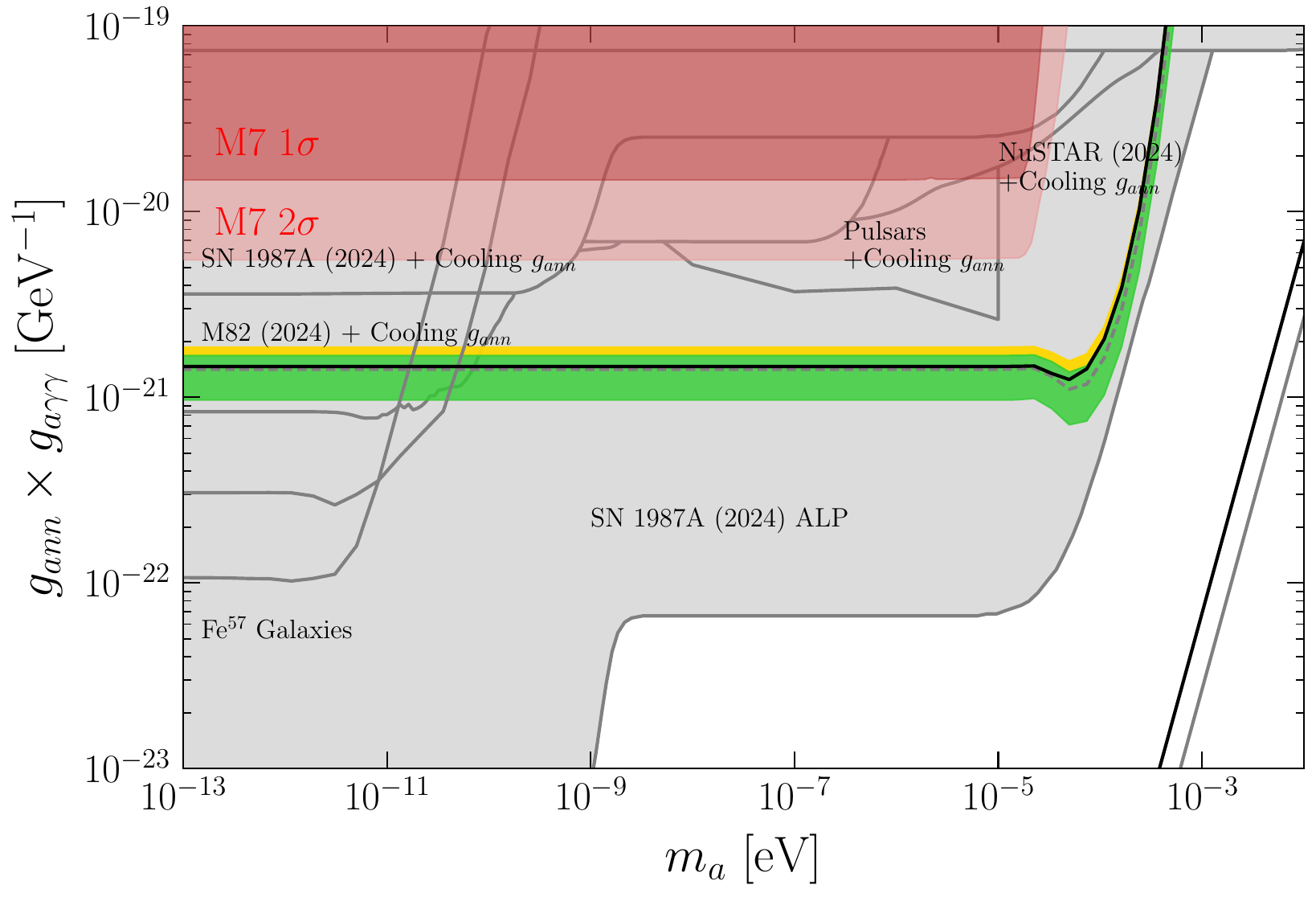}
\includegraphics[width=0.49\columnwidth]{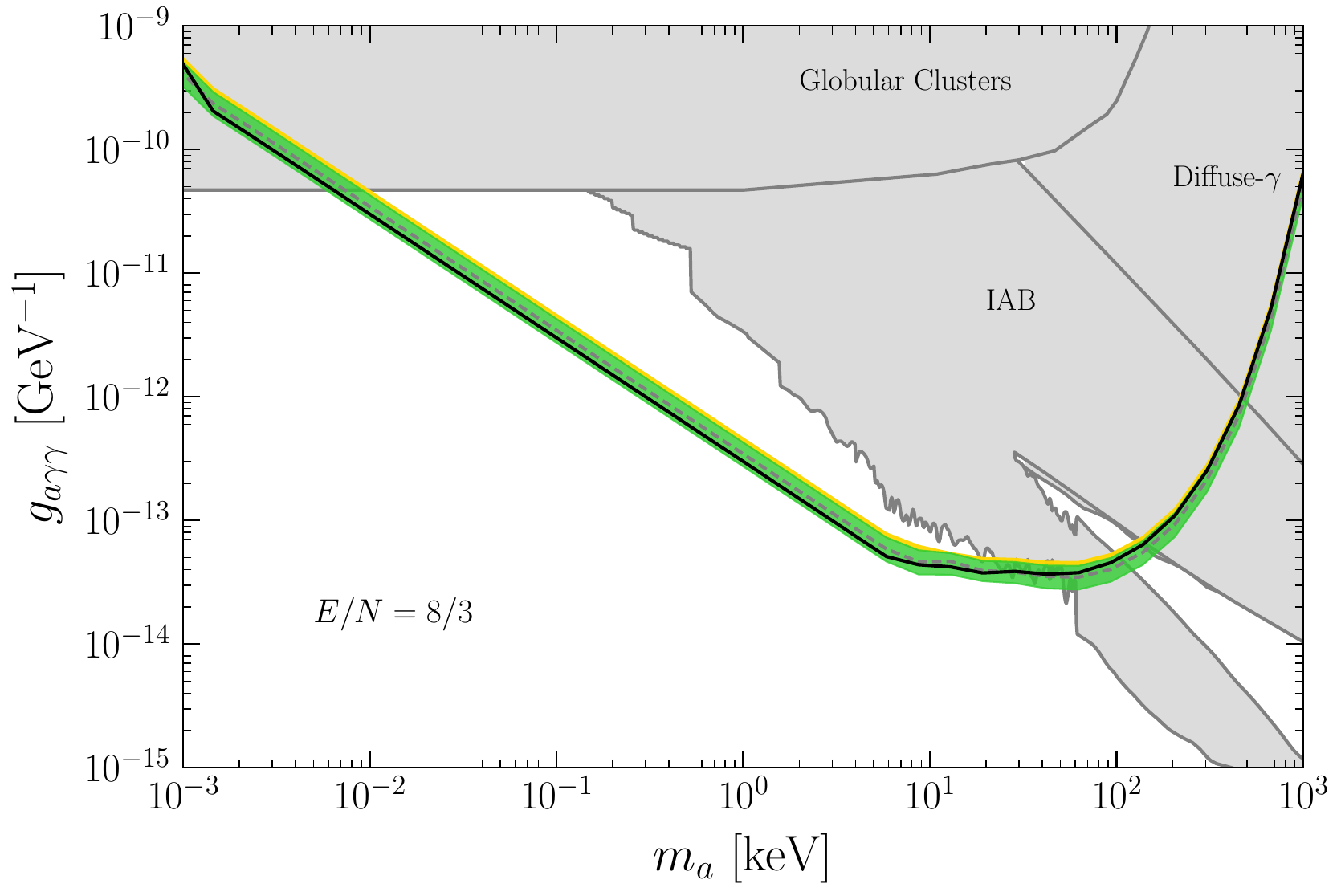}
\vspace{-0.4cm}
\caption{(Left) The derived 95\% upper limits (black) on $\ganngagg$ for ultralight axions in our fiducial model, compared to the $1\sigma/2\sigma$ expectations for the 95\% upper limits (green/gold bands). (Right) The same but for heavy axion limits. Note that in both cases our 95\% upper limits are power-constrained.}
\label{fig:asimov}
\end{figure}

We incorporate our total signal $\mu_i$ into a Gaussian spectral likelihood to compare to the CXB data. Fixing $m_a$, we proceed by profiling over the nuisance parameters at each $g_{\rm coupling}$. This likelihood is then used to construct the test statistic 
\es{eq:TS}{
    q(g_{\rm coupling} | m_a) \equiv 2 \ln p(\boldsymbol{d} | \{m_a, g_{\rm coupling}\}) - 2 \ln p(\boldsymbol{d} | \{ m_a, \Bar{g}_{\rm coupling}\})\,,
}
where $\boldsymbol{d}$ is the data, and $\bar{g}_{\rm coupling}$ is the signal strength which maximizes the likelihood (see, \textit{e.g.}~\cite{Cowan:2010js} and note this quantity could be negative). The 95\% power-constrained~\cite{Cowan:2011an} upper limit is derived through Wilks' theorem, and is given by the value $g_{\rm coupling} > \bar{g}_{\rm coupling}$ such that $q(g_{\rm coupling} | m_a) \approx 2.71$. From the Asimov procedure, we can also derive the $1\sigma$ and $2\sigma$ expectations for the 95\% upper limits under the null hypothesis, with details in~\cite{Cowan:2010js} (see also~\cite{Safdi:2022xkm} for a review). These expectations are illustrated as green and gold bands in Fig.~\ref{fig:asimov} for our fiducial model. 

As discussed in the main Letter, for our fiducial analysis we utilize the NuSTAR CXB data measured from over $\sim$7 Ms of extragalactic deep field observations~\cite{Krivonos:2020qvl}. This measurement is limited by instrumental systematics inhibiting further determination of the CXB spectra above $\sim$20 keV, which include systematics such as charged particle interactions, activation lines, stray light sources, cosmic rays, and Earth's own reflected hard X-ray emission~\cite{Wik:2014, 2007MNRAS.377.1726S, 2008MNRAS.385..719C}. Below $\sim$20 keV, however, the data are largely consistent with the canonical CXB model detailed in~\cite{1999ApJ...520..124G}, and which is thought to come primarily from AGN emissions (though see~\cite{Churazov:2006bk}). This CXB measurement is used to obtain our fiducial results highlighted in the main Letter.

On the other hand, we also examine how our constraints might change with a different choice of CXB data. Toward that end, we examine additional CXB measurements from the HEAO instrument~\cite{1999ApJ...520..124G}, Swift~\cite{Ajello:2008xb}, and INTEGRAL~\cite{Churazov:2006bk}. 

\begin{figure}[!t]
\centering
\includegraphics[width=0.49\textwidth]{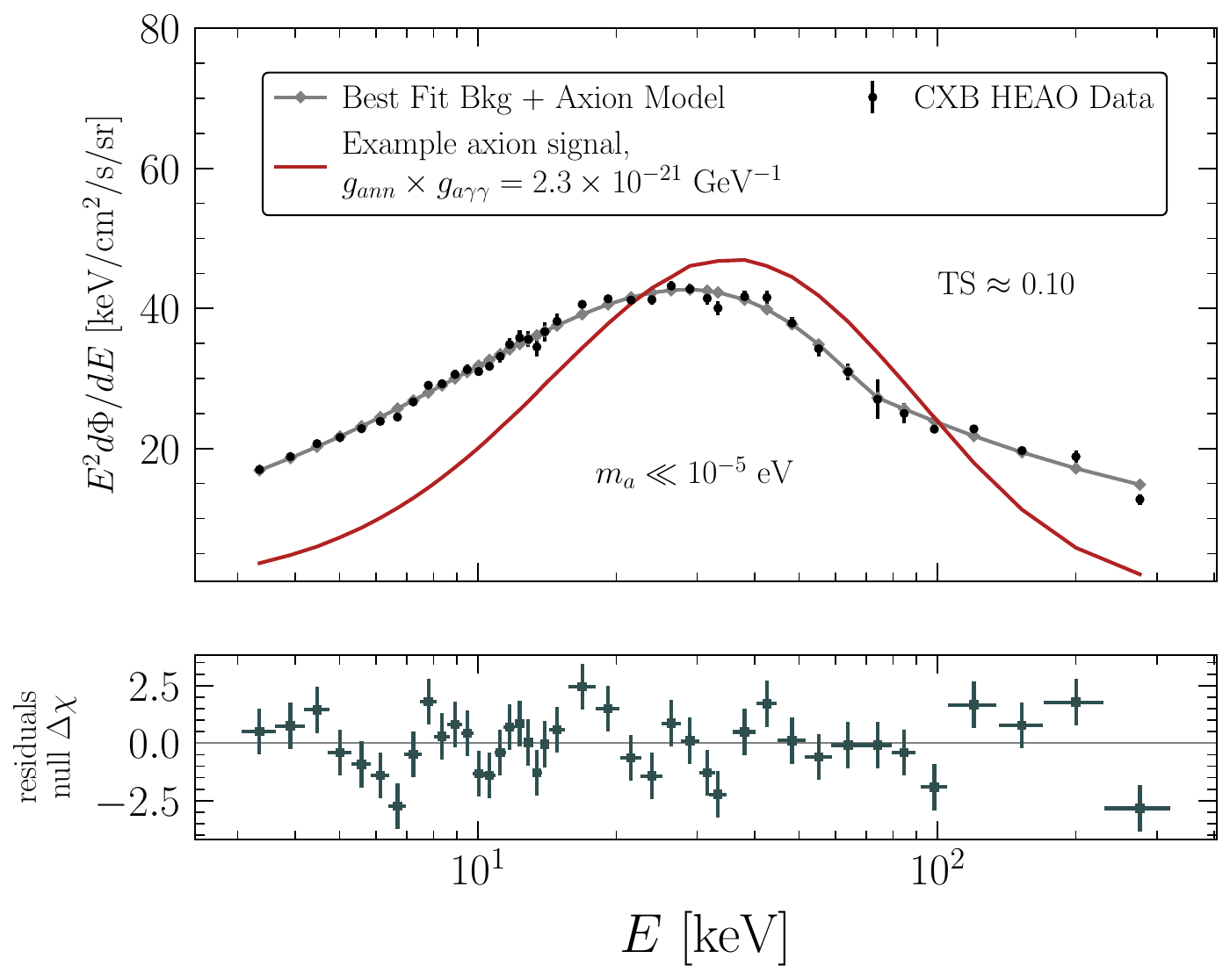}
\includegraphics[width=0.49\textwidth]{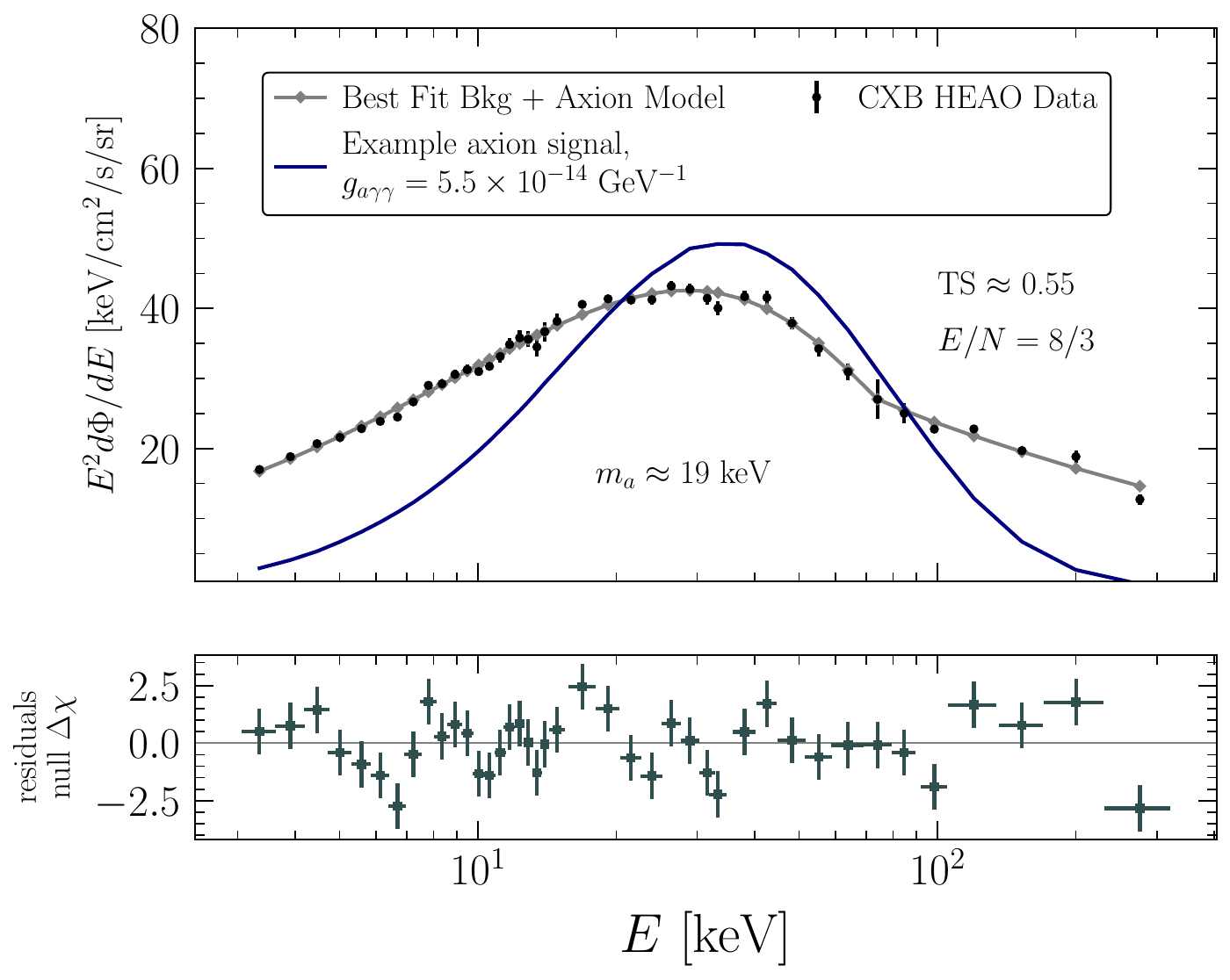}
\vspace{-0.4cm}
\caption{(Left) An illustration of our ultralight axion-induced signal compared to measured HEAO-1 CXB data~\cite{1999ApJ...520..124G}. We show our best-fit axion and background model (gray) as well as an example axion signal (red) with the indicated coupling, in the low $m_a$ limit. We also illustrate the residuals of the data compared to the fitted background model under the null hypothesis in the lower panel. (Right) The same but for our heavy axion search.}
\label{fig:data_HEAO}
\end{figure}

First, we examine the HEAO-1 run of the HEAO instrument taken from~\cite{1999ApJ...520..124G}. Although measured up to $\sim$200 keV, the HEAO measurement of the CXB comes at the cost of a lower overall effective area and greater systematic uncertainties, due to its older collimated detector optics which are non-focusing. Consequently, as pointed out in~\cite{1999ApJ...520..124G}, parts of the measured spectra are susceptible to artifacts of the spectral inversion procedure, which should be kept in mind when interpreting the data. For these reasons, we instead take the measured NuSTAR CXB discussed in the main Letter as our fiducial analysis. We show analogous constraints on $\ganngagg$ using this dataset in Figs.~\ref{fig:ultralight_limits} and~\ref{fig:heavy_limits}, which results in modestly stronger constraints. For ultralight axions, at low mass, $|\ganngagg| \lesssim 1.05 \times 10^{-22}$ GeV$^{-1}$, with a significance in favor of the axion model of $0.31\sigma$. However, here HEAO is able to probe slightly higher axion masses than NuSTAR due to its ability to probe higher energies coupled with the fact that higher-mass axions induce spectra peaked at higher energies (see, \textit{e.g.}, Fig.~\ref{fig:cosmo_axion_spec}). Similarly, for heavy axions, constraints on $\gagg$ from HEAO are slightly stronger than that for NuSTAR, and is more pronounced at higher masses. We show an illustration of our data analyses in Fig.~\ref{fig:data_HEAO}.

\begin{figure}[!t]
\centering
\includegraphics[width=0.49\textwidth]{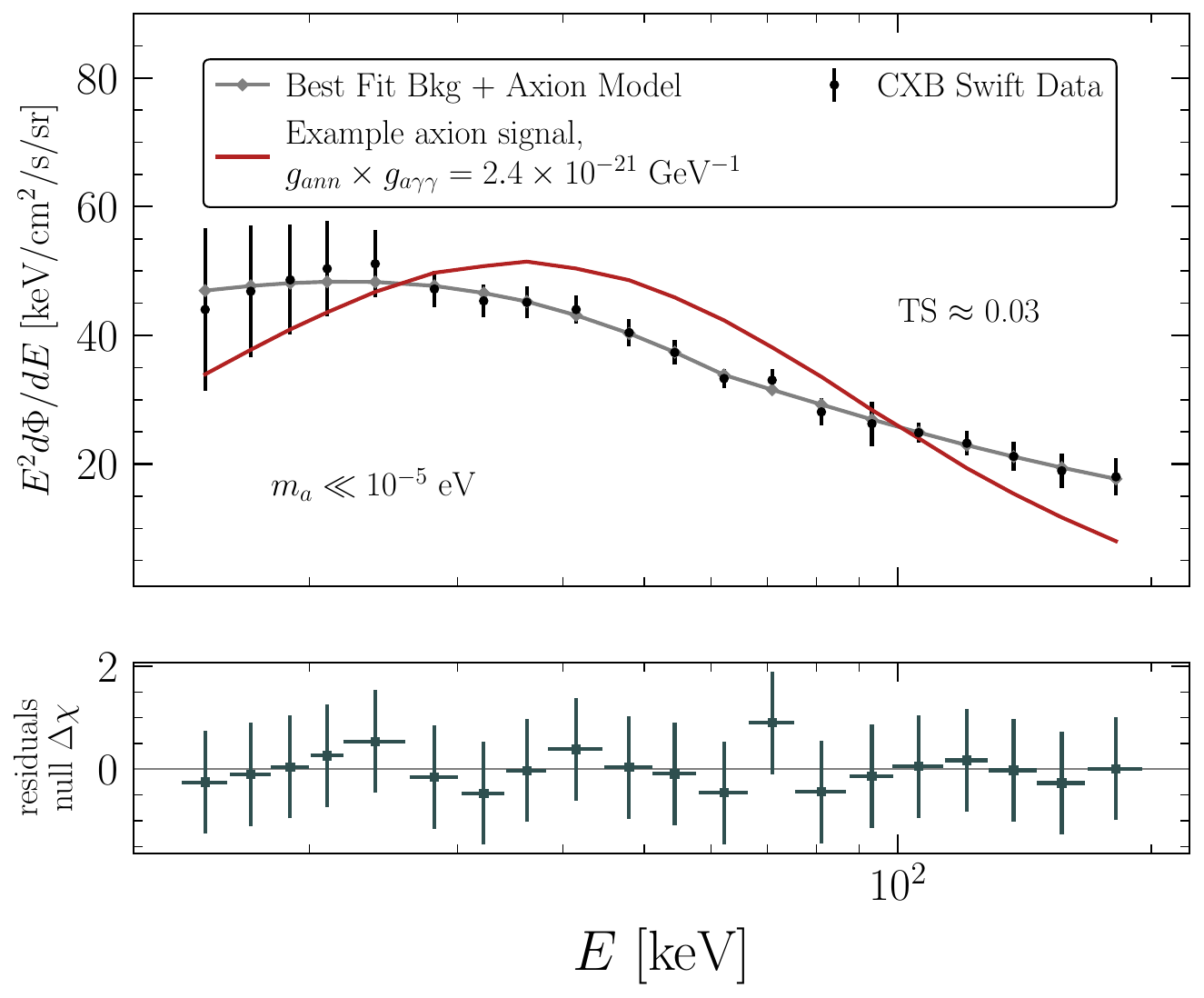}
\includegraphics[width=0.49\textwidth]{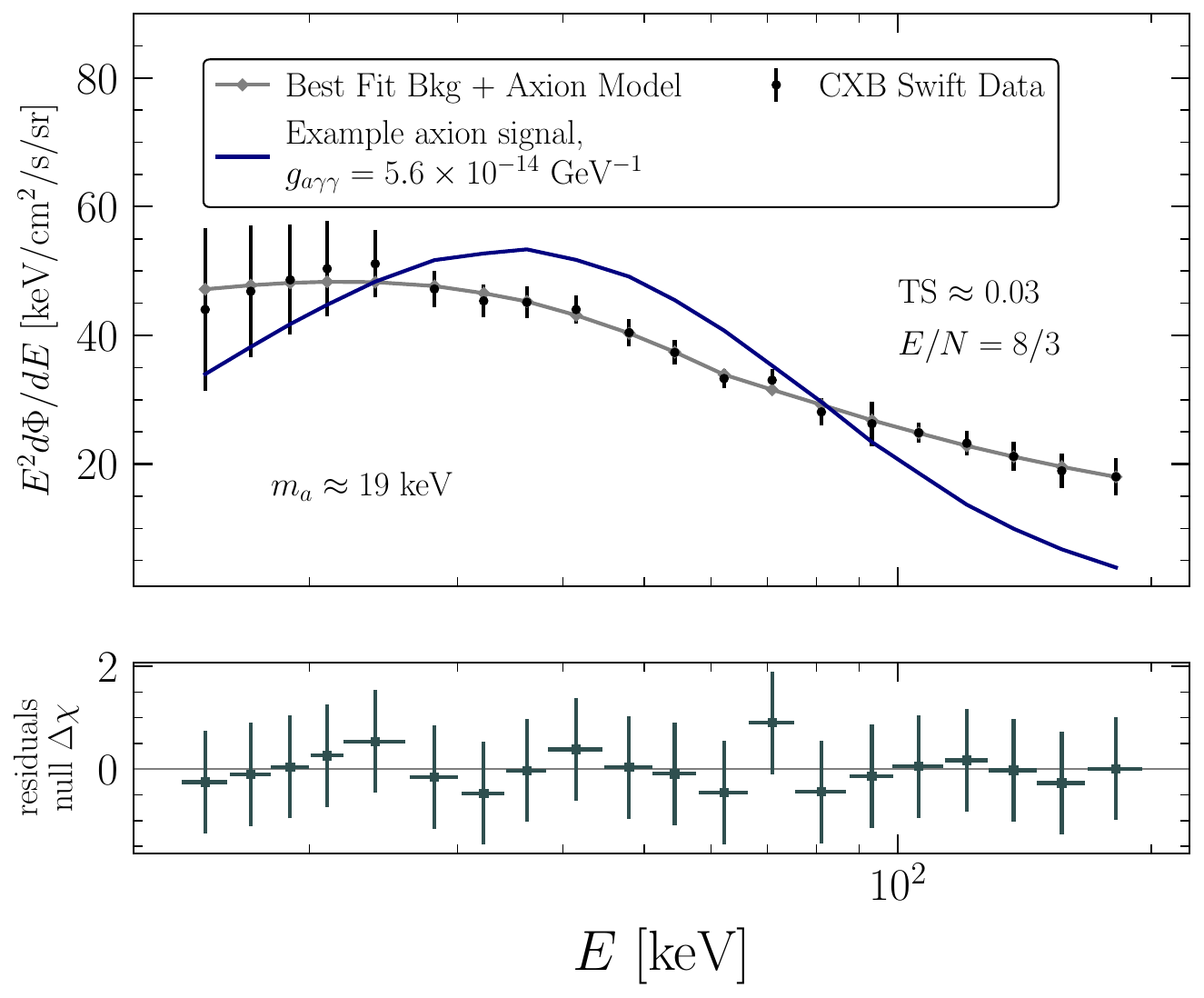}
\vspace{-0.4cm}
\caption{The same as Fig.~\ref{fig:data_HEAO} but for the measured Swift BAT CXB data~\cite{Ajello:2008xb}.}
\label{fig:data_SwiftBAT}
\end{figure}

\begin{figure}[!t]
\centering
\includegraphics[width=0.49\textwidth]{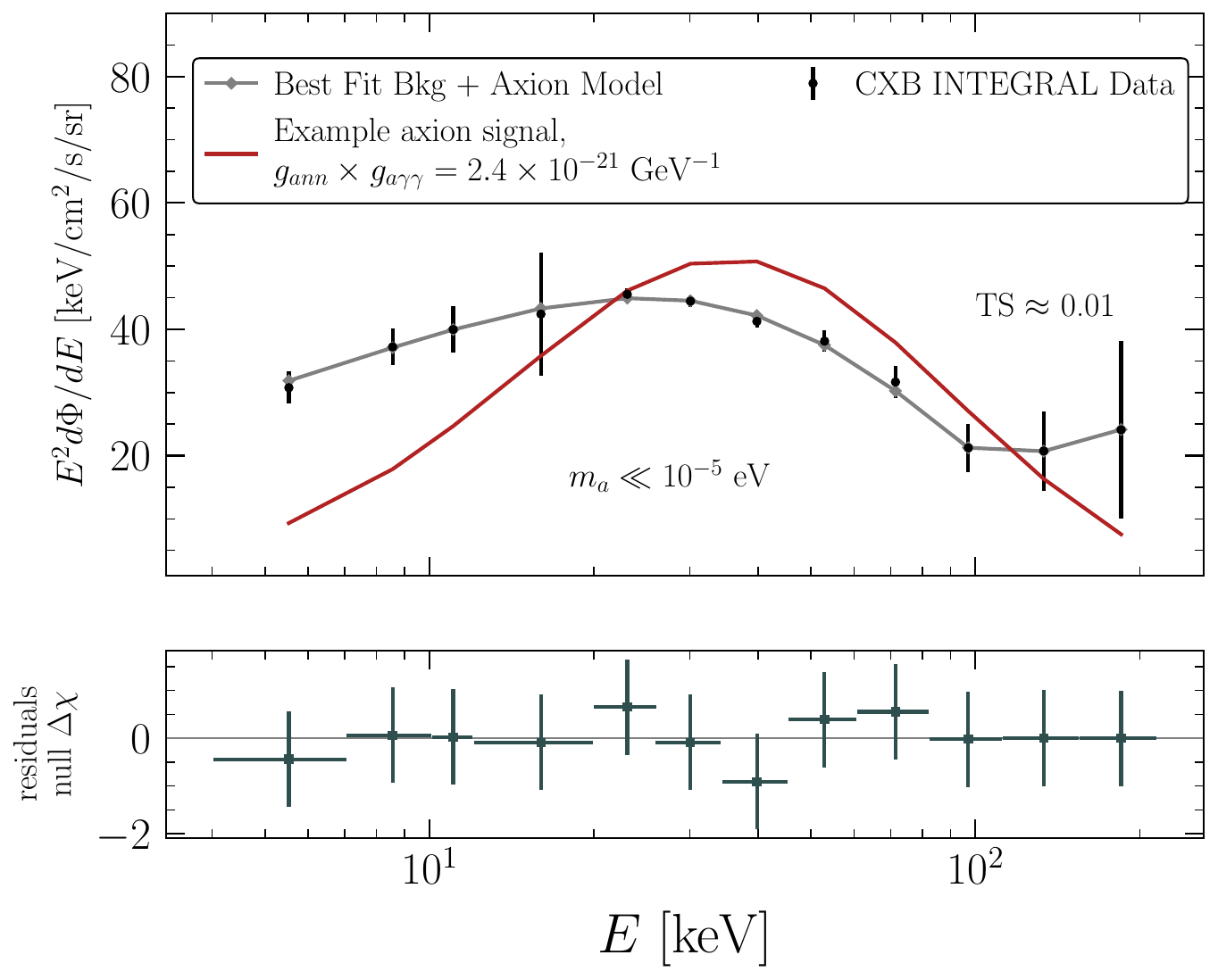}
\includegraphics[width=0.49\textwidth]{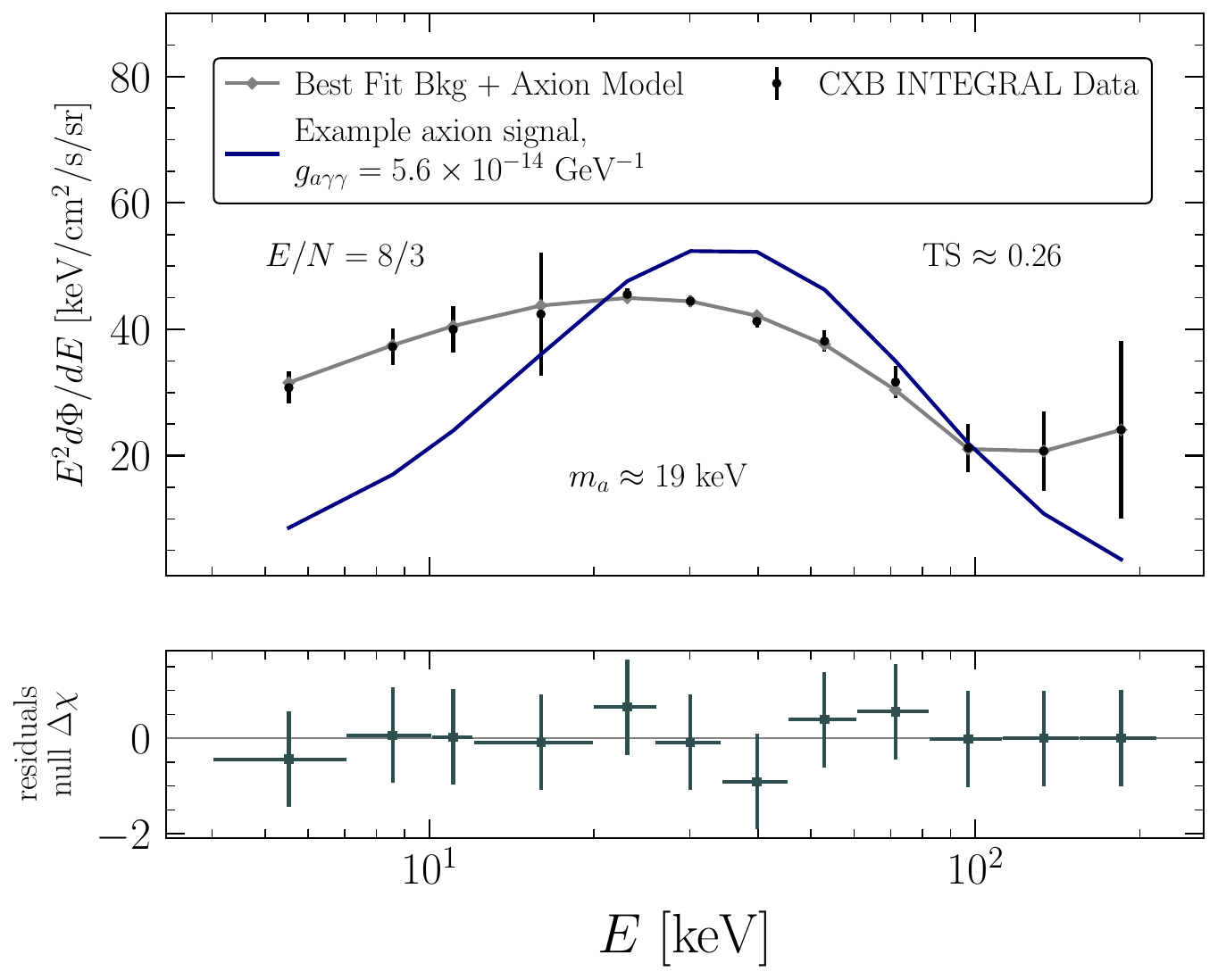}
\vspace{-0.4cm}
\caption{The same as Fig.~\ref{fig:data_HEAO} but for the measured INTEGRAL data derived using the JEM-X, IBIS/ISGRI, and SPI instruments~\cite{Churazov:2006bk}.}
\label{fig:data_INTEGRAL}
\end{figure}

Similarly, we find that the CXB measurements made by the BAT instrument on board the Swift observatory~\cite{Ajello:2008xb}, as well as that made by the JEM-X, IBIS/ISGRI, and SPI instruments on board INTEGRAL~\cite{Churazov:2006bk} (all going up to $\sim\mathcal{O}(100)$ keV), result in constraints comparable to that of HEAO for both our ultralight and heavy axion searches. We note that the HEAO constraints are overall marginally stronger than that by Swift and INTEGRAL, partially due to lower quoted systematic uncertainties, though we repeat that there are caveats regarding potential artifacts in some parts of the spectral reconstruction as discussed in~\cite{1999ApJ...520..124G}. We illustrate our data analyses for Swift and INTEGRAL in Fig.~\ref{fig:data_SwiftBAT} and~\ref{fig:data_INTEGRAL}, respectively.

\begin{figure}[!t]
\centering
\includegraphics[width=0.49\textwidth]{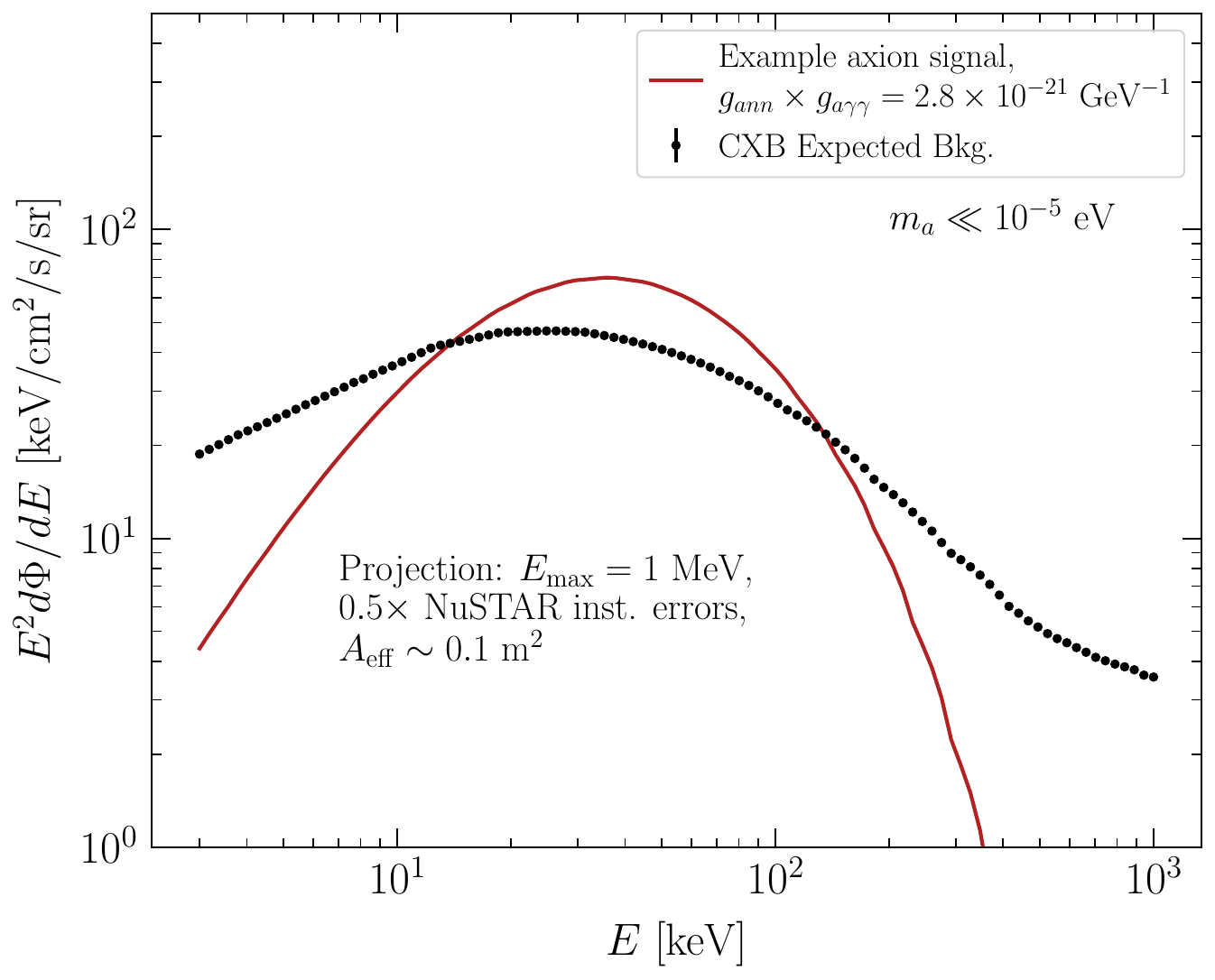}
\vspace{-0.4cm}
\caption{An illustration of the ultralight axion-induced signal and expected background for a future prospective telescope, corresponding to the red dashed-dot line in Fig.~\ref{fig:ultralight_limits}. This telescope, whose labeled properties were chosen to extend the constraints of our fiducial analysis, has the important property of extending to higher energies than the ones probed in our fiducial analysis with NuSTAR; these higher energies illustrate the contrast between a purported axion-induced signal (an example with the indicated coupling shown in red) and expected CXB data. See main text for details.}
\label{fig:data_proj}
\end{figure}

It is interesting, then, to consider how higher energy data are useful to further constrain the axion signal; given that our emitted axion signal, \textit{e.g.} in Fig.~\ref{fig:cosmo_axion_spec}, in principle has support up to $\mathcal{O}(100)$ keV and even $\mathcal{O}(1)$ MeV, a future telescope that is able to probe these scales -- where astrophysical backgrounds tend to be lower -- would be likely to even more stringently probe the axion parameters considered in this work. We make projections for the constraint on $\ganngagg$ that such a future telescope might be able to achieve, specifically considering the QCD axion as our theoretical prior in the case of ultralight axion searches. We construct an optimistic future telescope that can probe photons in a frequency band extending from $\sim$1 keV to roughly $\sim$1 or $\sim$10 MeV, and assume the same CXB background from~\cite{1999ApJ...520..124G} with an additional MeV-scale background component as prescribed in~\cite{2009ApJ...699..603A}, whose dominant source above $\sim$500 keV is diffuse emission from quasars (see also~\cite{2009ApJ...702..523I, 2011ApJ...733...66I}). We equip this telescope with a flat effective area of $\sim$0.1 m$^2$, partially inspired by Fermi-LAT, and we also reduce the instrumental errors on this hypothetical telescope by a factor of $\sim$2 and $\sim$20 compared to that of the NuSTAR CXB data used in our fiducial analysis~\cite{Krivonos:2020qvl}. The combination of these factors, driven mostly by the higher energy reach and the reduction of instrumental uncertainties, allow us to begin approaching the QCD axion line in the $\ganngagg$ vs. $m_a$ plane, as seen in Fig.~\ref{fig:ultralight_limits}. An illustration of an example axion-induced signal and the expected background is shown in Fig.~\ref{fig:data_proj}. While we caution that the parameters for this future telescope are highly optimistic, we highlight the fact that future searches for the diffuse axion signal from cosmological NS populations explored in this work can cover even larger portions of unprobed parameter space. 

\section{Astrophysical Uncertainties}
\label{app:systematics}
In this section, we explore how our results vary when considering uncertainties in our astrophysical modeling, which involves systematics at the level of individual NSs as well as the population and cosmology modeling parameters.

\noindent
{\bf NS Superfluidity and EOS---}At the level of individual NS modeling, we examine different configurations of the superfluidity within the NS, as well as the NS EOS. As discussed in Sec.~\ref{app:axion_production}, we primarily consider superfluidity from proton $^1S_0$ and neutron $^3P_2$ pairings in the NS core. To bracket the theoretical uncertainty on the nucleon pairing gaps and condensation temperatures, we follow Ref.~\cite{Sedrakian:2018ydt} and consider two pairing models for each type of superfluidity. For the proton $^1S_0$ gap, we consider a larger gap from Ref.~\cite{Baldo:2007jx} (BS07) and a smaller gap from Ref.~\cite{Baldo:1992kzz} (BCLL92). For the neutron $^3P_2$ gap, we consider a larger gap from Ref.~\cite{Baldo:1998ca} (BHF) and a smaller gap from Ref.~\cite{Ding:2016oxp} (SCGF). Since the neutron $^1S_0$ emission is suppressed and does not significantly change our results, we use only the model of Ref.~\cite{Schwenk:2002fq} (SFB) and do not bracket the uncertainty on this emission. In each model, the size of the pairing gap is a function of the Fermi momentum, and therefore density, so superfluidity occurs in a subfraction of the NS.

\begin{figure}[!t]
\centering
\includegraphics[width=0.49\textwidth]{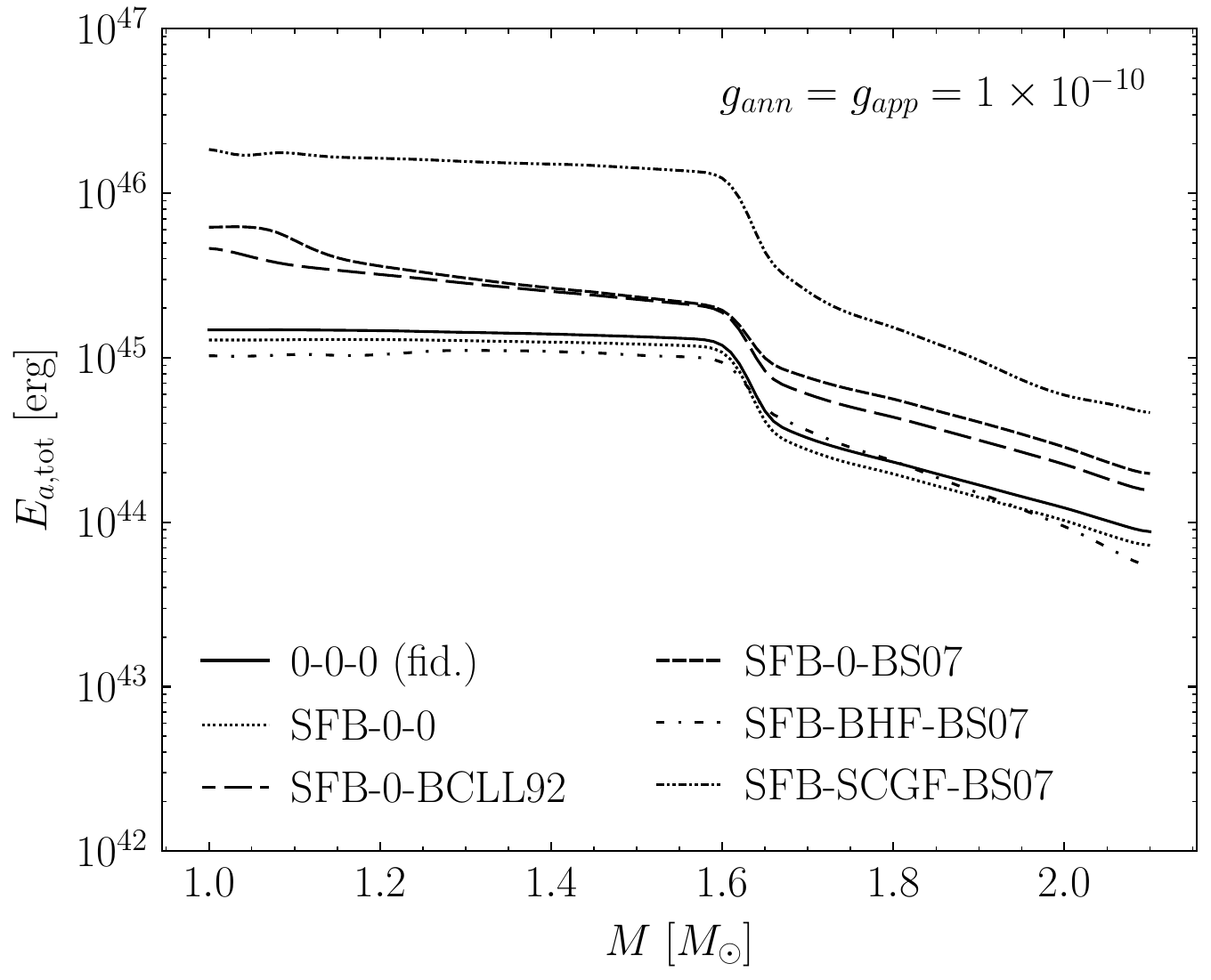}
\includegraphics[width=0.49\textwidth]{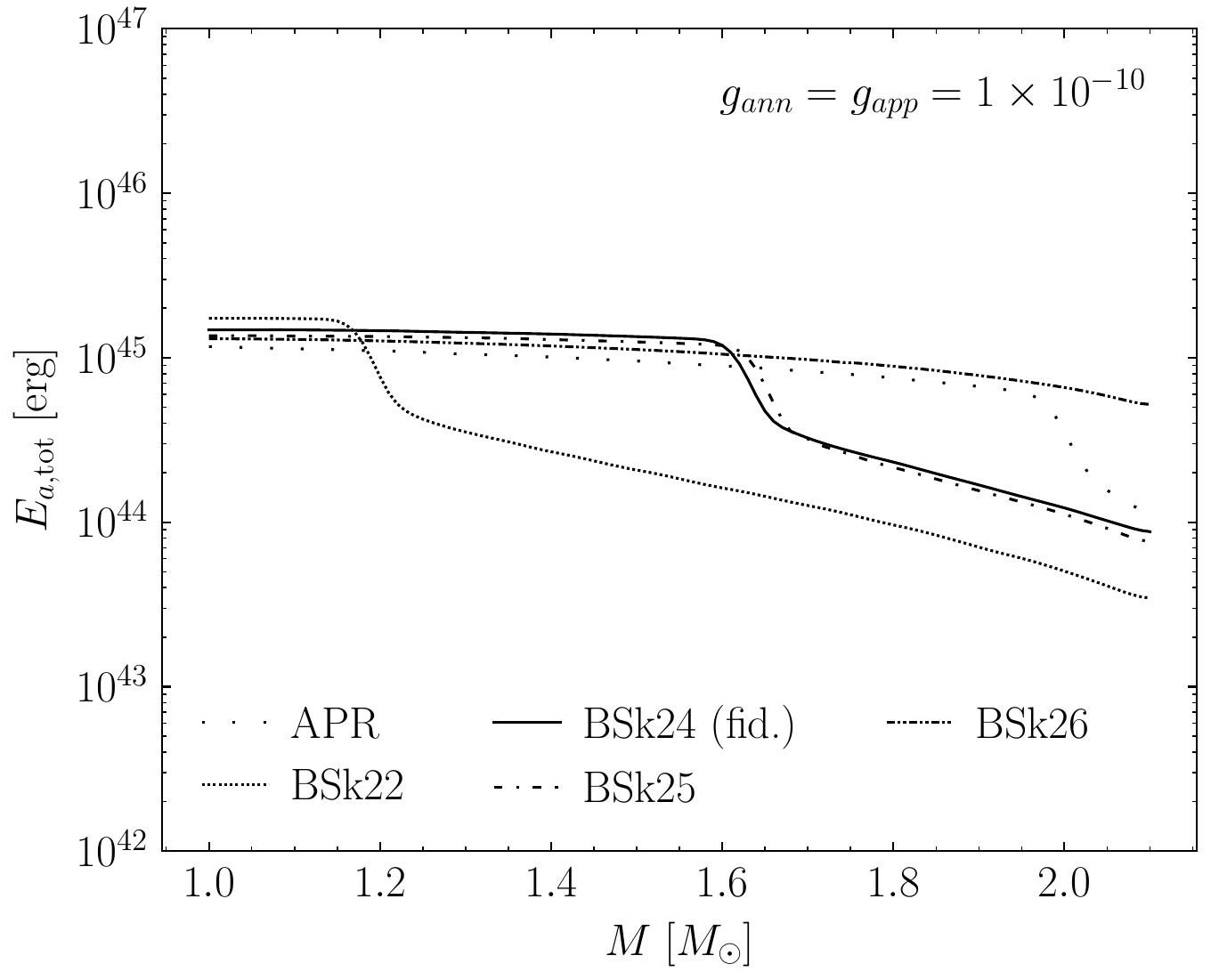}
\vspace{-0.4cm}
\caption{(Left) The total energy emitted as axions over the NS lifetime as a function of the NS mass for the full range of superfluidity models considered in this work. All superfluidity models shown here use the BSk24 EOS. (Right) The total energy emitted as axions over the NS lifetime as a function of the NS mass for the full range of EOS models considered in this work. All EOS models shown here contain no superfluidity.}
\label{fig:EOM_SF}
\end{figure}

\begin{figure}[!htb]
\centering
\begin{minipage}{0.49\linewidth}
\centering
\includegraphics[width=\linewidth]{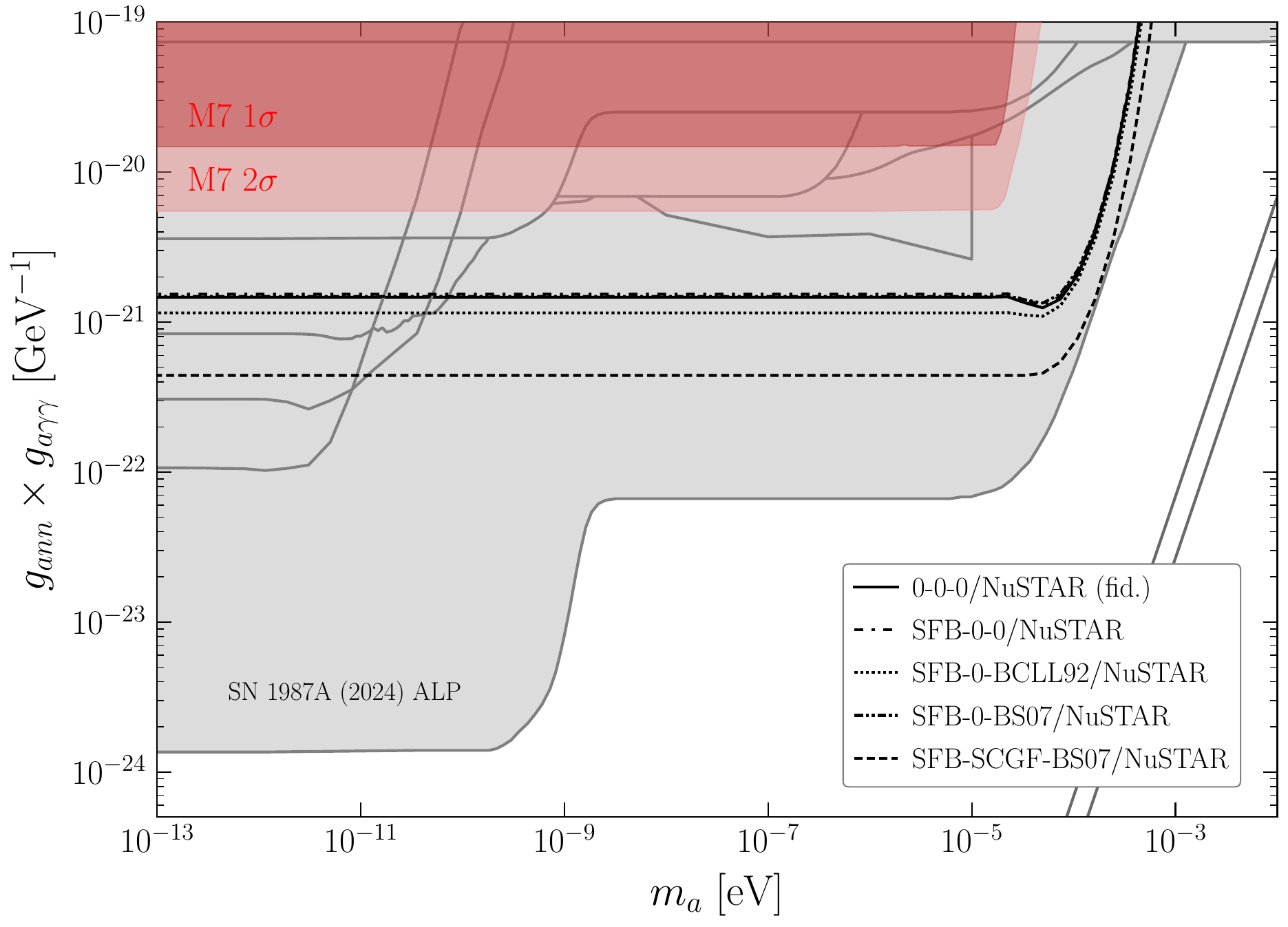}
\end{minipage}\hfill
\begin{minipage}{0.49\linewidth}
\centering
\includegraphics[width=\linewidth]{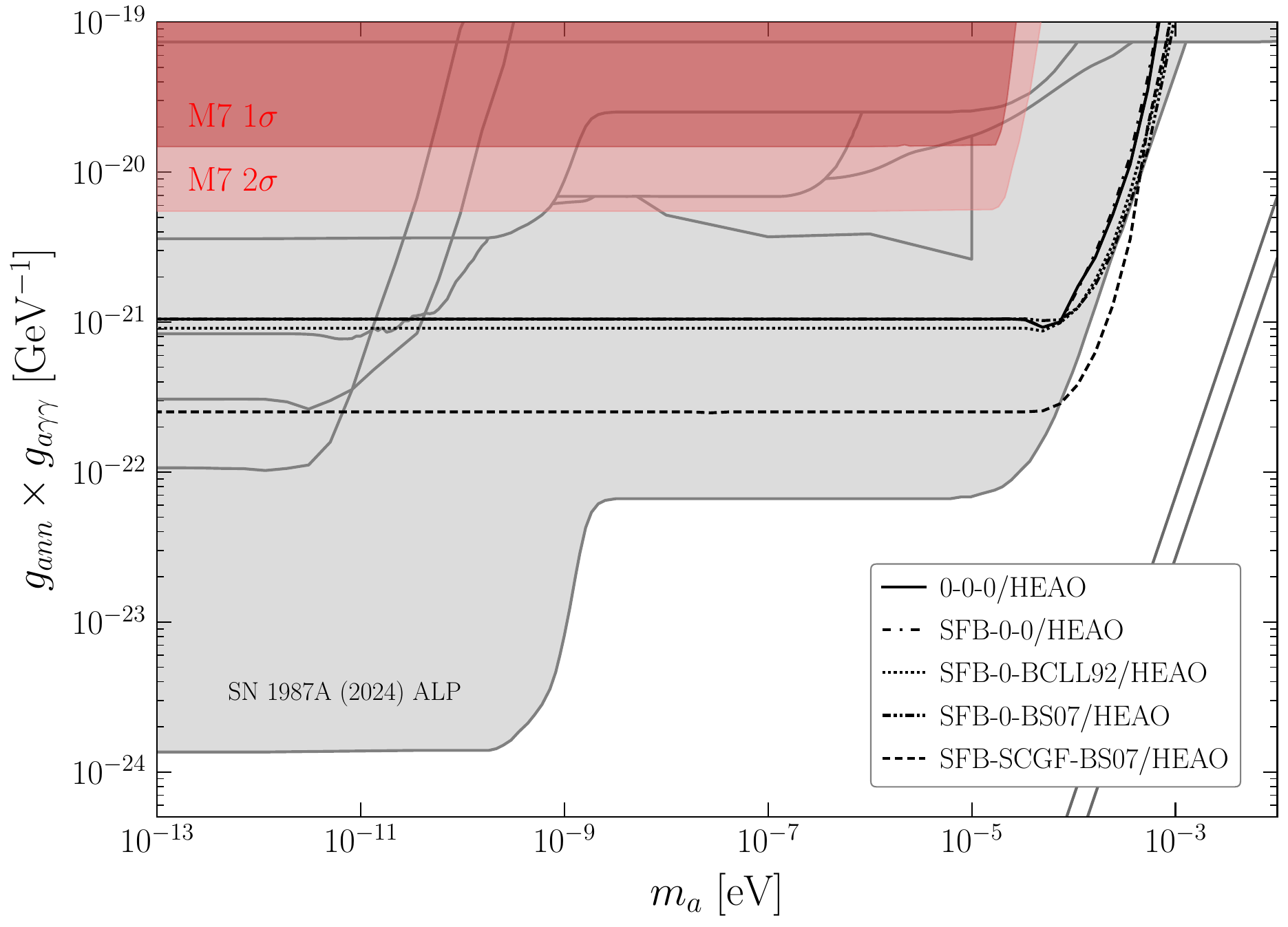}
\end{minipage}

\vspace{0.1cm}

\begin{minipage}{0.49\linewidth}
\centering
\includegraphics[width=\linewidth]{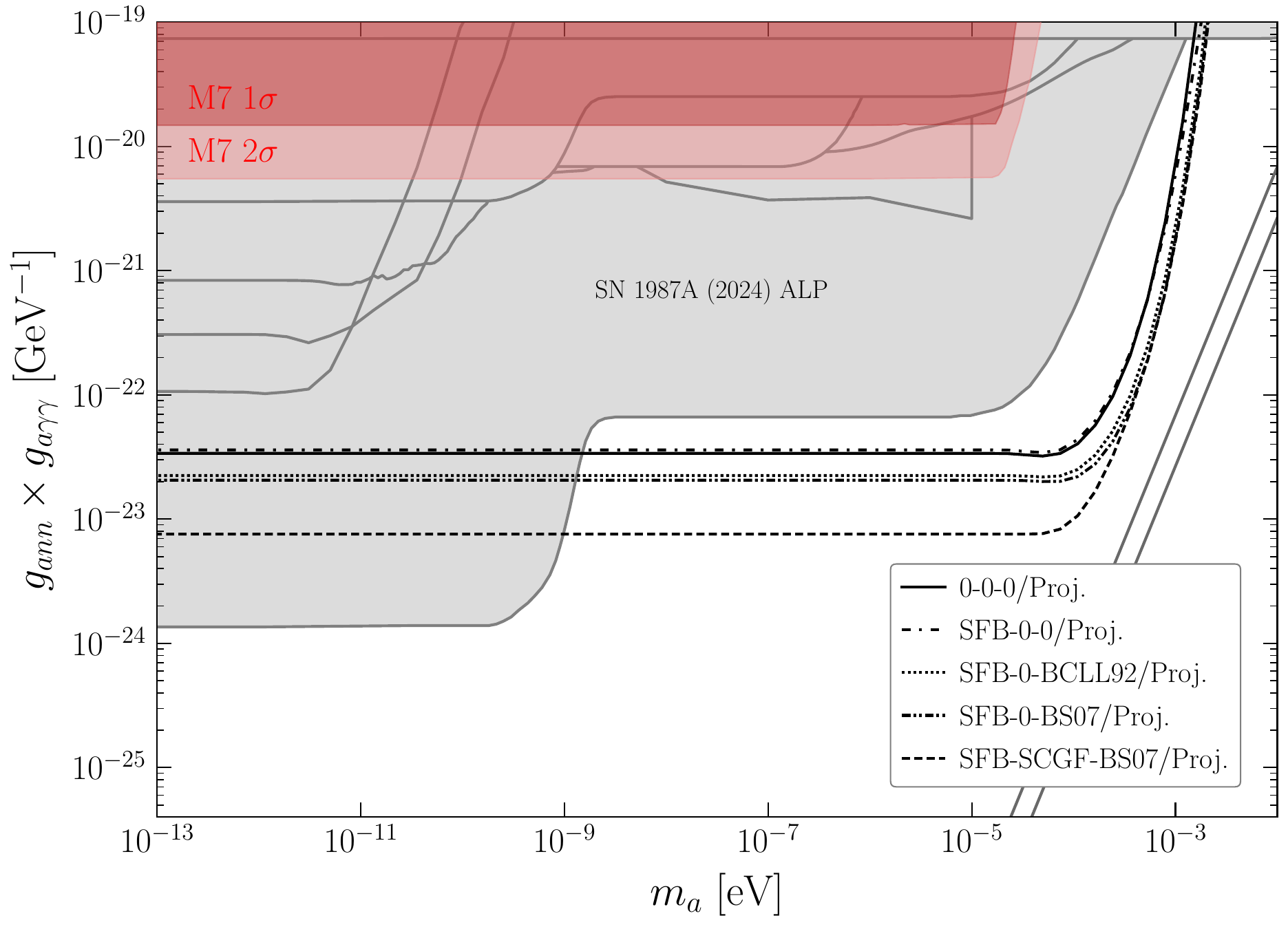}
\end{minipage}

\vspace{0.1cm}

\begin{minipage}{0.49\linewidth}
\centering
\includegraphics[width=\linewidth]{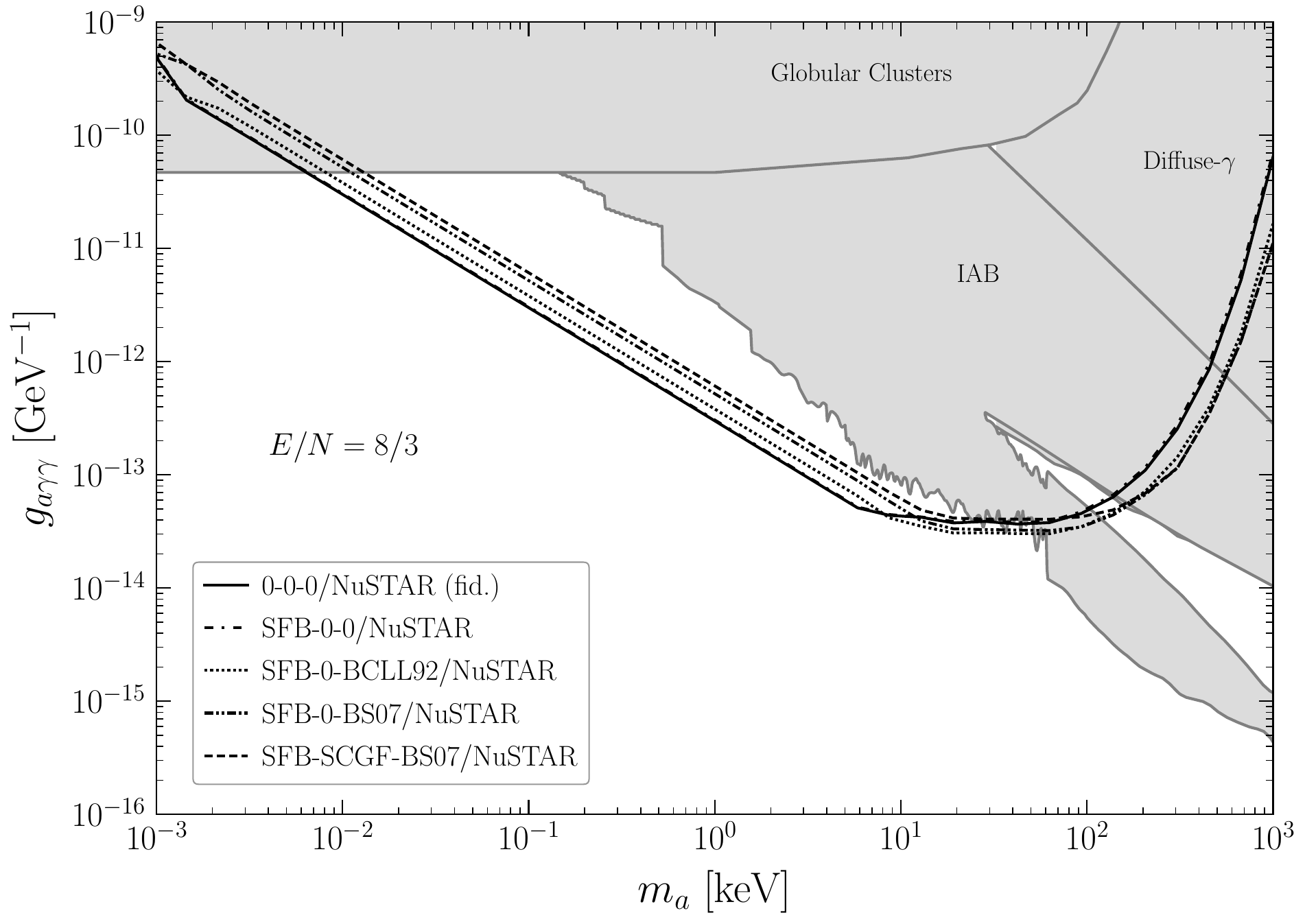}
\end{minipage}\hfill
\begin{minipage}{0.49\linewidth}
\centering
\includegraphics[width=\linewidth]{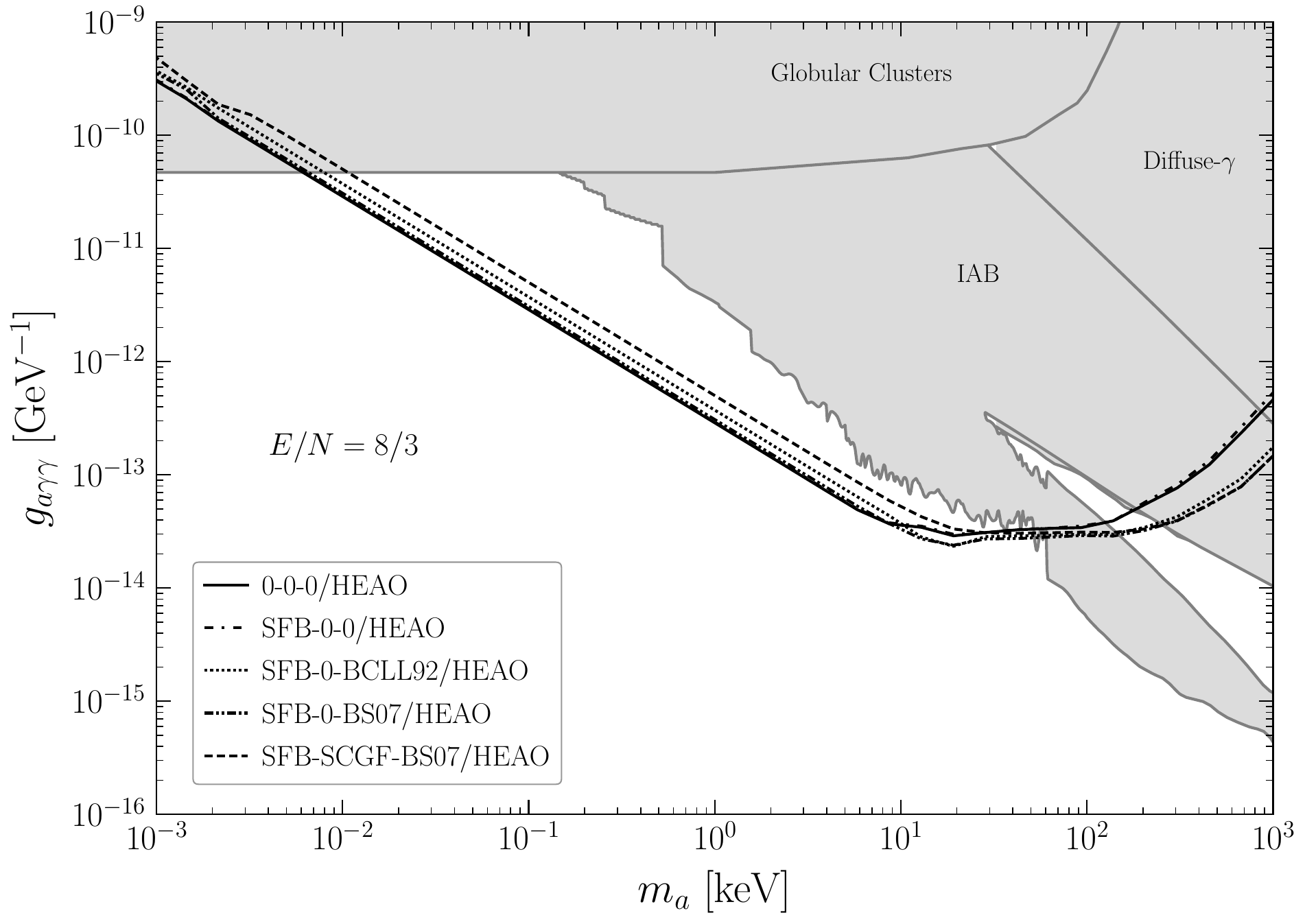}
\end{minipage}

\caption{(Top Left) An illustration of the effect of our main superfluidity models (including \texttt{0-0-0} indicating no superfluidity) on the inferred 95\% upper limit on $\ganngagg$ for ultralight axions using our fiducial NuSTAR CXB observations. (Top Right) The same but for our HEAO analysis, for ultralight axions. (Center) The same but for our most optimistic ultralight axion projection (assuming a telescope measuring up to 10 MeV and having 5\% of the instrumental errors from the NuSTAR CXB measurement). (Bottom Left) The same as top left but for heavy axions using NuSTAR CXB data. (Bottom Right) The same but for heavy axions and HEAO data.}
\label{fig:compare_SF}
\end{figure}

In the left panel of Fig.~\ref{fig:EOM_SF}, we show the total energy emitted as axions over the NS lifetime for a range of superfluidity models as a function of the NS mass. We see that the model with only neutron $^1S_0$ superfluidity (\texttt{SFB-0-0}) minimally affects the lifetime axion emission relative to the fiducial no superfluidity model (\texttt{0-0-0}). The two models including both proton and neutron $^1S_0$ pairing have enhanced axion emission (\texttt{SFB-0-BCLL92} and \texttt{SFB-0-BS07}). This enhancement occurs since the superfluidity occurs in a subfraction of the star and induces strong PBF axion emission without enough PBF neutrino emission to dramatically affect the cooling of star. We illustrate the resulting axion constraints derived from these four models in Fig.~\ref{fig:compare_SF}, where we see that the axion flux enhancement from the proton $^1S_0$ pairing gap indeed slightly strengthens the resulting axion coupling limits relative to the model without superfluidity and to the model where only neutron $^1S_0$ pairing is turned on. From Fig.~\ref{fig:EOM_SF}, we observe that the model that additionally contains $^3P_2$ superfluidity with the larger energy gap (\texttt{SFB-BHF-BS07}) has suppressed axion emission since superfluidity occurs throughout the entire core of the star and leads to rapid neutrino cooling. However, the model with the smaller $^3P_2$ energy gap (\texttt{SFB-SCGF-BS07}) produces superfluidity only in the lower density outer core, leading to less enhanced neutrino cooling. This keeps the NS at higher temperatures for longer, resulting in greater axion emission. We show the constraints assuming the \texttt{SFB-SCGF-BS07} model in Fig.~\ref{fig:compare_SF}. This particular $P$-wave superfluid model greatly strengthens the axion limits in the ultralight case, but slightly weakens the limit for heavy axions given the relative suppression of the axion-neutron coupling in our benchmark GUT model. However, we note that there exist large theoretical uncertainties on $P$-wave superfluidity in NSs and, as mentioned earlier, possible inconsistencies with the cooling data of isolated NSs~\cite{Buschmann:2021juv}. A better understanding of $P$-wave superfluidity will be helpful for improving related NS axion searches in the future.

Regarding the EOS, as discussed in the main Letter we use the BSk24 model~\cite{Pearson:2018tkr} as our fiducial model, which is the most consistent with mass-radius data of the NSs discussed in~\cite{Buschmann:2021juv}. Here, we test four additional choices for the EOS: APR~\cite{Akmal:1998cf}, BSk22, BSk25, and BSk26~\cite{Pearson:2018tkr}. We note that the APR EOS handles many-body interaction effects using variational methods informed by nucleon-nucleon scattering data. On the other hand, the BSk class of EOSs come from fitting atomic mass data to the Skyrme effective interaction, with different Skyrme symmetry energies determining the different BSk models. These models, including the APR model, broadly give a sense of the stiffness of the EOS through phenomenological prescriptions. 

\begin{figure}[!htb]
\centering
\includegraphics[width=0.49\textwidth]{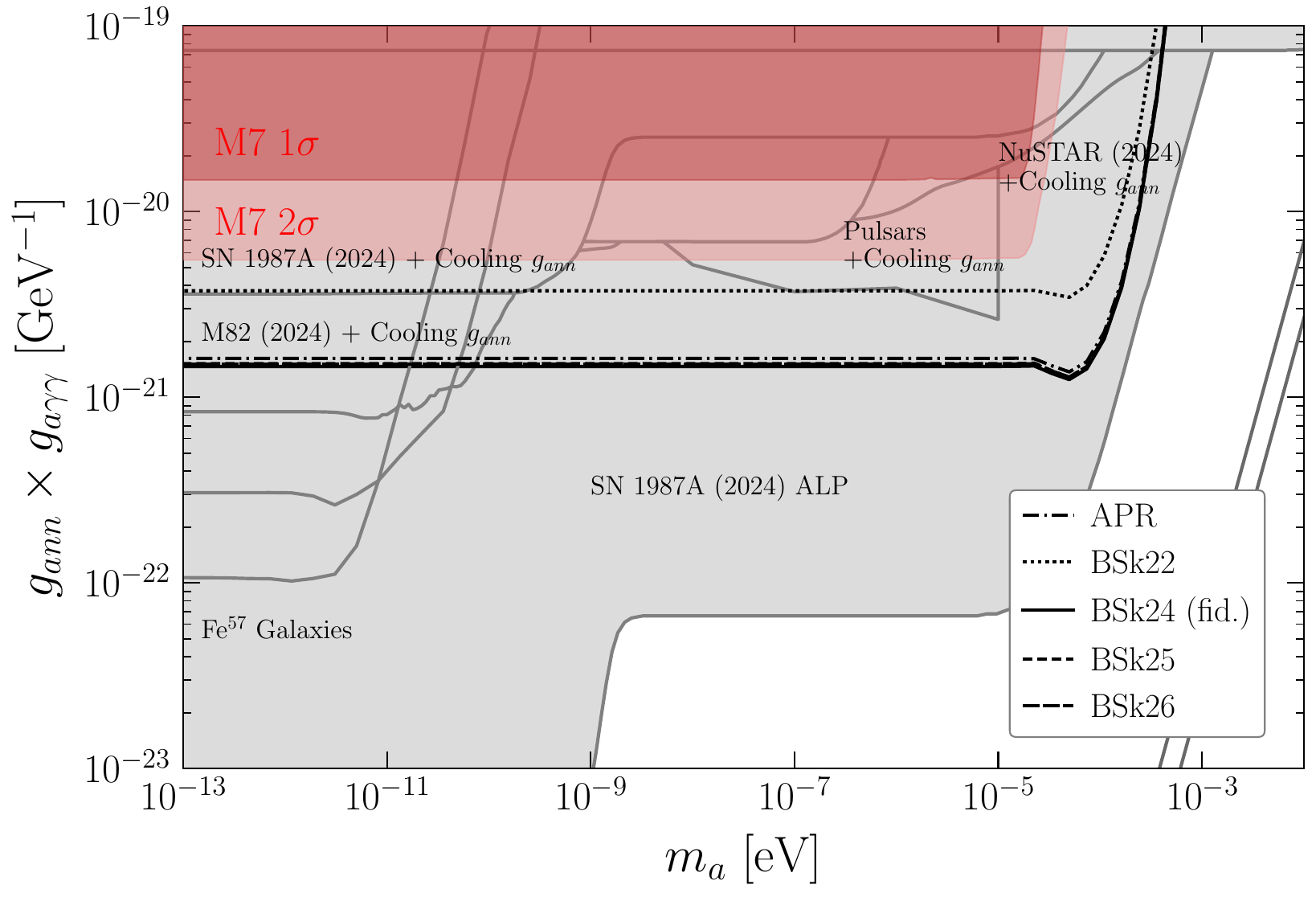}
\includegraphics[width=0.49\columnwidth]{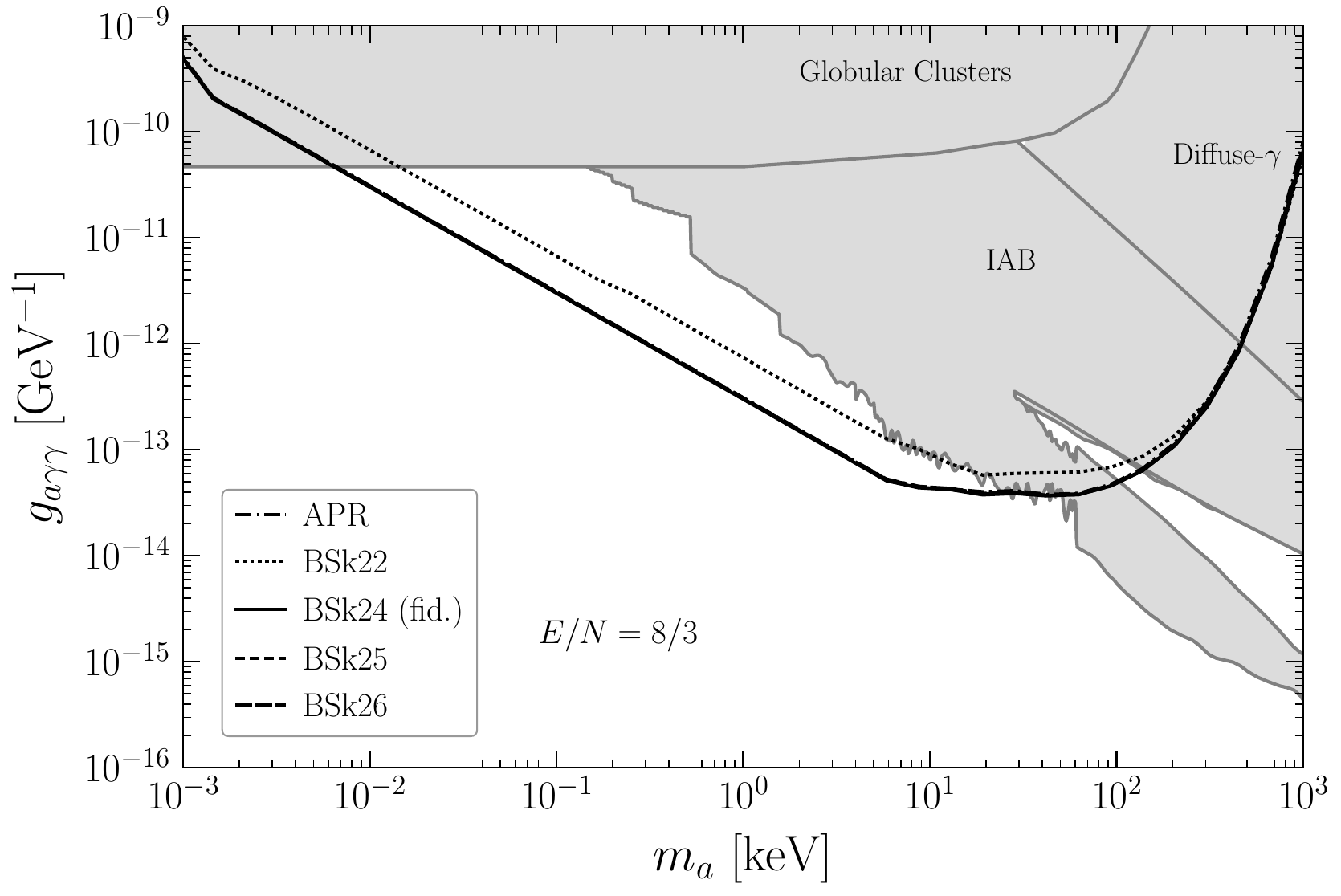}
\vspace{-0.4cm}
\caption{(Left) An illustration of variations on the 95\% upper limit on $\ganngagg$, for ultralight axion searches, resulting from changing the NS EOS. As in the main text, we note that the BSk22 model potentially has inconsistencies with observed NS cooling curves, although we include it for comparative purposes. (Right) The same but for heavy axion limits.}
\label{fig:compare_EOS}
\end{figure}

In the right panel of Fig.~\ref{fig:EOM_SF}, we show the total energy emitted as axions over the NS lifetime for each EOS model. For each model, the emission drops off steeply at the mass where the direct URCA process turns on. We illustrate the effects each EOS has on our final axion constraints in Fig.~\ref{fig:compare_EOS}, where we show that, with the exception of BSk22, all EOS models give roughly similar constraints. The constraints suggested by BSk22 are significantly weaker since the direct URCA process becomes active in the NS beginning at lower masses compared to other EOS models. In particular, NSs near the peak of the NS mass distribution (discussed later in this section) cool through the direct URCA process under the BSk22 model. We note however, that the analysis of Ref.~\cite{Buschmann:2021juv} suggests inconsistencies between the BSk22 model and the cooling curves of isolated NSs.

We finally note that modifying the mass fraction of light elements in the NS envelope, $\Delta M$, only directly affects the photon luminosity of the NS, therefore modifying only the late-time thermal evolution. These cooling modifications minimally affect the axion luminosity, and we adopt a mass fraction of $\Delta M=10^{-12}M_{\odot}$ in all of our simulations.

\begin{figure}[!htb]
\centering
\includegraphics[width=0.49\textwidth]{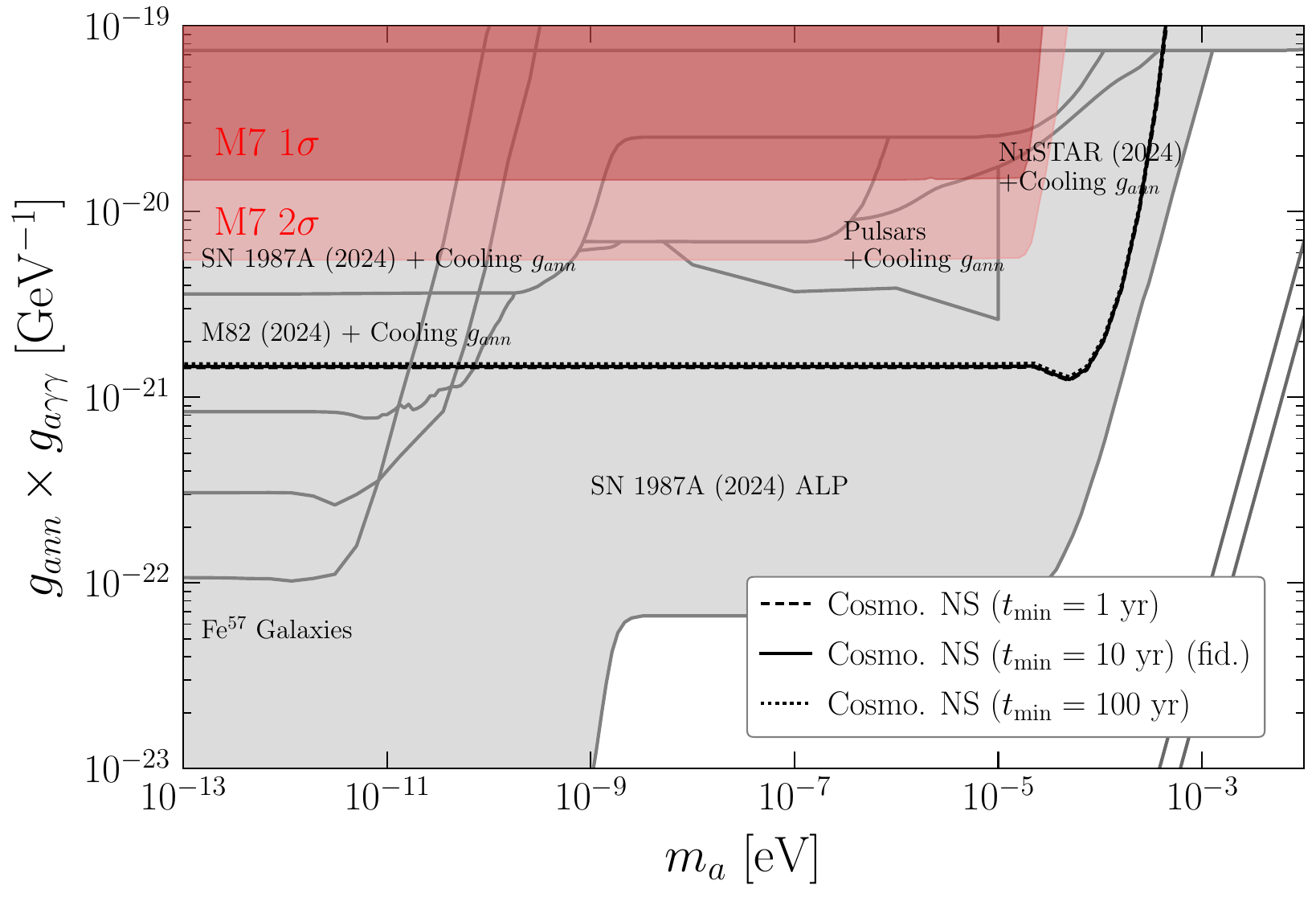}
\includegraphics[width=0.49\columnwidth]{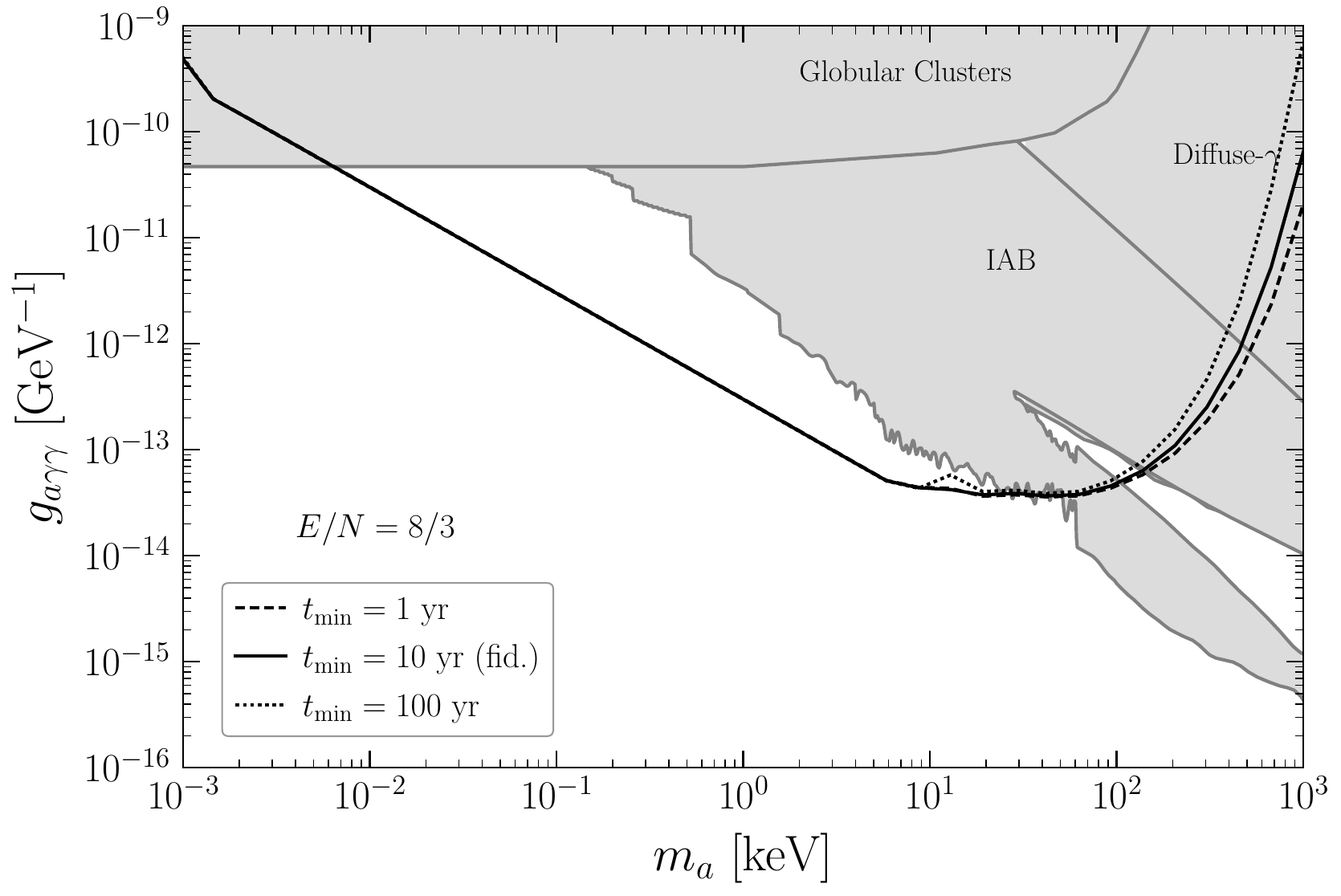}
\vspace{-0.4cm}
\caption{(Left) An illustration of variations on the 95\% upper limit on $\ganngagg$, for ultralight axion searches, resulting from the uncertainty on the minimum age a NS can have to contribute an axion-induced photon signal to the cosmological background searched for in this work. We illustrate the limits using minimum ages of $t_{\rm min} = 1, 10, 100$ years, and show this only has a marginal effect on the upper limit. (Right) The same but for heavy axion limits. Here, at high masses ($m_a \gtrsim100$ keV) the upper limits begin to more obviously deviate due to greater spectral differences in the high-mass regime of heavy axion emission and decay.}
\label{fig:compare_tmin}
\end{figure}

\noindent
{\bf NS Minimum Age---}We additionally vary the minimum age $t_{\rm min}$ at which we sample NS within our NS ensembles, as described in the main Letter. This source of systematic uncertainty comes from several uncertainties in the NS lifetime in the earliest stages of its evolution after SN; these include considerations such as ejecta debris after the initial SN burst which could potentially affect the opacity of the medium through which photon signals would propagate, as well as fallback accretion that could affect the magnetic field inside and around the NS, and other effects (see, \textit{e.g.},~\cite{Yakovlev:2004iq, Igoshev:2021ewx}). While these transient effects resulting from the immediate post-SN environment will eventually pass, it is unclear at what exact timescales they do, with some suggestions indicating at maximum $\mathcal{O}(1)$ months to $\mathcal{O}(1)$ years~\cite{Yakovlev:2004iq, Igoshev:2021ewx}. To account for this, in our fiducial analysis we opt to sample our NS starting from a fixed age $t_{\rm min} = 10$ years, which ensures that our NS are sufficiently evolved to ignore these early transient effects, at the cost of a lower core starting temperature. We examine the effects this uncertainty has on our final results by calculating the resulting upper limits on $\ganngagg$ and $\gagg$ for ultralight and heavy axions, respectively, by varying $t_{\rm min} = 1, 10, 100$ years. As seen in Fig.~\ref{fig:compare_tmin}, our ultralight axion constraints would only marginally improve by picking an order of magnitude lower minimum age, and analogously for the order of magnitude higher minimum age, demonstrating the fact that the exact NS distribution in the first $\sim$100 years of NS lifetimes has a minimal impact when accounting for the entire cosmological ultralight axion population signal. On the other hand, for axions that are heavy enough ($m_a \gtrsim 100$ keV), the variance in $t_{\rm min}$ has a greater impact on the shape of the heavy axion signal, mostly attributed to the fact that axion emission from younger NSs has support at higher energies which are especially important for the heavy axion decay kinematics.

\begin{figure}[!h]
\centering
\includegraphics[width=0.49\textwidth]{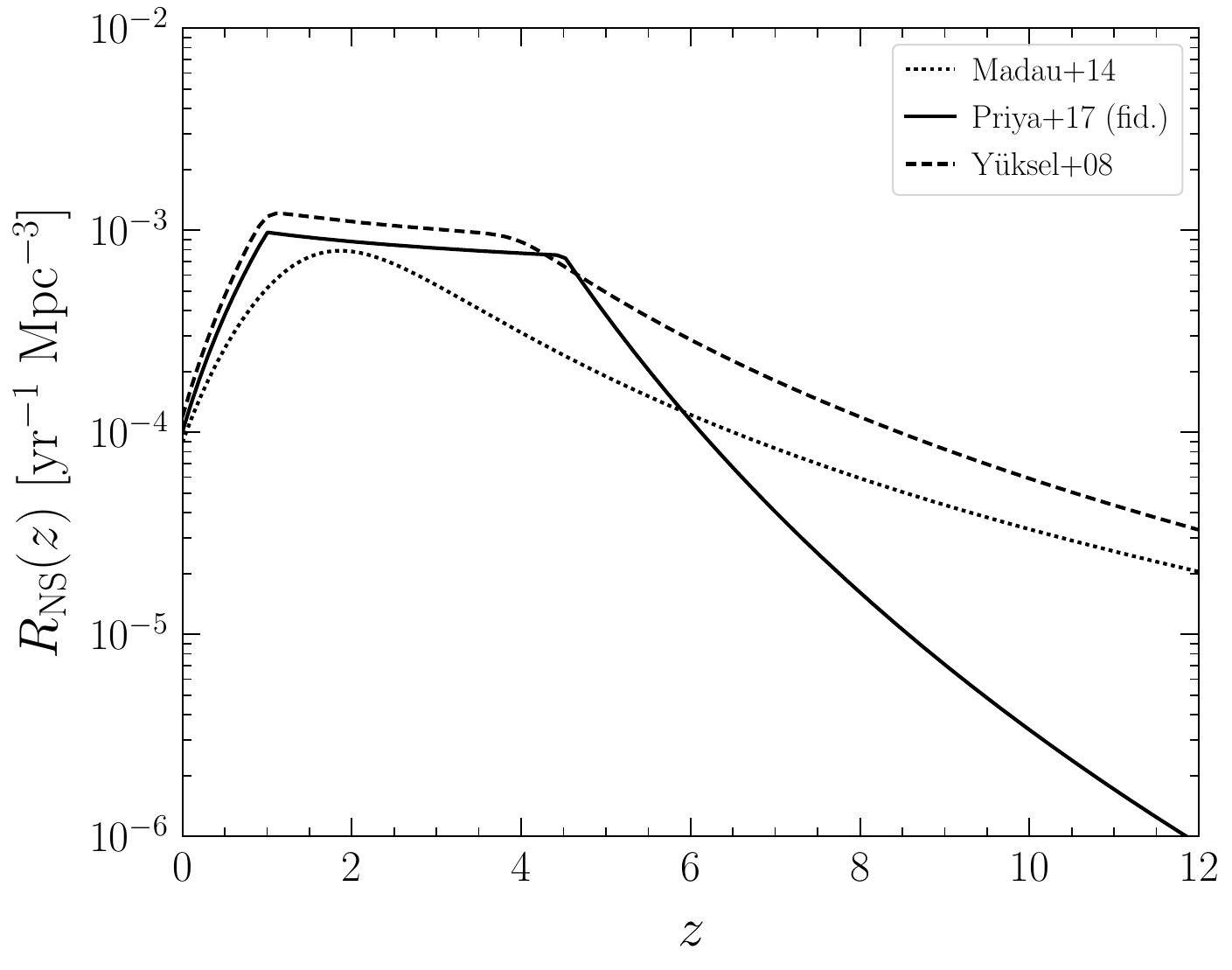}
\vspace{-0.4cm}
\caption{The NS birth rate $R_{\rm NS}(z)$ as a function of redshift $z$. The fiducial rate (black) is based on the SN formation rate, $R_{\rm SN}(z)$, described in~\cite{Priya:2017bmm}, with normalization following~\cite{Lien:2010yb, 2006ApJ...651..142H}. the fraction of SN which result in NS is then used to derive $R_{\rm NS}(z)$. We compare this with other parameterizations, see main text.} 
\label{fig:R_NS}
\end{figure}

\begin{figure}[!htb]
\centering
\includegraphics[width=0.49\textwidth]{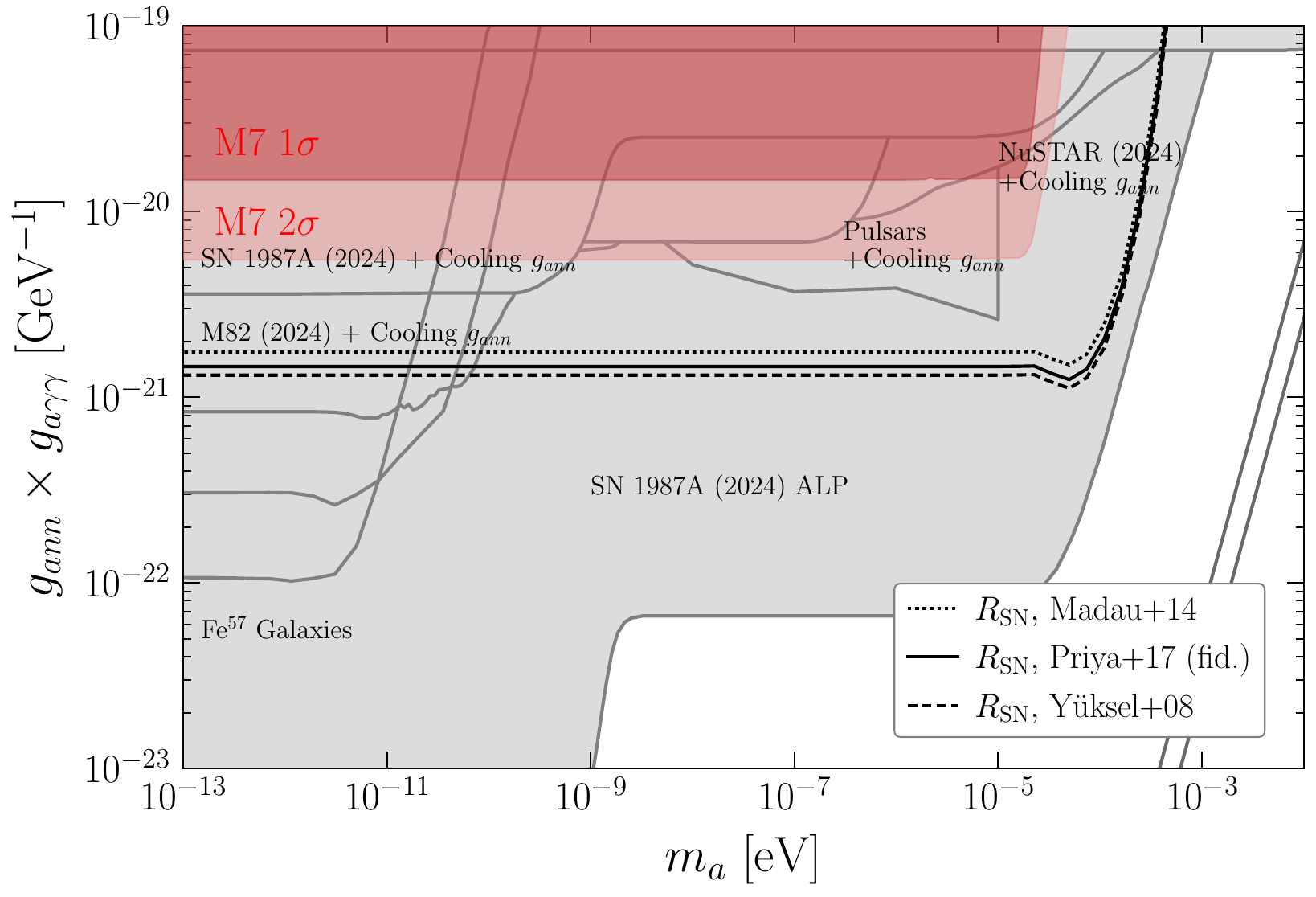}
\includegraphics[width=0.49\columnwidth]{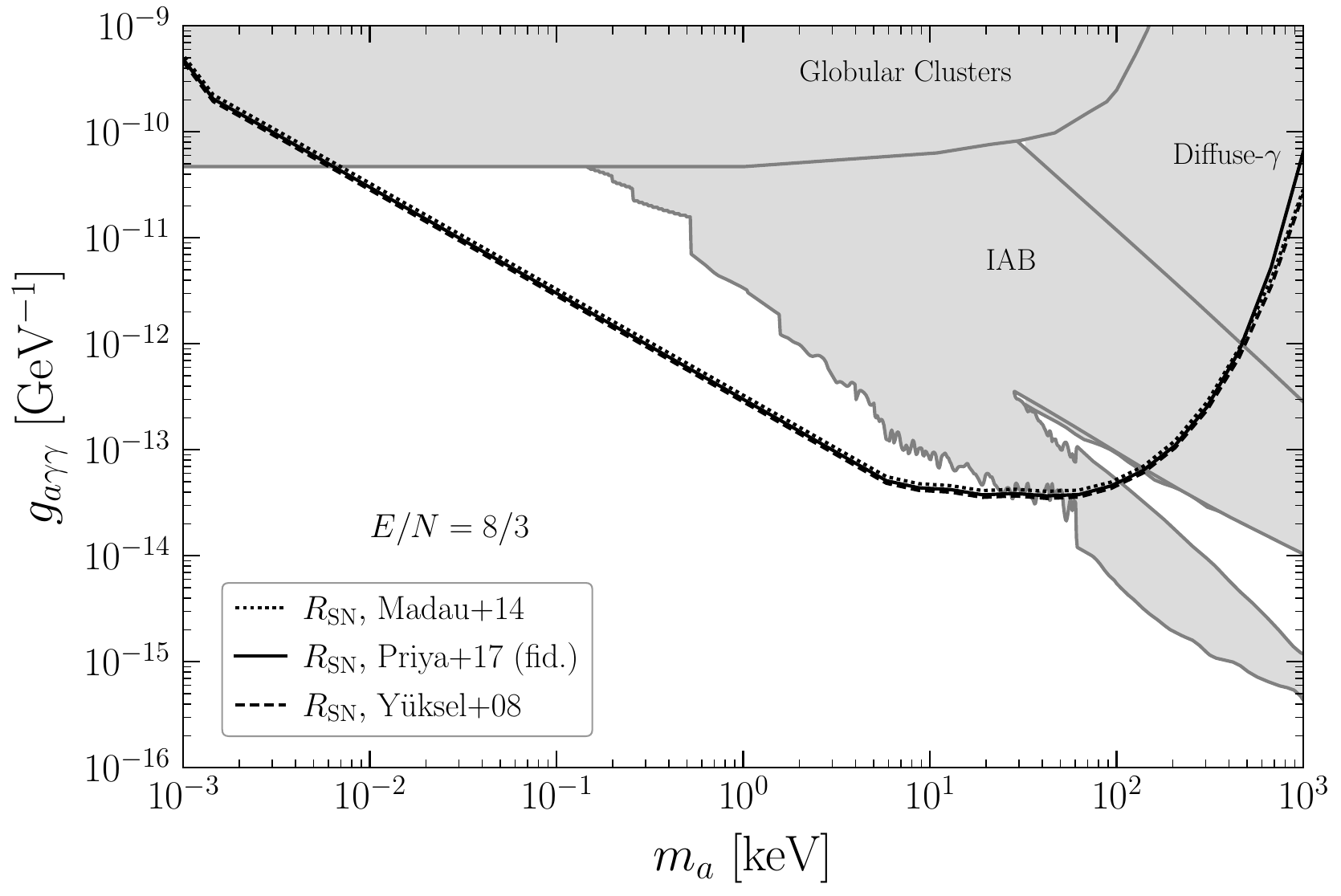}
\vspace{-0.4cm}
\caption{(Left) An illustration of variations on the 95\% upper limit on $\ganngagg$, for ultralight axion searches, resulting from the uncertainty on the NS formation rate. We see that the various parametric forms we explore (see main text) only has a marginal effect on the upper limit. (Right) The same but for heavy axion limits.}
\label{fig:compare_RNS}
\end{figure}

\noindent
{\bf NS Formation Rate---}Another source of uncertainty in our cosmological framework is the rate of NS formation, $R_{\rm NS}(z)$, which is the most crucial cosmological quantity determining the total population axion flux. The uncertainty is tied to the SN formation rate, $R_{\rm SN}(z)$. For our fiducial analysis, we adopt the form in~\cite{Priya:2017bmm} (Priya+17), which in turn utilizes data~\cite{Lien:2010yb} involving diffuse supernova neutrino background measurements. The fiducial $R_{\rm SN}$ model in~\cite{Priya:2017bmm} is essentially a power law with multiple breaks, largely consistent with SFR observations~\cite{2006ApJ...651..142H, Lien:2010yb}. We explore systematics associated with the SN formation rate by also analyzing two other common parameterizations. These include the parametric SFR in~\cite{Yuksel:2008cu} (Y\"uksel+08) which consists of piecewise terms of $(1+z)$ raised to various powers, as well as the canonical SFR in~\cite{Madau:2014bja} (Madau+14). The SFR in these parametric forms are related to the SN rate via normalization coefficients~\cite{Vitagliano:2019yzm}. We compare the $R_{\rm NS}$ from these forms in Fig.~\ref{fig:R_NS}, and the resulting axion coupling constraints in Fig.~\ref{fig:compare_RNS}, which illustrate that this uncertainty has only a minor effect on our overall limits. 

\begin{figure}[!h]
\centering
\includegraphics[width=0.49\textwidth]{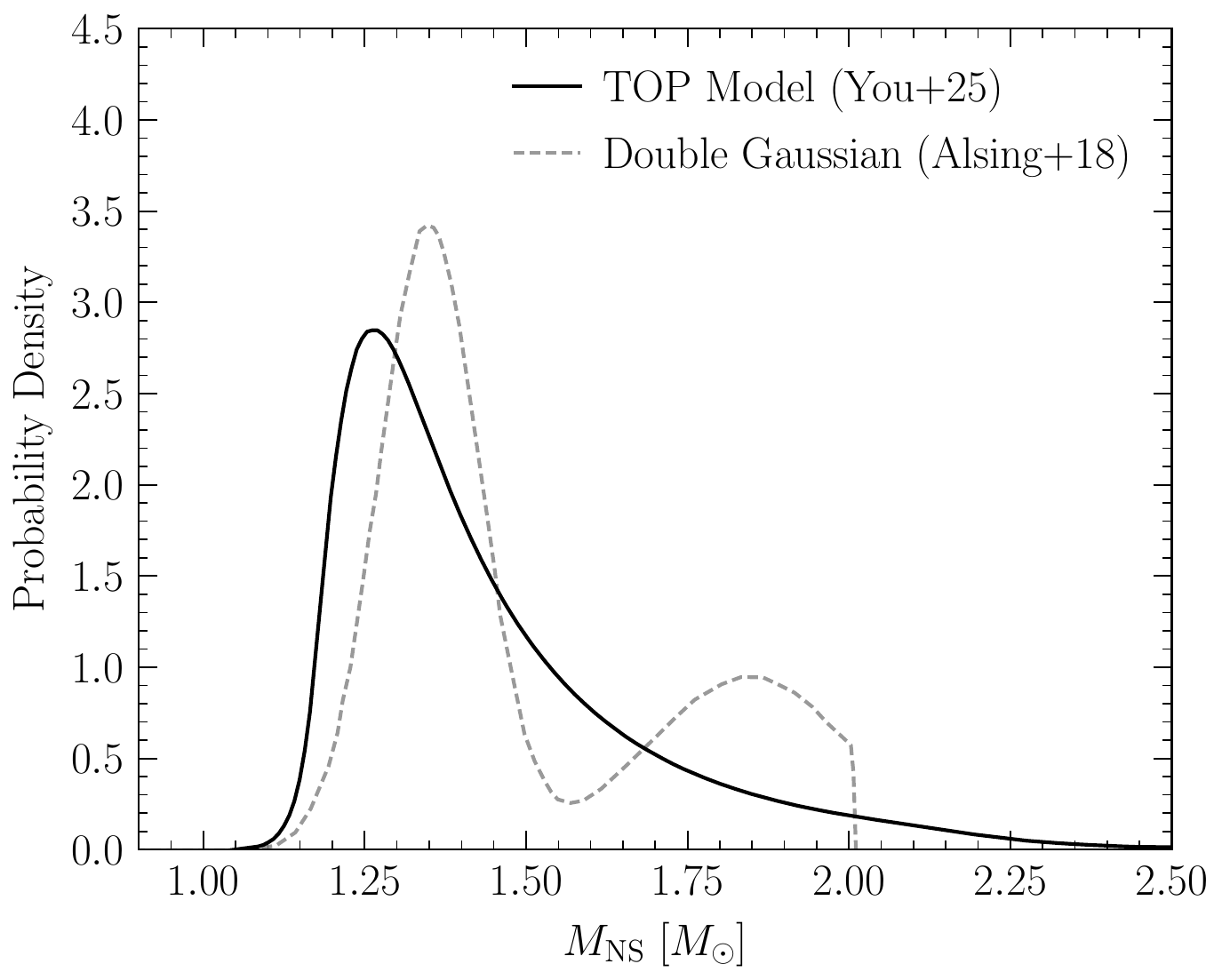}
\includegraphics[width=0.49\textwidth]{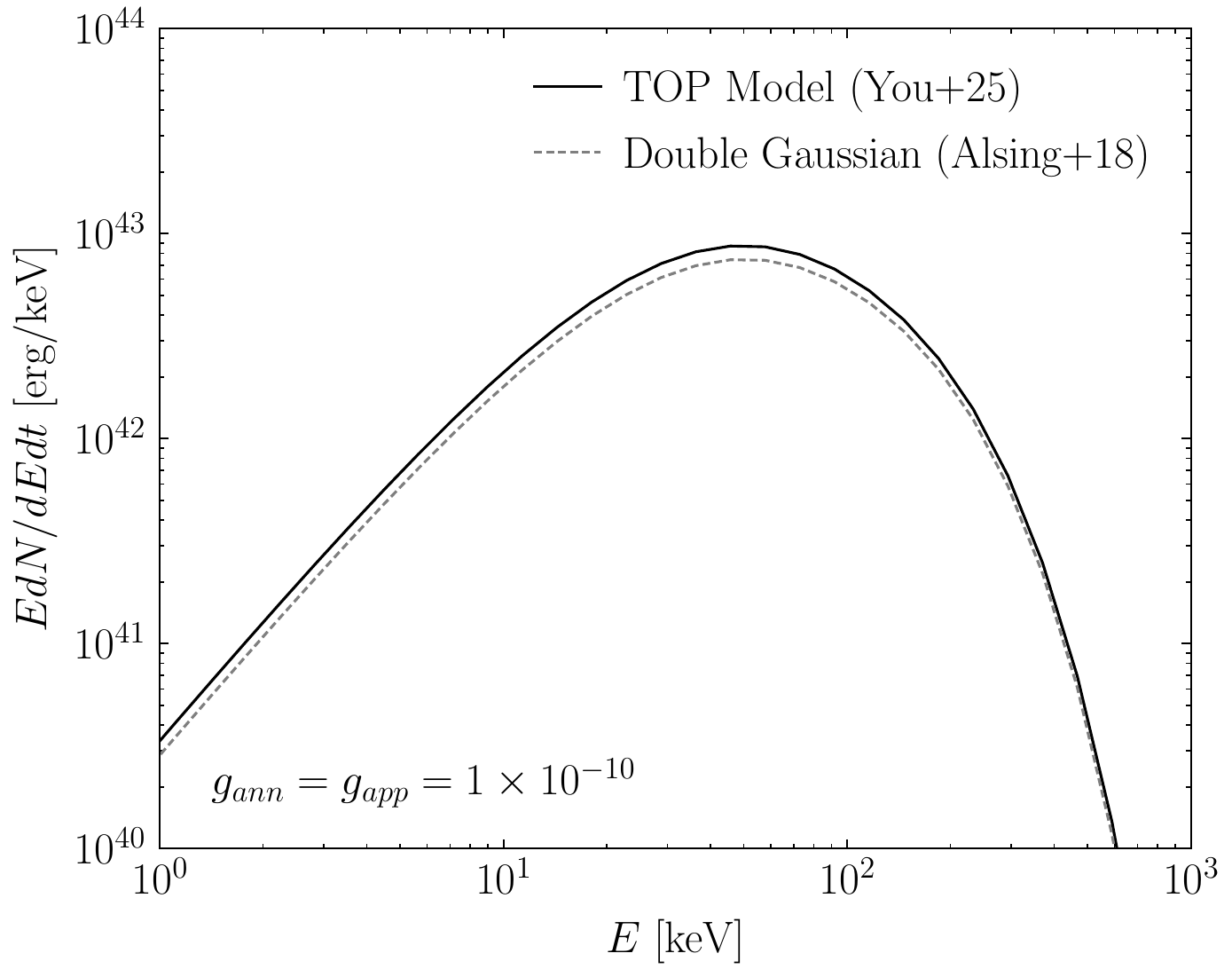}
\vspace{-0.4cm}
\caption{(Left) The fiducial (black) NS mass distribution used to construct our NS ensemble, taken from the favored turn-on power law (TOP) model in~\cite{You:2024bmk} (You+25). We also examine an alternate mass distribution taken from the historical class of double Gaussian models, for which we adopt the favored model in~\cite{Alsing:2017bbc} (Alsing+18). (Right) The NS mass-averaged lifetime axion emissivity for the fiducial \texttt{NSCool} model over the two mass distributions.} 
\label{fig:NS_mass_dist}
\end{figure}

\noindent
{\bf NS Mass Distribution---}As discussed in the main Letter, our NS population is drawn from a fiducial mass distribution, which we adopt from~\cite{You:2024bmk}, which fits an assortment of NS observations spanning radio pulsars, gravitational waves, and X-ray binaries to mass functions. Specifically, we adopt the turn-on power law model (TOP in~\cite{You:2024bmk}) which describes a distribution with a 1.1$M_{\odot}$ turn-on, a unimodal peak at 1.3$M_{\odot}$, and a subsequent power law which allows NS masses above 2$M_{\odot}$. We note that this distribution is largely consistent with the common 1.4$M_{\odot}$ assumption taken for NSs, as in, \textit{e.g.}~\cite{Buschmann:2019pfp}. However, the higher mass tail is in principle important in our analysis since heavier NSs cool faster through the direct URCA process, affecting axion emission through the changes in core temperature. We additionally test the class of double Gaussian models for the NS mass distribution following the parameterization and best fit in~\cite{Alsing:2017bbc}, which suggests a narrow and broad peak around 1.4 $M_{\odot}$ and 1.8 $M_{\odot}$, respectively. We illustrate these two main models and show the effect of the alternate double Gaussian distribution on the average NS emission in Fig.~\ref{fig:NS_mass_dist}, illustrating its subdominant effect on our overall analysis.

\section{Cosmological Regular Star Population Search}
\label{app:RS_cosmo}
We also consider the decay of heavy axions coming from the cosmological \textit{regular} star (RS) population, which we define as all stars that are not compact objects. Our work is similar to that in~\cite{Nguyen:2023czp}, which searched for heavy axion decays sourced from $\gagg$-only processes such as Primakoff production and photon coalescence (\textit{i.e.} $\gamma \gamma \to a$) in specific cosmological star populations. We adopt similar formalism to that in~\cite{Nguyen:2023czp} for the axion production through these $\gagg$ mechanisms in stars, with the important distinction that, unlike~\cite{Nguyen:2023czp} which only considers main sequence stars, we consider axions produced from all main stellar phases of our star populations, produced through the Modules for Experiments in Stellar Astrophysics (MESA) code~\cite{2011ApJS..192....3P, 2013ApJS..208....4P}. This generally allows for significantly higher axion emission at higher energies, which can come from, \textit{e.g.}, accessing hotter temperatures in evolved star populations; these evolved, often massive, stars are known to dominate axion emission over a variety of processes, see, \textit{e.g.},~\cite{Ning:2024eky, Ning:2025tit, Ning:2025kyu, Candon:2024eah}. We remark that in this overall search, as in~\cite{Nguyen:2023czp}, we only consider heavy axion decays as opposed to axion-to-photon conversion, since, unlike the NSs in our main Letter, the magnetic fields of regular star populations as well as the evolution and structure of magnetic fields at galactic and cosmological scales are far less known and subject to more systematic uncertainties (see, \textit{e.g.},~\cite{Mirizzi:2009nq, Mukherjee:2018zzg}).

We briefly outline the properties of the cosmological RS population here, for which more details can be found in~\cite{Nguyen:2023czp}. The stellar axion emission for this population can be seen, at a given redshift $z$, as the combination of various components:
\begin{equation}
    \frac{d\mathcal{N}}{dE}(E, z) = \int dM \, \xi(M) \int dt \, \psi(z(t)) \, \frac{dN}{dE}(E, M, t) \,,
\end{equation}
where $\xi(M)$ is the stellar initial mass function, $\psi(z)$ is the cosmic star formation rate density, and $dN (E, M, t)/dE$ is the axion luminosity spectrum for a single star with mass $M$ at age $t$. As discussed, we include $\gagg$-mediated processes like Primakoff production and photon coalescence for this analysis. Note that we simulate stars with a fixed metallicity set to the cosmic average, $\langle Z\rangle = 0.0175$, following~\cite{Calura:2004jc}.

\begin{figure}[!htb]
\centering
\includegraphics[width=0.47\columnwidth]{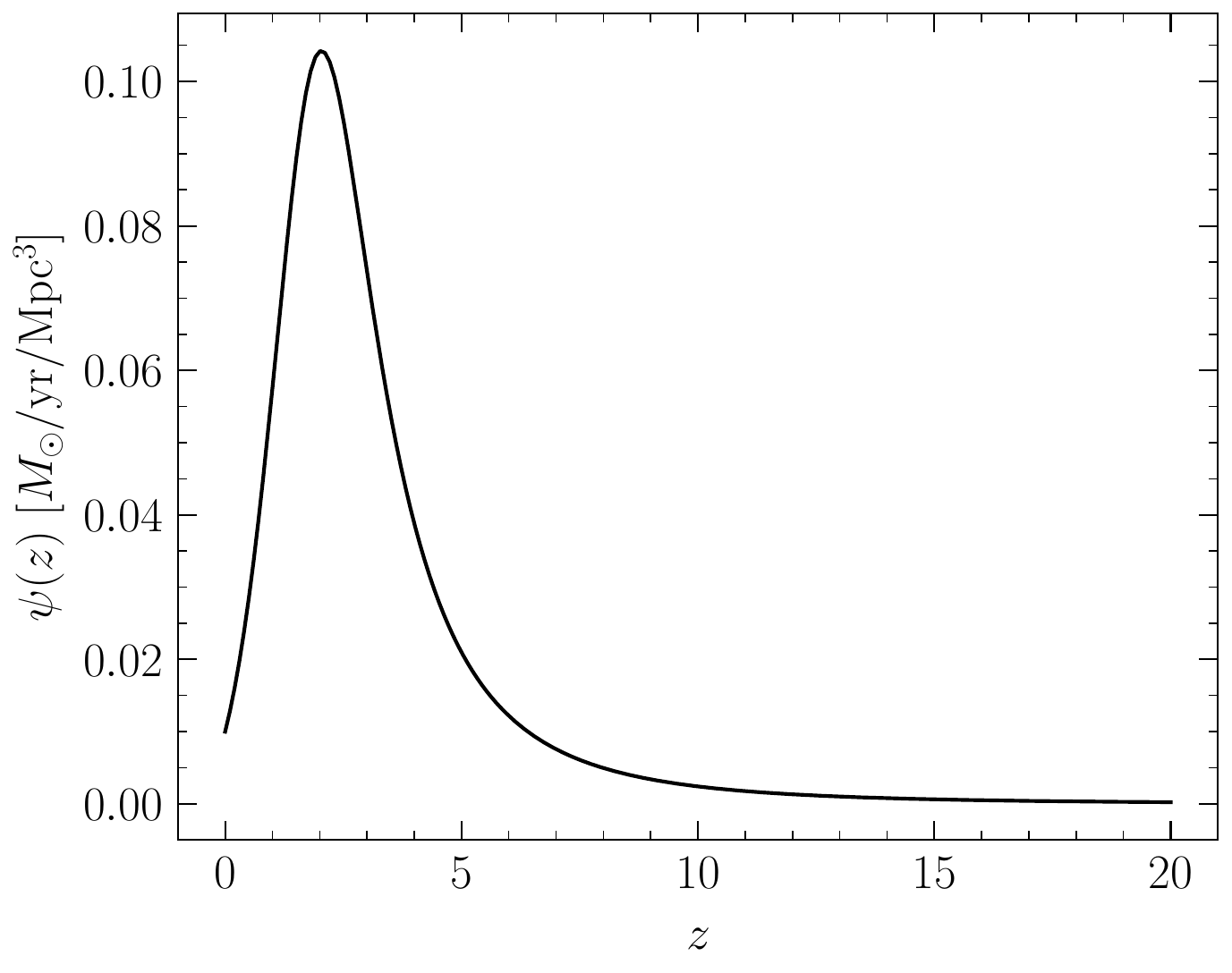}
\includegraphics[width=0.47\columnwidth]{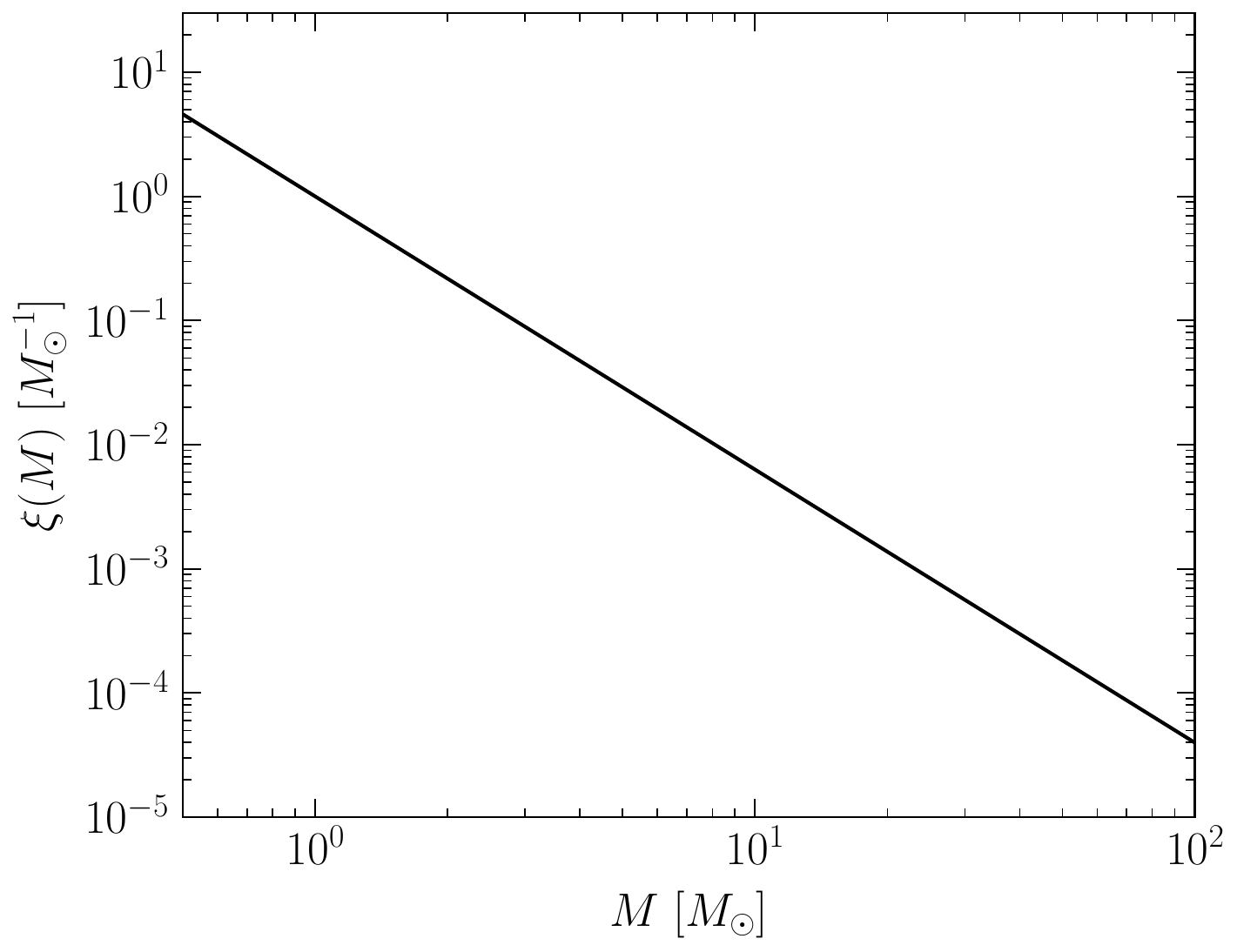}
\caption{(Left) The cosmic star formation density, $\psi(z)$, we use for our fiducial modeling of the cosmological axion signal. The model is taken from~\cite{Madau:2014bja, Madau:2016jbv} (Right) The fiducial Salpeter IMF $\xi(M)$ we take for our fiducial modeling of the cosmological axion signal.}
\label{fig:RS_cosmo}
\end{figure}

For this analysis, we take $\xi(M)$ as the canonical Salpeter initial mass function with $\alpha=-2.35$, although slightly shallower slopes are possible (see, \textit{e.g.}~\cite{2003ApJ...593..258B}). Then, we take the cosmic star formation rate density, $\psi(z)$, to follow the commonly adopted formulation in~\cite{Madau:2014bja, Madau:2016jbv}, which comes from multiwavelength analyses of galaxy survey observations. This reflects the total amount of stars formed per unit time per unit comoving volume at a redshift $z$, in units of, \textit{e.g.}, $M_{\odot}$/yr/Mpc$^3$, and is written as

\begin{equation}
    \psi(z) = \frac{0.01 (1 + z)^{2.6}}{1 + [(1+z)/3.2]^{6.2}} \, \, M_{\odot}{\rm/yr/Mpc}^3 \,.
\end{equation}
We note that $\psi(z)$ peaks at around $z \approx 2$, which also reflects when most stars are formed. We illustrate our fiducial cosmic SFH density and IMF in Fig.~\ref{fig:RS_cosmo}. 

\begin{figure}[!t]
\centering
\includegraphics[width=0.52\textwidth]{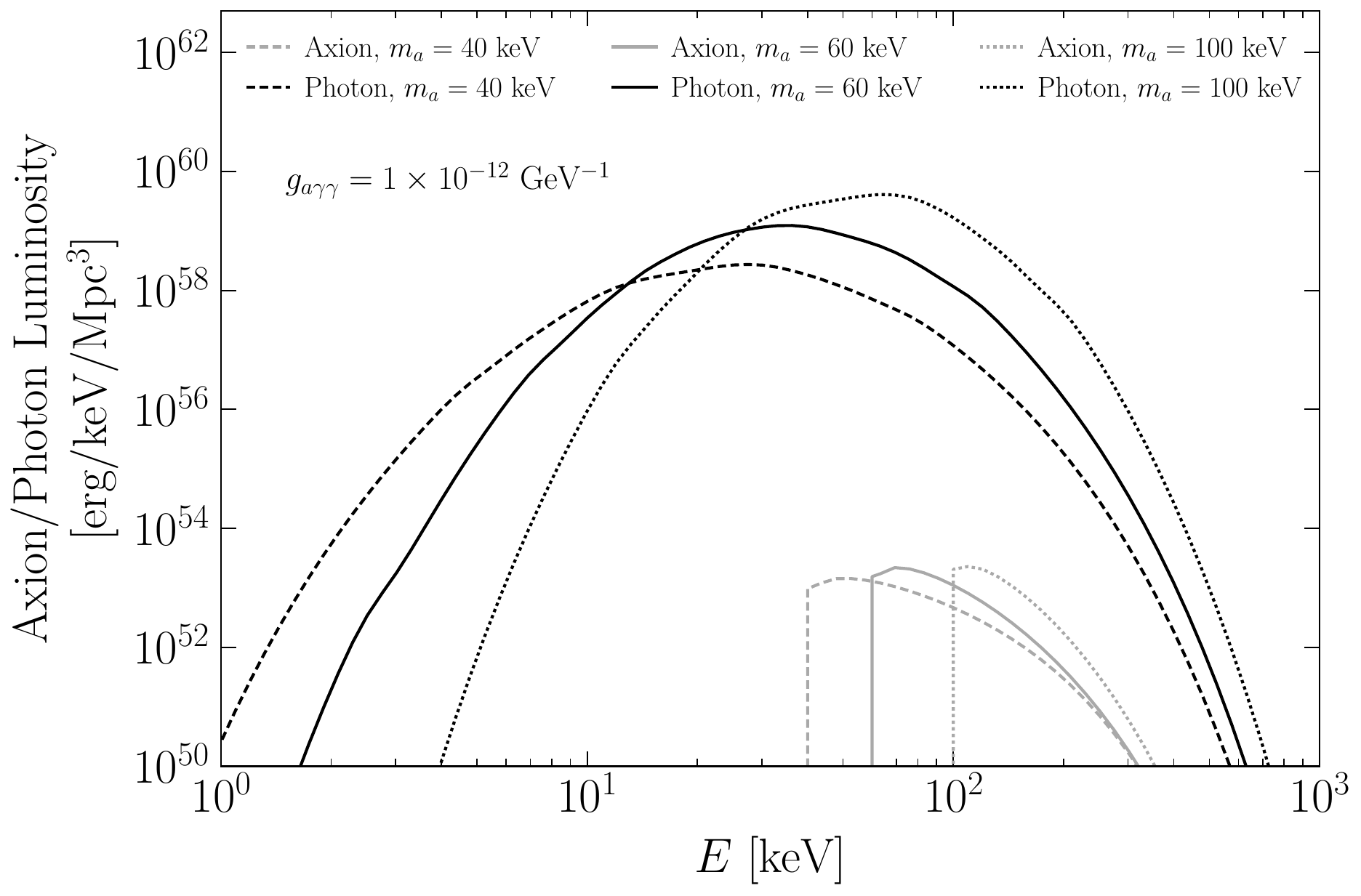}
\includegraphics[width=0.45\textwidth]{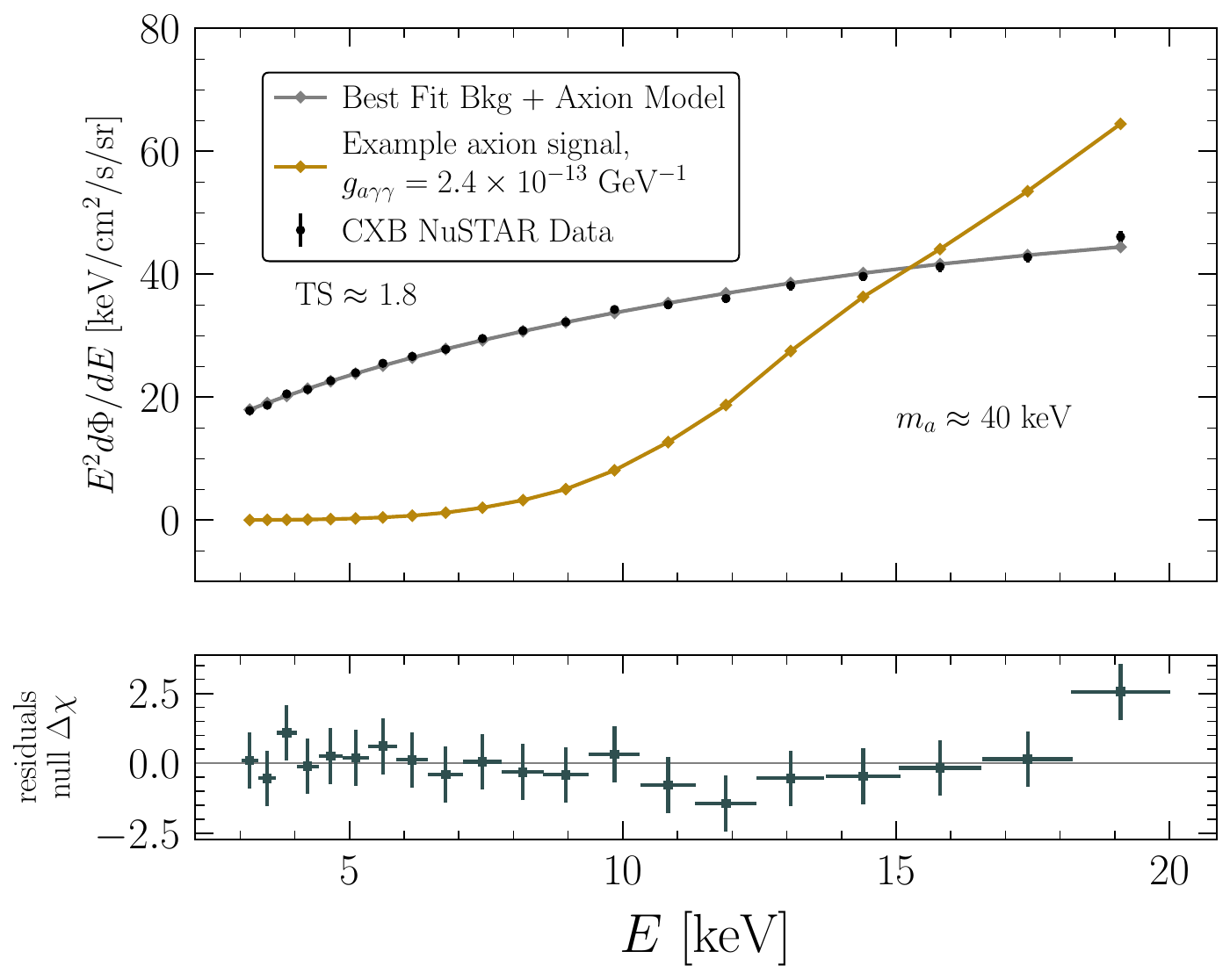}
\vspace{-0.4cm}
\caption{(Left) An illustration of the axion and resulting photon luminosity for our heavy axion-induced RS signal. Here, we choose three different relevant axion masses, broadly centered around where our total signal is strongest (resulting in the most stringent constraints on $\gagg$, see Fig.~\ref{fig:heavy_limits}), for the current epoch and for a coupling of $\gagg = 1\times 10^{-12}$ GeV$^{-1}$. We note that, given the NuSTAR CXB is only characterized up to $\sim$20 keV, only the photon spectrum below $\sim$20 keV is relevant in our final fiducial analysis (although the higher energies are important for HEAO). (Right) An illustration of our heavy axion-induced RS signal compared to fiducial NuSTAR CXB data~\cite{Krivonos:2020qvl}. We show our best-fit axion and background model (gray) in addition to an example axion signal (gold) with the indicated coupling and mass. In the lower panel we also illustrate the residuals of the data compared to the fitted background model under the null hypothesis.}
\label{fig:signal_data_RS}
\end{figure}

The population axion emission at a given redshift $z$, $\frac{d\mathcal{N}}{dE} (E, z)$, is then essentially inserted as a replacement for $R_{\rm NS}(z) \times \frac{dN(E)}{dE}$ in~\eqref{eq:NS_cosmo_heavy}, to reflect the rate of RS population formation and the corresponding total axion emission. The axions then decay according to~\eqref{eq:NS_cosmo_heavy} leading to a total photon-induced signal observable in the CXB, just as in the NS case in the main Letter. We illustrate examples of both the axion emission and resulting photon density signal in the left panel of Fig.~\ref{fig:signal_data_RS}. We remark that, at least for our fiducial NuSTAR search, only the spectra $\lesssim 20$ keV are relevant, although higher energies $\sim$100 keV are also relevant for HEAO, whose search we also discuss below.

Using the same fiducial CXB data from NuSTAR~\cite{Krivonos:2020qvl}, we find no evidence for axions, and set 95\% upper limits on $\gagg$, setting stringent constraints over the mass range $10^{-1}\, {\rm eV} \lesssim m_a \lesssim200 \,{\rm keV}$, seen as the blue solid line in Fig.~\ref{fig:heavy_limits}. We overlap and in some places lead in constraints over those obtained from existing freeze-in bounds~\cite{Langhoff:2022bij}, specifically where $5 \times 10^{1}\, {\rm keV} \lesssim m_a \lesssim 10^{2} \, {\rm keV}$. With HEAO-1 data~\cite{1999ApJ...520..124G}, we achieve slightly stronger bounds (see Fig.~\ref{fig:heavy_limits}) and also exclude new space from $10^{-1} \, {\rm keV} \lesssim m_a \lesssim 5 \times 10^{-1} \, {\rm keV}$, although we again caution the general systematic uncertainties associated with HEAO as discussed in Sec.~\ref{app:analysis}. We note that unlike the results in the main Letter for heavy axions from NS, here we do not rely on any specific UV completion of our axion model in order to constrain $\gagg$. On the other hand, assuming our benchmark GUT model with $E/N = 8/3$, our constraints here are largely subdominant compared to our fiducial NS axion constraint in the main Letter. We illustrate aspects of our data analysis, using our fiducial NuSTAR search, in the right panel of Fig.~\ref{fig:signal_data_RS}.

\end{document}